\documentclass[twocolumn,english,showpacs,preprintnumbers,amsmath,amssymb,floatfix]{revtex4-1}
\usepackage[T1]{fontenc}
\usepackage[latin9]{inputenc}
\usepackage{color}
\usepackage{array}
\usepackage{amstext}
\usepackage{graphicx}
\usepackage{esint}
\usepackage{rotating}
\usepackage[font=small,labelfont=bf]{caption}
\usepackage{xcolor}
\usepackage[colorlinks = true,
linkcolor = magenta,
urlcolor  = blue,
citecolor = red,
anchorcolor = blue]{hyperref}
\makeatletter

\@ifundefined{textcolor}{}
{%
 \definecolor{BLACK}{gray}{0}
 \definecolor{WHITE}{gray}{1}
 \definecolor{RED}{rgb}{1,0,0}
 \definecolor{GREEN}{rgb}{0,1,0}
 \definecolor{BLUE}{rgb}{0,0,1}
 \definecolor{CYAN}{cmyk}{1,0,0,0}
 \definecolor{MAGENTA}{cmyk}{0,1,0,0}
 \definecolor{YELLOW}{cmyk}{0,0,1,0}
 }

\@ifundefined{definecolor}
 {\usepackage{color}}{}
\@ifundefined{definecolor}
 {\usepackage{color}}{}
\makeatother
\usepackage{babel}

\begin{document}

%%%%%%%%%%%%%%%%%%%%%%%%%%%%%%

\title { Analytic derivation of the next-to-leading order proton structure function $F_2^p(x, Q^2)$ based on the Laplace transformation }

\author { Hamzeh Khanpour$^{1,2}$}
\email{Hamzeh.Khanpour@mail.ipm.ir}

\author{ Abolfazl Mirjalili$^{3}$}
\email{A.Mirjalili@yazd.ac.ir}

\author {S. Atashbar Tehrani$^{4}$}
\email{Atashbar@ipm.ir}

\affiliation {
$^{(1)}$Department of Physics, University of Science and Technology of Mazandaran, P.O.Box 48518-78195, Behshahr, Iran     \\
$^{(2)}$School of Particles and Accelerators, Institute for Research in Fundamental Sciences (IPM), P.O.Box 19395-5531, Tehran, Iran    \\
$^{(3)}$Physics Department, Yazd University, P.O.Box 89195-741, Yazd, Iran       \\
$^{(4)}$Independent researcher, P.O.Box 1149-8834413, Tehran, Iran    }

\date{\today}

%
%%%%%%%%%%%%%%%%%%%%%%%%%%%%%%%%%%%%%%%%%%%%%%%%%%%%%%%%%%%%%%%%%%%%%%%%%%%%%%%%%%%%%%%%%%%%%%%%%%%%%%%
\begin{abstract}\label{abstract}
An analytical solution based on the Laplace transformation technique for the Dokshitzer-Gribov-Lipatov-Altarelli-Parisi DGLAP evolution equations is presented at next-to-leading order accuracy in perturbative QCD. This technique is also applied to extract the analytical solution for the proton structure function, $F_2^p(x, Q^2)$, in the Laplace $s$-space. We present the results for the separate parton distributions for all parton species, including valence quark densities, the anti-quark and strange sea parton distribution functions (PDFs), and the gluon distribution.
We successfully compare the obtained parton distribution functions and the proton structure function with the results from {\tt GJR08} [Eur. Phys. J C \textbf{53} (2008) 355-366] and {\tt KKT12} [J. Phys. G \textbf{40} (2013) 045002] parametrization models as well as the $x$-space results using {\tt QCDnum} code. Our calculations show a very good agreement with the available theoretical models as well as the deep inelastic scattering (DIS) experimental data throughout the small and large values of $x$.
The use of our analytical solution to extract the parton densities and the proton structure function is discussed in detail to justify the analysis method considering the  accuracy and speed of calculations. Overall, the accuracy we obtain from the analytical solution using the inverse Laplace transform technique is found to be better than 1 part in 10$^{4}$ to 10$^{5}$.
We also present a detailed QCD analysis of non-singlet structure functions using all available DIS data to perform global QCD fits. In this regard we employ the Jacobi polynomial approach to convert the results from Laplace $s$ space to Bjorken $x$ space. The extracted valence quark densities are also presented and compared to the {\tt JR14}, {\tt MMHT14}, {\tt NNPDF} and {\tt CJ15} PDFs sets. We evaluate the numerical effects of target mass corrections (TMCs) and higher twist (HT) terms on various structure functions, and compare fits to data with and without these corrections.
\end{abstract}

\maketitle
\tableofcontents{}

%
%%%%%%%%%%%%%%%%%%%%%%%%%%%%%%%%%%%%%%%%%%%% Introduction %%%%%%%%%%%%%%%%%%%%%%%%%%%%%%%%%%%%%%%%%%%%%%%%%%%%%%%%%%%
%
\section{Introduction}\label{Introduction}

Dokshitzer-Gribov-Lipatov-Altarelli-Parisi (DGLAP) evolution equations~\cite{Dokshitzer:1977sg,Gribov:1972ri,Lipatov:1974qm,Altarelli:1977zs} are a set of an integro differential equations which can be used to evolve the parton distribution functions (PDFs) to an arbitrary energy scale, Q$^2$.
The solutions of the DGLAP evolution equations will provide us the gluon, valence quark and sea quark distributions inside the nucleon.
Consequently these equations can be used widely as fundamental tools to extract the deep inelastic scattering (DIS) structure functions (SFs) of the proton, neutron, and deuteron to enrich  our current information about the structure of hadrons. The standard procedure to obtain the $x$ dependence of the gluon and quark distributions is to solve numerically the DGLAP equations and compare the solutions with the data in order to fit the PDFs to some initial factorization scale, typically less than the square of the $c$-quark mass Q$_0^2 \approx $ ($m_c^2 \approx  $2 GeV$^2$).
The initial distributions for the gluon and quark are usually determined in a global QCD analysis including a wide variety of DIS data from HERA~\cite{Abramowicz:2016ztw,Abt:2016zth,Abramowicz:2015mha,Aaron:2009kv,Aaron:2009bp,Aaron:2009aa} and COMPASS~\cite{Adolph:2015saz}, hadron collisions at Tevatron~\cite{Aaltonen:2008eq,Abazov:2008ae,Abulencia:2007ez,Abbott:2000ew} fixed-target experiments over a large range of $x$ and Q$^2$, as well as $\nu (\bar{\nu}) {\rm N} \ xF_3$ data from CHORUS and NuTeV~\cite{Onengut:2005kv,Tzanov:2005kr}, and also the data for the longitudinal structure function $F_{\rm L}(x, Q^2)$~\cite{Collaboration:2010ry}.
Finally using the coupled integro-differential DGLAP evolution equations one can find the PDFs at higher energy scale, Q$^2$. For the most recent studies on global QCD analysis, see for instance~\cite{Harland-Lang:2014zoa,Khanpour:2012tk,Alekhin:2012ig,::2014uva,Buckley:2014ana,Ball:2014uwa,Martin:2009iq,JimenezDelgado:2008hf}.

Some analytical solutions of the DGLAP evolution equations using the Laplace transform technique, initiated by Block \textit{et al}., have been reported in recent years~\cite{Block:2010du,Block:2011xb,Block:2010fk,Block:2009en,Block:2010ti,Block:2007pg,Block:2008xc,Zarei:2015jvh,AtashbarTehrani:2013qea,Boroun:2015cta,Boroun:2014dka} with considerable phenomenological success.
In this paper, a detailed analysis has been performed, using repeated Laplace transforms, in order to find an analytical solutions of the DGLAP evolution equations at next-to-leading order (NLO) approximations. We also analytically calculate the individual gluon, singlet and non-singlet quark distributions from the initial distributions inside the nucleon. We present our results for the valence quark distributions $x u_v$ and $x d_v$, the anti quark distributions $x (\overline{d} + \overline{u})$ and  $x\Delta=x(\overline{d}-\overline{u})$, the strange sea distribution $x s = x \overline{s}$, and finally the gluon distribution $x g$. Using the Laplace transform technique, we also extract the analytical solutions for the proton structure function $F_2^p(x, Q^2)$ as the sum of  flavor singlet $F_2^{\rm S}(x, Q^2)$, $F_2^{\rm g}(x, Q^2)$  and flavor non singlet $F_2^{\rm NS}(x, Q^2)$ distributions. The obtained results indicate an excellent agreement with the DIS data as well as those obtained by other methods such as the fit to the $F_2^p$ structure function performed by {\tt KKT12}~\cite{Khanpour:2012tk} and {\tt GJR08}~\cite{Gluck:2007ck}.

In the present work, we also demonstrate once more the compatibility of the Laplace transform technique and the Jacobi polynomial expansion approach at the next-to-leading order and extract the valence quark densities as well as the values of the parameter $\alpha_s(M_Z^2)$ from the QCD fit to the recent DIS data. The effect of target mass corrections (TMCs), which are important especially in the high-$x$ and low-Q$^2$ regions, and the contribution from higher twist (HT) terms are also  considered in the analysis. To quantify the size of these corrections, we evaluate the structure functions at next-to-leading order in QCD, and compare the results with the DIS data used in our PDF fits.

The present paper is organized as follows:
In Sec.~\ref{Sec:Theoretical-formalism}, we provide a brief discussion of the theoretical formalism of the proton structure function $F_2^p(x, Q^2)$ at the NLO approximation of QCD. A detailed formalism to establish an analysis method for the solution of DGLAP evolution using the repeated Laplace transforms for the singlet sector have been presented in Sec.~\ref{Sec:Singlet}.
In Sec.~\ref{Sec:Nonsinglet}, we also review the method of the analytical solution of DGLAP evolution equations based on Laplace transformation techniques for the non singlet sector. In Sec.~\ref{Sec:Proton-structure-function}, we utilize this method to calculate the proton structure function $F_2^p(x, Q^2)$ by Laplace transformation.
We attempt a detailed comparison of our next-to-leading order results with recent results from the literature in Sec.~\ref{Sec:Results}.
We also discuss in detail the use of our analytical solution to justify the analysis method in terms of accuracy and speed.
A completed comparison between the obtained results and available DIS data is also presented in this section.
The application of the Laplace transformation techniques and Jacobi polynomial expansion machinery at the next-to-leading order are described in detail in Sec.~\ref{Sec:Jacobi-polynomials}. The method of the QCD analysis including the PDF parametrization, statistical procedures, and data selection are also presented in this section. The numerical effects of target mass corrections (TMCs) and higher twist terms (HT) on various structure functions are also discussed.
Finally, we give our summary and conclusions in Sec.~\ref{Sec:Summary}. In Appendix~{\bf A}, we render the results for the different splitting functions in the Laplace transformed $s$ space, and Appendix~{\bf B} includes the analytical expression for the coefficient functions of the singlet and gluon distribution in $s$ space.

%
%%%%%%%%%%%%%%%%%%%%%%%%%%%%%%%%%%%%%%%%%%%% Theoretical formalism %%%%%%%%%%%%%%%%%%%%%%%%%%%%%%%%%%%%%%%%%%%%%%%%%%%%%%%%%%%
%
\section{Theoretical formalism}\label{Sec:Theoretical-formalism}

The present DIS and hadron collider data provide the best determination of quark and gluon distributions in a wide range of $x$~\cite{Abramowicz:2015mha,Aaron:2009bp,Aaron:2009aa}. In this article we will be concerned specifically with the proton structure function at next-to-leading order accuracy in perturbative QCD.
In the common $\overline{\rm MS}$ renormalization scheme the $F_2(x,Q^2)$ structure function, extracted from the DIS $ep$ process, can be written as the sum of a flavour singlet $F_{2,{\rm S}}(x,Q^2)$, $F_{2,g}(x,Q^2)$ and a flavour non-singlet $F_{2,{\rm NS}}(x,Q^2)$ distributions in which we will have,
%--------------------------------------
\begin{eqnarray}\label{eq:F2p-SF}
	\frac{F_2(x,Q^2)}{x} & = & \frac{1}{x} \left( F_{2,{\rm S}}(x,Q^2)+ F_{2,g}(x,Q^2) + F_{2,{\rm NS}}(x,Q^2) \right)   \nonumber   \\
	& = & <e^2> C_{2,{\rm S}}(x,Q^2) \ \otimes \ q_{\rm S}(x,Q^2)    \nonumber   \\
	& + & <e^2> C_{2,{\rm g}}(x,Q^2) \ \otimes \ g(x,Q^2)    \nonumber   \\
	& + & C_{2,{\rm NS}}(x,Q^2) \ \otimes \ q_{\rm NS}(x,Q^2)  \,,
\end{eqnarray}
%--------------------------------------
here $g$ and $q_i$ represent the gluon and quark distribution functions respectively.
The $q_{\rm NS}$ stands for the usual flavour non-singlet combination, $xu_v = x (u - \bar{u})$, $xd_v = x (d - \bar{d})$   and $q_{\rm S}$ stand for the flavour-singlet quark distribution, $$x q_{\rm S} = \sum_{i = 1}^{N_f} x (q_i + \bar q_i)\,,$$
where $N_f$ denotes the number of active massless quark flavours. In Equation \eqref{eq:F2p-SF}
the $\otimes$ symbol denotes the convolution integral which turns into a simple multiplication in Mellin $N$ space and $<e^2>$ represents the average squared charge. $C_{2,{\rm S}}$ and $C_{2,{\rm NS}}$ are the common next-to-leading order Wilson coefficient functions~\cite{Vermaseren:2005qc}.
The analytical expression for the  additional next-to-leading order gluonic coefficient function $C_{2,{\rm g}}$ can be found in Ref.~\cite{Vermaseren:2005qc}.
As we already mentioned the gluon and quark distribution functions at the initial state $Q_0^2 $ can be determined by fit to the precise experimental data over a large numerical range for $x$ and Q$^2$. The individual quark and gluon distributions are parametrized with the pre-determined shapes as a standard functional form. This function is given in terms of $x$ and a chosen value for the input scale Q$_0^2$. The gluon distribution $x g(x,Q_0^2)$ is a far more difficult case for PDF parametrizations to obtain precise information due to the small constraints provided by the recent data~\cite{Khanpour:2012tk,Martin:2009iq}.

In the following, we will present our analytic method based on the newly developed Laplace transform technique to determine the non singlet $F_{\rm NS} (x, Q^2)$ and singlet $F_{\rm S} (x, Q^2)$ and $G (x, Q^2)$ structure functions using the input distributions  $F_{\rm NS0} (x, Q_0^2)$,  $F_{\rm S0} (x, Q_0^2)$ and $G_0 (x, Q_0^2)$ at Q$_0^2$ = 2 GeV$^2$. We use the {\tt KKT12}~\cite{Khanpour:2012tk} and {\tt GJR08}~\cite{Gluck:2007ck} input parton distributions to determine the individual parton distribution functions at an arbitrary Q$^2$ > Q$_0^2$, which can be obtained, using the DGLAP evolution equations. Having the parton distribution functions and using the inverse Laplace transform, one can extract the proton structure function $F_2^p(x, Q^2)$ as a function of $x$ at any desired Q$^2$ value.

%
%%%%%%%%%%%%%%%%%%%%%%%%%%%%%%%%%%%%%%%%%%%% Singlet solution %%%%%%%%%%%%%%%%%%%%%%%%%%%%%%%%%%%%%%%%%%%%%%%%%%%%%%%%%%%
%
\section{Singlet solution in Laplace space at the next-to-leading order approximation}\label{Sec:Singlet}

For the most important high energy processes the next-to-leading order approximation is the standard one which we also consider it in our analysis.
The DGLAP evolution equations can describe the perturbative evolution of the singlet $x q_{\rm S}(x,Q^2)$ and gluon $x g(x,Q^2)$ distribution functions.
The coupled DGLAP evolution equations at the next-to-leading order approximation, using the convolution symbol $\otimes$, can be written as
~\cite{Block:2007pg,Block:2008xc}
%--------------------------------------
\begin{eqnarray}\label{eq:singletF}
	&& \frac{4\pi}{\alpha_s(Q^2)} \frac{\partial F_{\rm S}} {\partial lnQ^2}(x,Q^2)
	= F_{\rm S}\otimes\left(P_{qq}^0 + \frac{\alpha_s(Q^2)} {4\pi}P_{qq}^1\right)(x,Q^2)  \nonumber \\
	&& +\ G \otimes \left(P_{qg}^0 + \frac{\alpha_s(Q^2)}{4\pi} P_{qg}^1\right)(x,Q^2)\, ,
\end{eqnarray}
%--------------------------------------
\begin{eqnarray}\label{eq:singletG}
	&& \frac{4\pi}{\alpha_s(Q^2)}\frac{\partial G}{\partial lnQ^2}(x,Q^2)
	= F_{\rm S}\otimes\left(P_{gq}^0 + \frac{\alpha_s(Q^2)} {4\pi}P_{gq}^1\right)(x,Q^2) \nonumber \\
	&& +\ G\otimes\left(P_{gg}^0 + \frac{\alpha_s(Q^2)} {4\pi} P_{gg}^1\right)(x,Q^2)\, ,
\end{eqnarray}
%--------------------------------------
where $\alpha_s(Q^2)$ is the running coupling constant and the splitting functions $P_{ij}^0(x,\alpha_s(Q^2))$ and $P_{ij}^1(x,\alpha_s(Q^2))$ are the Altarelli-Parisi splitting kernels at one and two loop corrections respectively as~\cite{Altarelli:1977zs,Curci:1980uw,Furmanski:1980cm},
%--------------------------------------
\begin{eqnarray}\label{eq:splitting}
	P_{ij}(x,\alpha_s(Q^2)) = P_{ij}^{\rm LO}(x) + \frac{\alpha_s(Q^2)}{2\pi}  P_{ij}^{\rm NLO}(x)\,.
\end{eqnarray}
%--------------------------------------
In the evolution equations, we take N$_f$ = 4 for $m^2_c < \mu^2 < m^2_b$ and N$_f$ = 5 for $m^2_b < \mu^2 < m^2_t$ and adjust the QCD parameter $\Lambda$ at each heavy quark mass threshold, $\mu^2 = m^2_c$ and $m^2_b$. Consequently the renormalized coupling constant $\alpha_s(Q^2)$ can be run continuously when the N$_f$ changes at the $c$ and $b$ mass thresholds~\cite{Botje:2010ay}.

We are now in a position to briefly review the method of extracting the parton distribution functions via analytical solution of DGLAP evolution equation using the Laplace transformation technique.
By considering the variable changes $\nu \equiv \ln (1/x)$ and  $w \equiv \ln (1/z)$, one can rewrite the evolution equations presented in Eqs.(\ref{eq:singletF}) and (\ref{eq:singletG})  in terms of the convolution integrals and  with respect to $v$ and $\tau$ variables as~\cite{Block:2010du,Zarei:2015jvh}
%--------------------------------------
\begin{eqnarray}\label{eq:singletGlaplace}
	&&\frac{\partial {\hat F}_{\rm S}}{\partial \tau}(\nu,\tau)  =    \nonumber \\
	&&\int_0^\nu  \left({\hat K}_{qq}(\nu-w) \ + \ \frac{\alpha_s(\tau)}{4\pi}{\hat K}_{qq}^1(\nu-w)\right)  {\hat F}_{\rm S}(w,\tau) d\,w   \nonumber \\
	&&+ \int_0^\nu  \left({\hat K}_{qg}(\nu-w) \ + \ \frac{\alpha_s(\tau)}{4\pi}{\hat K}_{qg}^1(\nu-w)\right)  {\hat G}(w,\tau) d\,w   \nonumber \\
\end{eqnarray}
%--------------------------------------
\begin{eqnarray}\label{eq:singletflaplace}
	&&\frac{\partial {\hat G}}{\partial \tau}(\nu,\tau) = \nonumber \\
	&&\int_0^\nu  \left({\hat K}_{gq}(\nu-w) \ + \ \frac{\alpha_s(\tau)}{4\pi}{\hat K}_{gq}^1(\nu-w)\right) {\hat F}_{\rm S}(w,\tau) d\,w   \nonumber \\
	&&+ \int_0^\nu \left({\hat K}_{gg}(\nu-w) \ + \ \frac{\alpha_s(\tau)}{4\pi}{\hat K}_{gg}^1(\nu-w)\right)  {\hat G}(w,\tau)  d\,w,   \nonumber \\
\end{eqnarray}
%--------------------------------------
where the Q$^2$ dependence of  above evolution equations is expressed entirely thorough the variable $\tau$ as $\tau(Q^2, Q_0^2) \equiv {1 \over 4\pi} \int_{Q^2_0}^{Q^2} \alpha_s({Q^{\prime}}^2) d \ \ln {Q^{\prime}}^2$. Note that we used the notation $\hat F_{\rm S}(\nu, \tau) \equiv F_{\rm S}(e^{-\nu},Q^2)$ and $\hat G(\nu, \tau) \equiv G(e^{-\nu},Q^2)$.
The above convolution integrals show that using one-loop ${\hat K}_{ij}^0(\nu)  \equiv e^{-\nu} P_{ij}^0 (e^{-\nu})$ and two-loop ${\hat K}_{ij}^1(\nu) \equiv  e^{-\nu}P_{ij}^1 (e^{-\nu})$ kernels where the $i$ and $j$ are a combination of quark $q$ or gluon $g$, one can obtain the singlet ${\hat F}_{\rm S} (\nu,\tau)$  and gluon ${\hat G} (\nu,\tau)$ sectors of distributions.

Defining the Laplace transforms $f(s,\tau) \equiv  {\cal L} [ \hat F_{\rm S}(\nu,\tau); s]$ and $g(s,\tau)\equiv {\cal L}[\hat G(\nu,\tau); s]$ and using this fact that the Laplace transform of a convolution factors is simply the ordinary product of the Laplace transform of the factors, which have been presented in ~\cite{Block:2010du,Block:2011xb}, the Laplace transforms of Eqs.(\ref{eq:singletGlaplace}), and (\ref{eq:singletflaplace}) convert to ordinary first-order differential equations in Laplace space $s$ with respect to variable $\tau$. Therefore we will arrive at
%--------------------------------------
\begin{eqnarray}\label{eq:flaplacepartial1sspace}
	{\partial f\over \partial \tau }(s,\tau) & = & \left( \Phi_f^{\rm LO}(s) \ + \ \frac{\alpha_s(\tau
		)}{4\pi} \Phi_f^{\rm NLO}(s)\right) f(s,\tau) \nonumber \\
	& + & \left(\Theta_f^{\rm LO}(s) \ + \ \frac{\alpha_s(\tau)}{4\pi} \Theta_f^{\rm NLO}(s)\right)g(s,\tau) \,,
\end{eqnarray}
%--------------------------------------
\begin{eqnarray}\label{eq:flaplacepartial2sspace}
	{\partial g\over \partial \tau }(s,\tau) & = &\left( \Phi_
	g^{\rm LO}(s) \ + \ \frac{\alpha_s(\tau)}{4\pi} \Phi_g^{\rm NLO}(s)\right) g(s,\tau) \nonumber \\
	& + &\left(\Theta_g^{\rm LO}(s) \ + \ \frac{\alpha_s(\tau)}{4\pi} \Theta_g^{\rm NLO}(s)\right)f(s, \tau)\, ,
\end{eqnarray}
%--------------------------------------
whose the leading-order splitting functions for the structure function $F_2$, presented in~\cite{Altarelli:1977zs,Floratos:1981hs} in Mellin space, are given by $\Phi_{(f,g)}^{\rm{LO}}$ and $\Theta_{(f,g)}^{\rm{LO}}$ at Laplace $s$ space by

%--------------------------------------
\begin{eqnarray}
	\Phi_f^{\rm{LO}} = 4 - \frac{8}{3}\left(\frac{1}{s + 1} + \frac{1}{s + 2} + 2\left(\gamma_E + \psi (s + 1)\right)\right)\,,
\end{eqnarray}
%--------------------------------------
%--------------------------------------
\begin{eqnarray}
	\Theta_f^{\rm{LO}} = 2 N_f \left(\frac{1}{1 + s} - \frac{2}{2 + s} + \frac{2}{3 + s}\right)\,,
\end{eqnarray}
%--------------------------------------
%--------------------------------------
\begin{eqnarray}
	\Phi_g^{\rm{LO}} &=& 12 \left(\frac{1}{s} - \frac{2 }{1 + s} + \frac{1 }{2 + s} - \frac{1 }{3 + s} - \left(\gamma_E + \psi (s + 1)\right)\right)\nonumber\\ &&+ \frac{33 - 2 N_f}{3}\,,
\end{eqnarray}
%--------------------------------------
and
%--------------------------------------
\begin{eqnarray}
	\Theta_g^{\rm{LO}} = \frac{8}{3} \left(\frac{2}{s} - \frac{2}{1 + s} + \frac{1}{2 + s}\right)\,,
\end{eqnarray}
%--------------------------------------
where the N$_f$ is the number of active quark flavors, $\gamma _E$ is the Euler's constant and  $\psi$ is the digamma function.
The next-to-leading order splitting functions  $\Phi_{(f,g)}^{\rm NLO}$ and  $\Theta_{(f,g)}^{\rm NLO}$ are too lengthy to be include here and we present them in Appendix~{\bf A}.
One can easily determine these next-to-leading order splitting functions in Laplace $s$ space using the next-to-leading order results derived in Ref.~\cite{Altarelli:1977zs,Curci:1980uw,Furmanski:1980cm}.
The leading-order solution of the coupled ordinary first order differential equations in Eqs.(\ref{eq:flaplacepartial1sspace}) and  (\ref{eq:flaplacepartial2sspace}) in terms of
the initial distributions are straightforward. Considering the initial distributions for the gluon, $g^0(s)$, and singlet distributions, $f^0(s)$, at the input scale Q$_0^2 = 2$ GeV$^2$, the evolved solutions in the Laplace $s$ space are given by~\cite{Block:2010du,Block:2011xb},
%--------------------------------------
\begin{eqnarray}\label{eq:gluonandsigma}
	f(s,\tau)& = & k_{ff}(s, \tau) f^0(s) \ + \ k_{fg}(s, \tau) g^0(s) \nonumber \\
	g(s,\tau)& = & k_{gg}(s, \tau) g^0(s) \ + \ k_{gf}(s, \tau) f^0(s),
\end{eqnarray}
%--------------------------------------
The inverse Laplace transform of coefficients  $k$ in the above equations are defined as kernels $K_{ij} (\nu,\tau) \equiv  {\cal L}^{-1} [ k_{ij}(s,\tau); \nu]$ and the input distributions by
$\hat{F}_{\rm S}^0 (\nu) \equiv  {\cal L}^{-1} [ f^0(s); \nu]$  and $\hat{G}^0 (\nu) \equiv  {\cal L}^{-1} [ g^0(s); \nu]$. Then the following decoupled solutions with respect to  $\nu$ and Q$^2$ variables and in terms of the convolutions integrals can be written as,
%--------------------------------------
\begin{eqnarray}\label{eq:convolutions-final}
	\hat F_{\rm S}(\nu, Q^2)  = \int_0^v  && K_{\rm FF} (\nu-w, \tau) \hat F_{\rm S}^0 (w) \,dw  \nonumber \\
	+  &&\int_0^\nu K_{\rm FG}(\nu-w, \tau) \hat G_0(w)\, dw,  \\
	\hat G(\nu,Q^2)  =  \int_0^\nu  && K_{\rm GG}(\nu-w, \tau) \hat G_0(w) \, dw  \nonumber \\
	+ &&\int_0^\nu K_{\rm GF} (\nu-w, \tau) \hat F_{\rm S}^0 (w) \,dw. \label{eq:convolutions-final-1}
\end{eqnarray}
%--------------------------------------
Considering the $\nu \equiv ln(1/x)$, one can finally arrive at the solutions of the DGLAP evolution equations with respect to $x$ and Q$^2$ variables. As we mentioned earlier, the $Q^2$ dependence of the distributions functions $\hat F_{\rm S}(\nu, Q^2)$ and $\hat G(v,Q^2)$  are specified by $\tau$ variable. Clearly knowledge of the initial distributions $F_{\rm S}^0 (x)$ and $G^0 (x)$ at $Q_0^2$ is needed to obtained the distributions at any arbitrary energy scale $Q^2$

Now we intend to extend our calculations to the next-to-leading order approximation for gluon and singlet sectors of unpolarized
parton distributions. In this case, to decouple and to solve DGLAP evolutions in Eqs.(\ref{eq:flaplacepartial1sspace}) and (\ref{eq:flaplacepartial2sspace}) we need an extra
Laplace transformation from $\tau$ space to $U$ space. The $U$ will be a parameter in this new space.
In the rest of the calculation, the $\alpha_s(\tau)/4\pi$ is replaced for brevity by $a(\tau)$. Therefore the solution of the first-order differential equations in Eqs.(\ref{eq:flaplacepartial1sspace}) and (\ref{eq:flaplacepartial2sspace})  can be converted to
%--------------------------------------
\begin{eqnarray}\label{eq:FandF}
	U&&{\cal F}(s,U) -  f^0(s) = \Phi_f^{\rm LO}(s) {\cal F}(s,U) \nonumber \\
	&& + \Phi_f^{\rm NLO}(s) \ {\cal L}[a(\tau)  f(s,\tau); U]      \nonumber\\
	&& + \Theta_f^{\rm LO}(s){\cal G}(s,U) + \Theta_f^{\rm NLO}(s) \ {\cal L}[a(\tau) g(s,\tau);U]\,, \nonumber\\
\end{eqnarray}
%--------------------------------------
%--------------------------------------
\begin{eqnarray}\label{eq:FandG}
	U&&{\cal G} (s,U) - g^0(s) = \Phi_g^{\rm LO}(s){\cal G}(s,U) \nonumber \\
	&& + \Phi_g^{\rm NLO}(s) \ {\cal L}[a(\tau)  g(s,\tau);U]          \nonumber\\
	&& + \Theta_g^{LO}(s) {\cal F}(s,U) + \Theta_g^{\rm NLO}(s) \ {\cal L}[a(\tau) f(s,\tau);U]\,. \nonumber\\
\end{eqnarray}
%--------------------------------------
We can consider a very simple parametrization for $a(\tau)$ as $a(\tau) = a_0$ .
Generally to do a more precise calculation at the next-to-leading order approximation, one can consider the following expression for the $a(\tau)$ as~\cite{Zarei:2015jvh}
%--------------------------------------
\begin{eqnarray}\label{eq:atau}
	a(\tau)  \approx a_0+ a_1e^{-b_1\tau}\,.
\end{eqnarray}
%--------------------------------------
This expansion involves excellent accuracy to a few parts in $10^4$. Using $a(\tau)$ defined in the above equation and the conventions which were presented in~\cite{Block:2010du,Block:2011xb}, the following simplified notations for the splitting functions in $s$ space can be introduced by:
%--------------------------------------
\begin{eqnarray}\label{eq:spliLOandNLO}
	\Phi_{f,g}(s)\equiv \Phi_{f,g}^{\rm LO}(s) + a_0 \Phi_f^{\rm NLO}(s), \nonumber \\
	\Theta_{f,g}(s)\equiv \Theta_g^{\rm LO}(s) + a_0 \Theta_g^{\rm NLO}(s)\;.
\end{eqnarray}
%--------------------------------------
Equations.(\ref{eq:FandF}) and (\ref{eq:FandG}) can be solved simultaneously to get the desired coupled algebraic equations for singlet ${\cal F}(s,U)$ and gluon ${\cal G}(s,U)$ distributions arriving at,
%--------------------------------------
\begin{eqnarray}\label{eq:FU}
	&&\left[ U -\Phi_f(s)\right]{\cal F}(s,U)- \Theta_f(s){\cal G}(s,U) =    \nonumber \\
	f^0(s) && + a_1\left[\Phi_f^{\rm NLO}(s){\cal F}(s, U + b_1)
	+ \Theta_f^{\rm NLO}(s) {\cal G}(s, U + b_1)\right]\,,   \nonumber\\
\end{eqnarray}
%--------------------------------------
%--------------------------------------
\begin{eqnarray}\label{eq:GU}
	&&-\Theta_g(s){\cal F}(s,U) + \left[ U - \Phi_g(s)\right]{\cal G}(s,U) =   \nonumber \\
	g^0(s) && + a_1\left[\Theta_g^{\rm NLO}(s){\cal F}(s, U + b_1)
	+ \Phi_g^{\rm NLO}(s){\cal G}(s, U + b_1)\right]\,.    \nonumber \\
\end{eqnarray}
%--------------------------------------
The simplified solutions of above equations can be obtained by setting $a_1 = 0$ in Eq.\eqref{eq:atau}. For $a(\tau) = a_0$, the Eqs.(\ref{eq:FU}) and (\ref{eq:GU}) lead us to,
%--------------------------------------
\begin{eqnarray}\label{eq:FUa1=0}
	\left[ U -\Phi_f(s)\right]{\cal F}_1(s,U)- \Theta_f(s){\cal G}_1(s,U) =  f^0(s) \,,
\end{eqnarray}
%--------------------------------------
%--------------------------------------
\begin{eqnarray}\label{eq:GUa1=0}
	- \Theta_g(s){\cal F}_1(s,U) + \left[ U - \Phi_g(s)\right]{\cal G}_1(s,U) =  g^0(s) \,.
\end{eqnarray}
%--------------------------------------
One can easily solve these equations and extract the ${\cal F}_1(s,U)$ and ${\cal G}_1(s,U)$ distributions. The results are clearly based on the input quarks $f^0(s)$ and $g^0(s)$ gluon distribution functions at Q$_0^2$. Using the Laplace transform technique, it is possible to go back from $U$ space to $\tau$ space, leading to the desired $f(s,\tau)$ and $g(s,\tau)$ expressions.
The complete solutions of Eqs.~(\ref{eq:FU}) and (\ref{eq:GU}) can be obtained via iteration processes. The iteration can be continued to any required order but we will restrict ourselves to getting a sufficient convergence of the solutions. Our results show that the second order of iterations is sufficient to get a reasonable convergence. Using the iterative solution of Eqs.~(\ref{eq:FU}) and (\ref{eq:GU}) and  the inverse Laplace transform technique to get back from  $U$ space to $\tau$ space, the following expressions for the singlet and gluon distributions can be obtained~\cite{Block:2010du,Block:2011xb,AtashbarTehrani:2013qea}:
%--------------------------------------
\begin{eqnarray}\label{eq:NLODGLAPSspace}
	f(s,\tau) & = & k_{ff}(a_1, b_1, s, \tau) \, f^0(s) + k_{fg}(a_1, b_1, s, \tau) \, g^0(s)\,, \nonumber \\
	g(s,\tau) & = & k_{gg}(a_1, b_1, s, \tau) \, g^0(s) + k_{gf}(a_1, b_1, s, \tau) \, f^0(s)\,, \nonumber \\
\end{eqnarray}
%--------------------------------------
The analytical expressions for the next-to-leading order approximation of coefficients $k_{ff}$, $k_{fg}$, $k_{gf}$ and $ k_{gg}$ up to the desired steps of iteration are given in Appendix~{\bf B}.
Using Laplace inversion in Eq.~\eqref{eq:NLODGLAPSspace} from $s$ to $\nu$ space, we can arrive to the decoupled solutions ($\nu$, $\tau$) space as the result of convolution defined by the Eqs.~\eqref{eq:convolutions-final} and \eqref{eq:convolutions-final-1}.

As a brief description, we have used the Laplace transform algorithm presented in Refs.~\cite{Block:2009en,Block:2010ti} for the numerical inversion of Laplace transformations and convolutions to obtain the required parton distribution functions. The analytical result at the LO approximation is given by Eq.~\eqref{eq:gluonandsigma}. Employing the iterative numerical method through Eq.~\eqref{eq:FandF}  -\eqref{eq:GUa1=0}, up to desired order to achieve a sufficient convergence, will yield us the analytical expressions for the patron densities in $s$ space at the NLO approximation given by Eq.~\eqref{eq:NLODGLAPSspace}.
To return the distributions to the $\nu$ space we need to convolution integral, Eqs.~\eqref{eq:convolutions-final} and \eqref{eq:convolutions-final-1} in both LO and the NLO approximations.
The Q$^2$ dependence of the solutions are determined  by the $\tau$ variable and recalling that $\nu \equiv \ln (1/x)$, the solutions can be transformed back into the usual $x$ space. Consequently, one can obtain the singlet and gluon distributions as $F_S(x,Q^2)$ and $G(x,Q^2)$ respectively.

We have used the numerical Laplace transform algorithm presented in Refs.~\cite{Block:2009en,Block:2010ti} for the numerical inversion of Laplace transformations and convolutions to obtain the parton distribution functions and structure function in $x$ and $Q^2$ space.

%
%%%%%%%%%%%%%%%%%%%%%%%%%%%%%%%%%%%%%%%%%%%% Nonsinglet solution %%%%%%%%%%%%%%%%%%%%%%%%%%%%%%%%%%%%%%%%%%%%%%%%%%%%%%%%%%%
%
\section{Non-singlet solution in Laplace space at the next-to-leading order approximation}\label{Sec:Nonsinglet}

Here we wish to extend our calculations to the next-to-leading order approximation for the non singlet sector of the parton distributions.
For the non singlet distribution $F_{\rm NS}(x,Q^2)$, one can schematically write the logarithmic derivative of $F_{\rm NS}$ as a convolution of non-singlet distribution $F_{\rm NS}(x,Q^2)$ with the non-singlet splitting functions, $p_{qq}^{\rm LO, NS}$ and $p_{qq}^{\rm NLO, NS}$ \cite{Altarelli:1977zs,Curci:1980uw,Furmanski:1980cm}. Therefore the next-to-leading order contributions for the $F_{\rm NS}(x,Q^2)$ can be written as:
%--------------------------------------
\begin{eqnarray}\label{eq:nonsinglet}
	\frac{4\pi}{\alpha_s(Q^2)} && \frac{\partial F_{\rm NS}}{\partial ln Q^2} (x,Q^2) \nonumber \\
	&& = F_{NS}\otimes \left(p_{qq}^{\rm LO, NS} + \frac{\alpha_s(Q^2)}{4\pi}p_{qq}^{\rm NLO, NS}\right)(x,Q^2)\,. \nonumber \\
\end{eqnarray}
%--------------------------------------
Again changing to the required variable, $\nu \equiv \ln(1/x)$, and going to the Laplace space $s$, we arrive at the simple solution as,
%--------------------------------------
\begin{eqnarray}\label{eq:nonsinglet-laplace}
	\frac{\partial \hat{F}_{\rm NS}}{\partial\tau}(\nu,\tau) &&   =  \int_0^\nu  \left (p_{qq}^{\rm LO, NS}(\nu-w)\right.   +  \nonumber \\
	&& \left.\frac{\alpha_s(\tau)}{4\pi}  p_{qq}^{\rm NLO, NS}(\nu-w) \right)  \hat{F}_{\rm NS}(w,\tau)e^{-(\nu-w)}\,dw \,. \nonumber \\
\end{eqnarray}
%--------------------------------------
Going to Laplace $s$ space, we can obtain the first-order differential equations in Laplace space $s$ with respect to the $\tau$ variable for the non-singlet distributions
$f_{\rm NS}(s,\tau)$:
%--------------------------------------
\begin{eqnarray}\label{eq:nonsinglet-laplace-space}
	\frac{\partial f_{\rm NS}}{\partial \tau}(s,\tau) = \left ( \Phi_{\rm NS}^{\rm LO} + \frac{\alpha_s(\tau)}{4\pi}\Phi_{\rm NS,qq}^{\rm NLO} \right ) f_{\rm NS}(s,\tau)\,.
\end{eqnarray}
%--------------------------------------
The above equation has a very simplified solution,
%--------------------------------------
\begin{eqnarray}\label{eq:solve-nonsinglet}
	f_{\rm NS}(s,\tau) = e^{\tau\Phi_{\rm NS}(s)}f^0_{\rm NS}(s)\,,
\end{eqnarray}
%--------------------------------------
where $\Phi_{\rm NS}(s)$ is contains  the next-to-leading order contributions of the splitting functions at $s$ space, defined as
%--------------------------------------
\begin{eqnarray}\label{eq:fi-nonsinglet}
	\Phi_{\rm NS}(s)\equiv\Phi_{\rm NS}^{\rm LO}(s)+\frac{\tau_2}{\tau}\Phi_{\rm NS,qq}^{\rm NLO}(s)\,.
\end{eqnarray}
%--------------------------------------
The evaluation of $\Phi_{\rm NS, qq}^{\rm NLO}(s) = {\cal L} \left[e^{-\nu}p_{qq}^{\rm NLO, NS}(e^{-\nu}); s \right]$ is straightforward but too lengthy to present here.
The analytical results for the unpolarized splitting functions in the transformed Laplace $s$ space at the next-to-leading order approximation are given in Appendix~{\bf A}.
The Q$^2$ dependence of the evolution equations is represented by $\tau$ at the leading order approximation and by $\tau_2$ at the next-to-leading order approximation which the latter one defined as~\cite{Block:2010du,Block:2011xb,AtashbarTehrani:2013qea},
%--------------------------------------
\begin{eqnarray}\label{eq:tau2}
	\tau_2 \equiv \frac{1}{4\pi} \int_0^{\tau} \alpha(\tau^{\prime}) d\tau^{\prime} \ = \ (\frac{1}{4\pi})^2 \int_{Q_0^2}^{Q^2} \alpha_s^2 (Q^{\prime 2}) \ d \ln Q^{\prime 2}\,. \nonumber \\
\end{eqnarray}

Since all parts of the current analysis are done at the next-to-leading order approximation, we should use the $\tau_2$ variable as well. However to simplify in notation, the $\tau$ variable is used insteadly through out the whole paper.

Similar to the singlet case, any non-singlet solution, $F_{\rm NS}(x, Q^2)$, can be obtained using the non-singlet kernel $K_{\rm NS} \equiv {\cal L}^{-1}[e^{\tau \Phi_{\rm NS}(s)}; \nu]$ which is defined by,
%--------------------------------------
\begin{eqnarray}\label{eq:convolation-nonsinglet}
	\hat{F}_{\rm NS}(\nu,\tau) = \int_0^\tau K_{\rm NS}(\nu-w)\hat{F}_{\rm NS}^0 (w) dw\,.
\end{eqnarray}
%--------------------------------------
Using again the appropriate change of variable, $\nu \equiv ln(1/x)$, the solution of Eq.\eqref{eq:convolation-nonsinglet} can be converted to the usual $(x, Q^2)$ space. The iterative numerical method of Laplace transformations at the NLO approximation is followed by the convolutions, based on  Eqs.~\eqref{eq:convolutions-final} and \eqref{eq:convolutions-final-1}.
For the numerical inversion of Laplace transformations and convolutions to obtain the appropriate PDFs and SF in $x$ and $Q^2$ space, we again used the numerical inversion routine presented in Refs.~\cite{Block:2009en,Block:2010ti}.

%
%%%%%%%%%%%%%%%%%%%%%%%%%%%%%%%%%%%%%%%%%%%% Proton structure function %%%%%%%%%%%%%%%%%%%%%%%%%%%%%%%%%%%%%%%%%%%%%%%%%%%%%%%%%%%
%
\section{Proton structure function $F_2^p(x, Q^2)$ in Laplace space}\label{Sec:Proton-structure-function}

We perform here an next-to-leading order analytical analysis for the proton structure function $F_2^p(x, Q^2)$ using the Laplace transform technique.
The result for singlet, gluons and non singlet parton distributions which we obtained in previous sections are used to extract the nucleon structure function.
The next-to-leading order proton structure function $F_2^p(x, Q^2)$ for massless quarks can be written as~\cite{Dokshitzer:1977sg,Gribov:1972ri,Lipatov:1974qm,Altarelli:1977zs}
%--------------------------------------
\begin{eqnarray}\label{eq:F2p-DGLAP}
	F_2(x, Q^2) &=& \sum_{i=1}^{n_f} e_i^2 x  (  C_q (x, \alpha_s) \otimes [ q_i (x, Q^2) + \bar{q}_i (x, Q^2) ] \nonumber \\
	&+& C_g (x, \alpha_s) \otimes g(x, Q^2)  ) \,,
\end{eqnarray}
%--------------------------------------
where $C_q$ and $C_g$ are the next-to-leading order quarks and gluon Wilson coefficients, and $q_i$, $\bar{q}_i$, and $g(x, Q^2)$ are the quark, anti-quark and gluon distributions, respectively.
We exactly follow the method that we introduced before to solve the DGLAP evolution equations analytically to drive the  proton structure function at the  next-to-leading order approximation first in Laplace $s$ space and then in Bjorken $x$ space.
As we already mentioned, only the initial knowledge of singlet $F_{\rm S}^0(x)$, gluon $G^0(x)$, and non singlet $F_{\rm NS}^0(x)$  distributions is required to solve the DGLAP evolution equations via the Laplace transform technique.

For our numerical investigation, we use the  {\tt KKT12}~\cite{Khanpour:2012tk} and
{\tt GJR08}~\cite{Gluck:2007ck} parton distribution functions at Q$_0^2$ = 2 GeV$^2$.
The valence quark distributions $x u_v$ and $x d_v$, the anti-quark distributions $x (\overline{d} + \overline{u})$ and  $x\Delta=x(\overline{d}-\overline{u})$,
the strange sea distribution $x s = x \overline{s}$ and the gluon distribution $x g$ of the {\tt KKT12} and {\tt GJR08} models
are generically parameterized via the following standard functional form
%--------------------------------------
\begin{eqnarray}\label{eq:prton}
	xq = a_q x^{b_q} (1-x)^{c_q}(1 + d_q x^{f_q} + e_q x),
\end{eqnarray}
%--------------------------------------
subject to the constraints that $\int_0^1 u_v \ dx = 2$, $\int_0^1 d_v \ dx = 1$, and the total momentum sum rule
%--------------------------------------
\begin{eqnarray}\label{eq:constrain}
	\int_0^1 x[u_v + d_v + 2(\bar{u} + \bar{d} + \bar{s}) + g] \ dx = 1\,.
\end{eqnarray}
%--------------------------------------
After changing to the variable $\nu \equiv \ln (1/x)$ and using the Laplace transform $q(s) = {\cal L} [e^{-\nu} q(e^{-\nu});s]$, one can easily obtain Eq.\eqref{eq:prton} in Laplace $s$ space,
%--------------------------------------
\begin{eqnarray}\label{eq:prtonlaplacespace}
	q(s) & = & a_q\left(B[1+ c_q, b_q + s] + e_q \ B[1 + c_q, 1 + b_q + s]\right. \nonumber \\
	&& \left. + d_q \ B[1 + c_q, b_q + f_q + s]\right)\,.
\end{eqnarray}
%--------------------------------------
We use the following standard parametrizations in Laplace $s$ space at the input scale Q$_0^2$=2 GeV$^2$
for all parton types $x q_i$, obtained from {\tt GJR08} set of the free parton distribution functions~\cite{Gluck:2007ck}:
%--------------------------------------
\begin{eqnarray}\label{eq:u_v(s)}
	u_v(s) & = & 0.5889 \left( \ {\rm B} [4.7312, 0.3444 + s] \right.      \nonumber  \\
	&-& 0.175\ {\rm B} [4.7312, 0.8444 + s]      \nonumber   \\
	&+&\left. 17.997 \ {\rm B} [4.7312, 1.3444 + s]\right)\,,
\end{eqnarray}
%--------------------------------------
%--------------------------------------
\begin{eqnarray}\label{eq:d_v(s)}
	d_v(s) & = & 0.2585 \left( \ {\rm B} [5.8682, 0.295 1 + s] \right.     \nonumber  \\
	&-& 1.0552 \ {\rm B} [5.8682, 0.7951 + s]     \nonumber  \\
	&+& \left.26.536 \ {\rm B} [5.8682, 1.2951 + s]\right)\,,    \nonumber  \\
\end{eqnarray}
%--------------------------------------
%--------------------------------------
\begin{eqnarray}\label{eq:Delta(s)}
	\bar{d}(s) - \bar{u}(s)  & = & 7.2874 \left( \ {\rm B} [19.756, 1.2773 + s] \right.   \nonumber \\
	&-& 6.3187 \ {\rm B} [19.756, 1.7773 + s]  \nonumber\\
	&+& \left.18.306 \ {\rm B} [19.756, 2.2773 + s]\right)\,,
\end{eqnarray}
%--------------------------------------
%--------------------------------------
\begin{eqnarray}\label{eq:dbar-ubar}
	\bar{d}(s) + \bar{u}(s) & = & 0.2295 \left( \ {\rm B} [9.8819, -0.1573 + s] \right.   \nonumber  \\
	&+& 0.8704 \ {\rm B} [9.8819, 0.3427 + s]    \nonumber  \\
	&+& \left. 8.2179 \ {\rm B} [9.8819, 0.8427 + s]\right)\,,
\end{eqnarray}
%--------------------------------------
%--------------------------------------
\begin{eqnarray}\label{eq:g(s)}
	g(s)& = & 1.3667 \ {\rm B}  [4.3258, -0.105 + s]\,,
\end{eqnarray}
%--------------------------------------
where ${\rm B}$ is the common Euler beta function. The strange quark distribution function is assumed to be symmetric ( $x s = x \overline{s}$ ) and it is proportional to the isoscalar light quark sea which parameterized as
%--------------------------------
\begin{eqnarray} \label{ssbar}
	s(s) = \bar{s} (s) = \frac{k}{2} \ \left(  \bar{d}(s) + \bar{u}(s)  \right)\,,
\end{eqnarray}
%--------------------------------
where in practice $k$ is a constant fixed to $k$ = $0.5$~\cite{Khanpour:2012tk,Gluck:2007ck}.

The proton structure function $F_2^p(x, Q^2)$ in Laplace $s$ space, up to the next-to-leading order approximation,
can be written as
%--------------------------------
\begin{eqnarray}\label{eq:F2pLaplacespacelight}
	{\cal F}_2^{\rm p, light} (s, \tau) =  {\cal F}_2^{\rm S} (s, \tau) + {\cal F}_2^G(s, \tau) + {\cal F}_2^{\rm NS}(s, \tau) \,,
\end{eqnarray}
%--------------------------------
where the flavour singlet ${\cal F}_2^{\rm S}$ and gluon ${\cal F}_2^G$ contribution read
%--------------------------------
\begin{eqnarray}\label{eq:F2pLaplacespace-singlet}
	{\cal F}_2^{\rm S}(s,\tau) & = & \left(\frac{4}{9}2\bar{u}(s,\tau)  +  \frac{1}{9}2\bar{d}(s,\tau)  + \frac{1}{9}2\bar{s}(s,\tau)\right)  \nonumber  \\
	&& \times \left(1 + \frac{\tau}{4\pi}C^{(1)}_q(s) \right) \,,
\end{eqnarray}
%--------------------------------
%--------------------------------
\begin{eqnarray}\label{eq:F2pLaplacespace-gluon}
	{\cal F}_2^{\rm G}(s,\tau)& = &  \frac{2}{9} g(s,\tau)\left(\frac{\tau}{4\pi}C^{(1)}_g(s)\right)\,.
\end{eqnarray}
%--------------------------------
Finally the non-singlet contribution for three active (light) flavours is given by
%--------------------------------
\begin{eqnarray}\label{eq:F2pLaplacespace-nonsinglet}
	{\cal F}_2^{\rm NS}(s,\tau) &=&\left(\frac{4}{9} u_v(s,\tau) + \frac{1}{9}d_v(s,\tau)\right)\left(1+ \frac{\tau}{4\pi}C^{(1)}_q(s)\right)\,, \nonumber\\
\end{eqnarray}
%--------------------------------
where the $C_q^{(1)}(s)$ and $C_g^{(1)}(s)$ are the common next-to-leading order approximation of Wilson coefficients functions, derived in Laplace $s$ space by $c_q(s) = {\cal L} [e^{-\nu} c_q(e^{-\nu}); s]$ and $c_g(s) = {\cal L} [e^{-\nu} c_g(e^{-\nu}); s]$,
%-------------------------------
\begin{widetext}
\begin{eqnarray}
	&& C_q^{(1)}(s)= \nonumber\\
	&& C_F \left(-9-\frac{2 \pi ^2}{3}-\frac{2}{(1+s)^2}+\frac{6}{1+s}-\frac{2}{(2+s)^2}+\frac{4}{2+s}+\right. \nonumber\\
	&&3 \left(\gamma _E+\psi (s+1)\right)+ \frac{2 \left(\gamma _E+\psi (s+2)\right)}{1+s}+\frac{2 \left(\gamma _E+\psi (s+3)\right)}{2+s}+ \nonumber\\
	&& \left.\frac{1}{3} \left(\pi ^2+6 \left(\gamma _E+\psi (s+1)\right){}^2-6 \psi '(s+1)\right)+4 \psi '(s+1)\right)\,,\nonumber\\
\end{eqnarray}
%--------------------------------
%--------------------------------
\begin{eqnarray}
	&& C_g^{(1)}(s) =  \nonumber\\
	&& f \left(\frac{2}{(1+s)^2}-\frac{2}{1+s}-\frac{4}{(2+s)^2}+\frac{16}{2+s}+\frac{4}{(3+s)^2}-\frac{16}{3+s}-\right. \nonumber\\
	&& \left.\frac{2 \left(\gamma _E+\psi (s+2)\right)}{1+s}+\frac{4 \left(\gamma _E+\psi (s+3)\right)}{2+s}-\frac{4 \left(\gamma _E+\psi (s+4)\right)}{3+s}\right)\,.\nonumber\\
\end{eqnarray}
\end{widetext}
Once again the Q$^2$ dependence of  proton structure function in Eq.\eqref{eq:F2pLaplacespacelight} is evaluated by $\tau(Q^2, Q_0^2) \equiv {1 \over 4\pi} \int_{Q^2_0}^{Q^2} \alpha_s({Q^{\prime}}^2) d \, \ln {Q^{\prime}}^2$. The final desired solution of the proton structure functions in Bjorken $x$ space, $F_2^{\rm p, light}(x,Q^2)$, are readily found using the inverse Laplace transform and the appropriate change of variables.

The next-to-leading order contribution of heavy quarks, $F_i^{c,b} (x,Q^2)$, to the proton structure function can be calculated in the fixed flavor number scheme (FFNS) approach~\cite{Khanpour:2012tk,Riemersma:1994hv,Laenen:1992zk,Gluck:2006pm,Gluck:2004fi,Gluck:2008gs,Gluck:1993dpa,Laenen:1992cc} and will yield the total structure functions as $F^{\rm p, total}_2 (x,Q^2) = F_2^{\rm p, light}(x,Q^2) + F_i^{\rm heavy}(x,Q^2)$ where the $F_2^{\rm p, light}(x,Q^2)$ refers to the common $u, \, d, \, s$ (anti) quarks and gluon initiated contributions, and $F_i^{\rm heavy}(x,Q^2) = F_2^c(x,Q^2) + F_2^b(x,Q^2)$  are the charm and bottom quarks structure functions. We should mention that for the  $F^{\rm p, total}_2$ only its light contribution is derived by Laplace transform technique. Its heavy contribution results from the usual Mellin transform technique. In the present analysis we use the {\tt GJR08} values for $m_c$ = 1.30 GeV and $m_b$ = 4.20, GeV which slightly differ from the {\tt KKT12} default values of $m_c$ = 1.41 GeV and $m_b$ = 4.50 GeV.

%--------------------------------
\begin{figure}[htb]
	%\vspace*{0.5cm}
	\begin{center}
		\includegraphics[clip,width=0.45\textwidth]{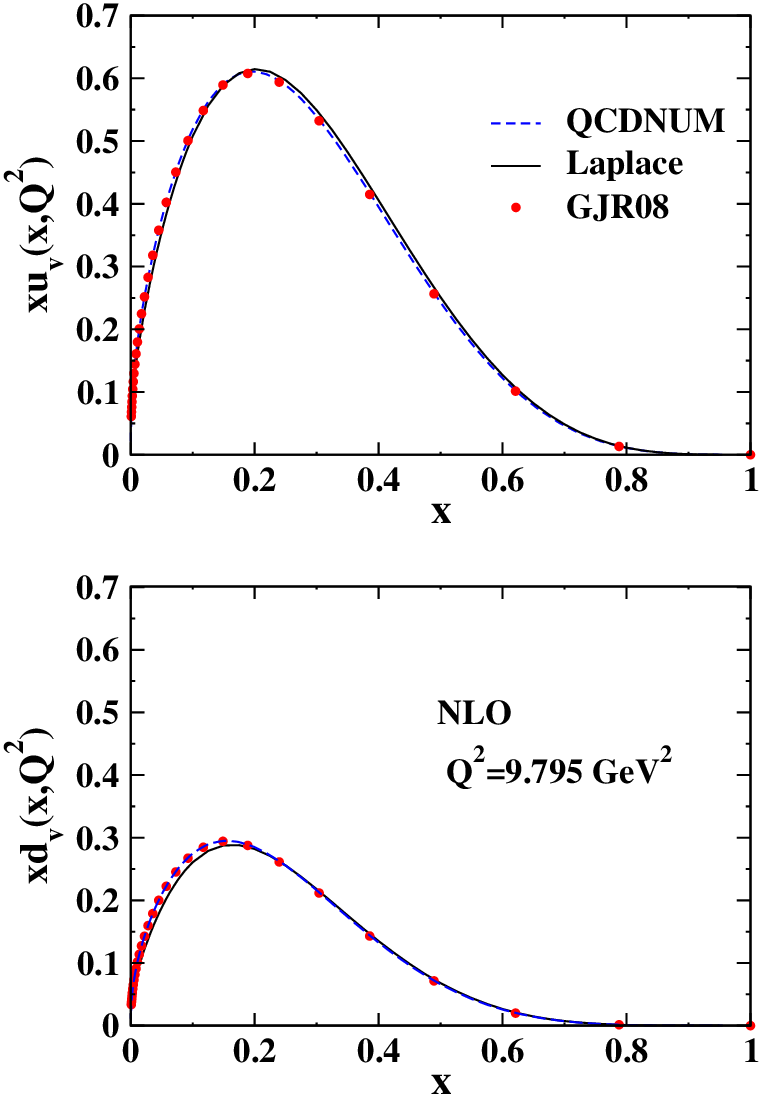}
		%\vspace{-1cm}
		\caption{(Color online) Our results for the non-singlet distribution, $x u_v (x,Q^2)$ and $x d_v (x,Q^2)$, using  Eq.\eqref{eq:convolation-nonsinglet} and  comparison with the global QCD analysis of {\tt GJR08}. The $x$-space results from the QCD evolution package, {\tt QCDnum}, are also presented (dashed line). }\label{fig:nonsingletQ9p795}
	\end{center}
\end{figure}
%--------------------------------

%--------------------------------
\begin{figure}[htb]
	\begin{center}
		\includegraphics[clip,width=0.45\textwidth]{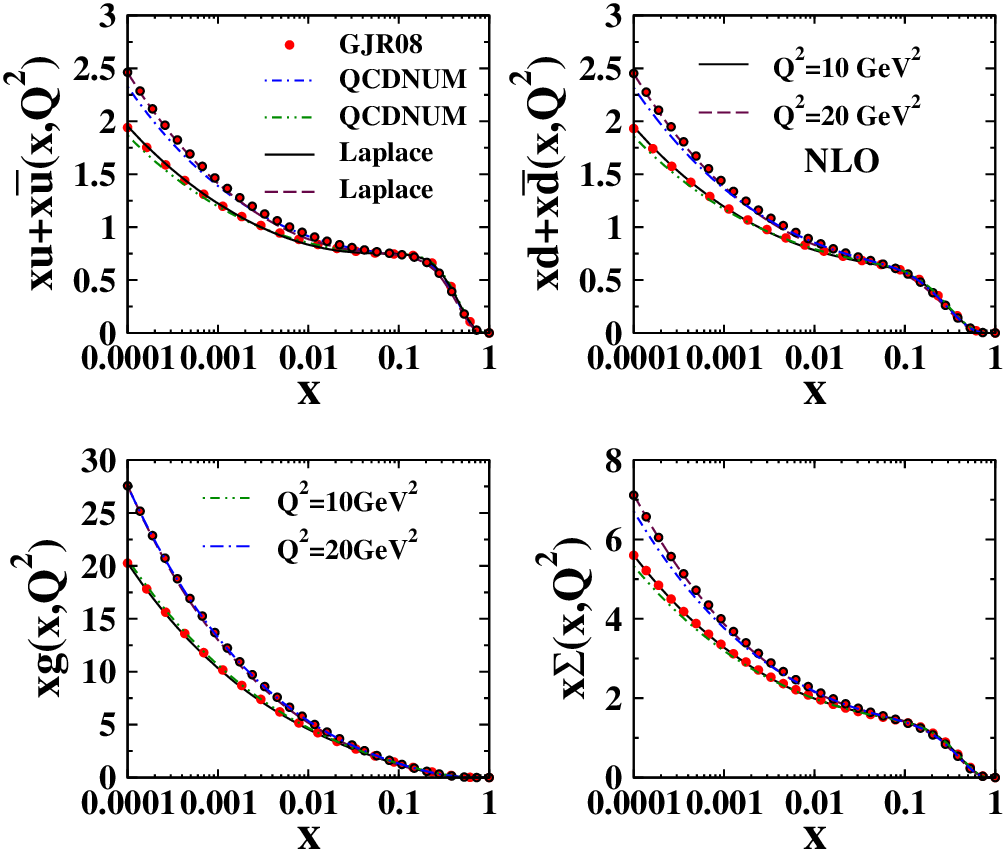}
		%\vspace{-1cm}
		\caption{(Color online) Sea quarks and singlet distributions in comparison with the next-to-leading order results of {\tt GJR08} model. The gluon distribution are also shown. The solid-line correspond to Q$^2$ = 10 GeV$^2$ and the dashed-line correspond to Q$^2$ = 20 GeV$^2$. The results from the QCD evolution package, {\tt QCDnum}, are also presented (dash-dotted and dash double-dotted lines).  }\label{fig:singletQ9p795}
	\end{center}
\end{figure}
%--------------------------------

%--------------------------------
\begin{figure}[htb]
	\begin{center}
		\includegraphics[clip,width=0.45\textwidth]{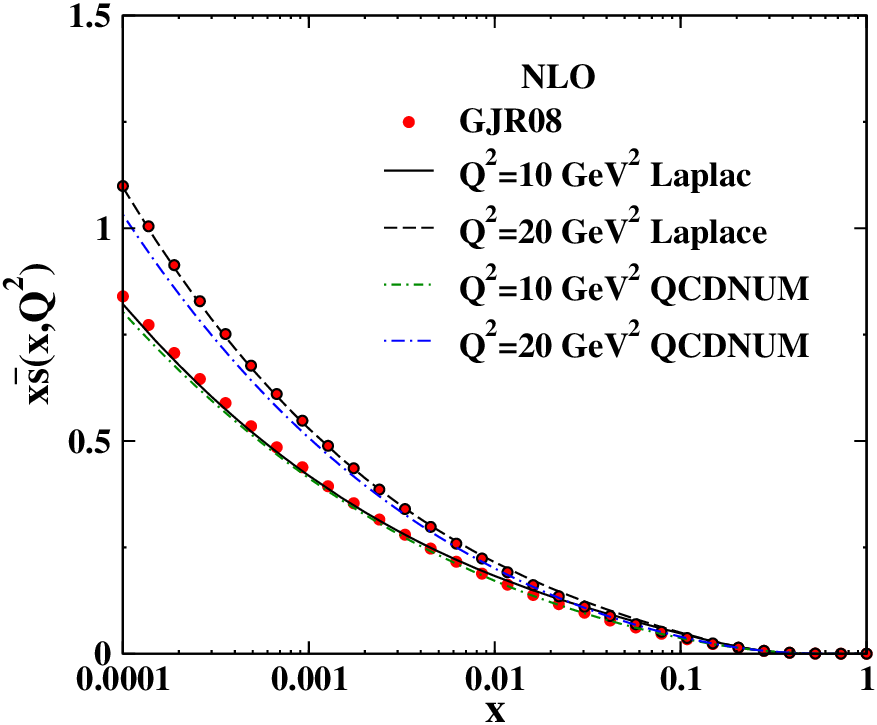}
		%\vspace{-1cm}
		\caption{(Color online) The strange sea distribution $x s = x \overline{s}$ in comparison with the next-to-leading order results of {\tt GJR08} model. The solid line correspond to Q$^2$ = 10 GeV$^2$ and the dashed-line corresponds to Q$^2$ = 20 GeV$^2$. The results from the QCD evolution package, {\tt QCDnum}, are also presented (dash-dotted and dash-double-dotted lines).  }\label{fig:strangelaplace}
	\end{center}
\end{figure}
%--------------------------------
%--------------------------------
\begin{figure}[htb]
	\begin{center}
		\includegraphics[clip,width=0.45\textwidth]{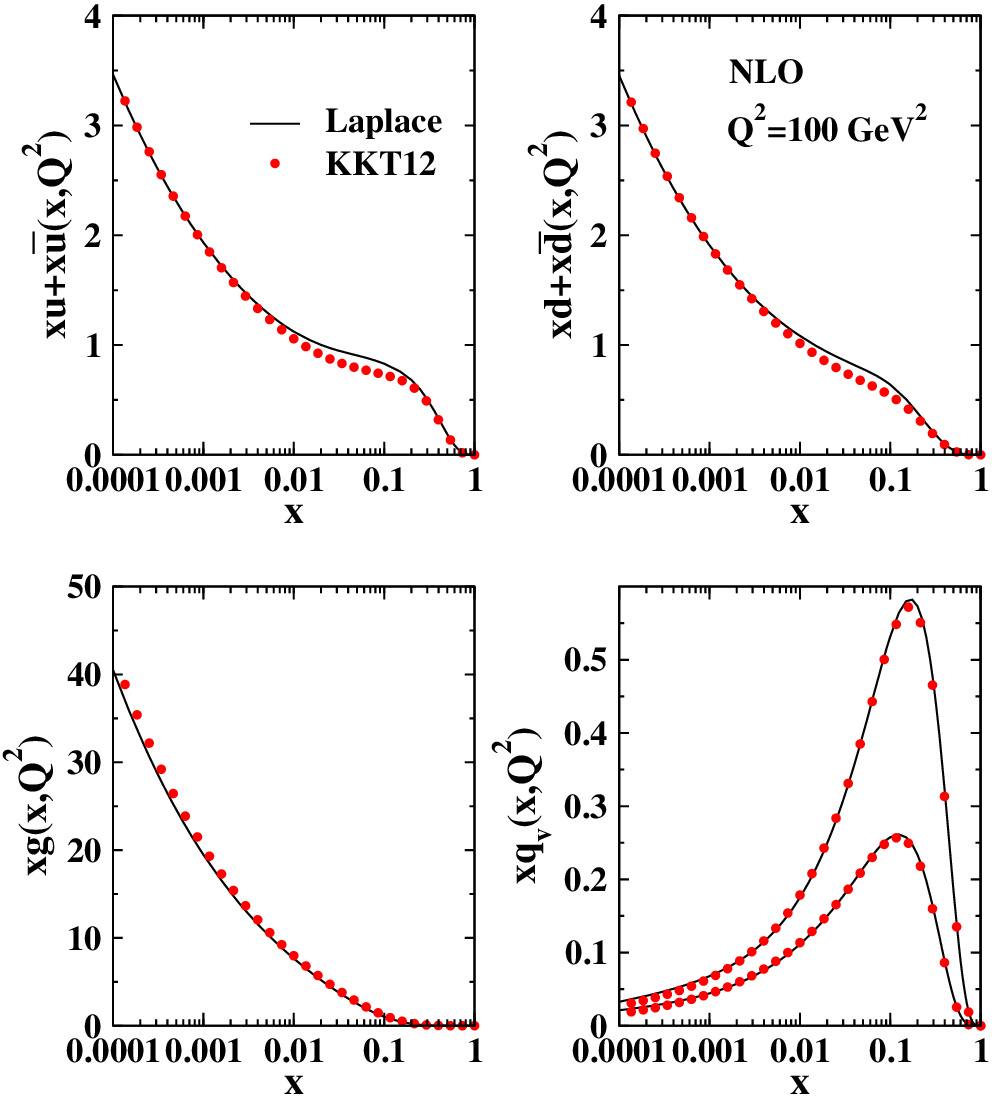}
		%\vspace{-1cm}
		\caption{(Color online) Sea quarks, gluon and non-singlet distributions and comparison with the results from next-to-leading order {\tt KKT12} global QCD analysis at Q$^2$ = 100 GeV$^2$.}\label{fig:kktQ100}
	\end{center}
\end{figure}
%--------------------------------
%--------------------------------
\begin{figure}[htb]
	\begin{center}
		\includegraphics[clip,width=0.45\textwidth]{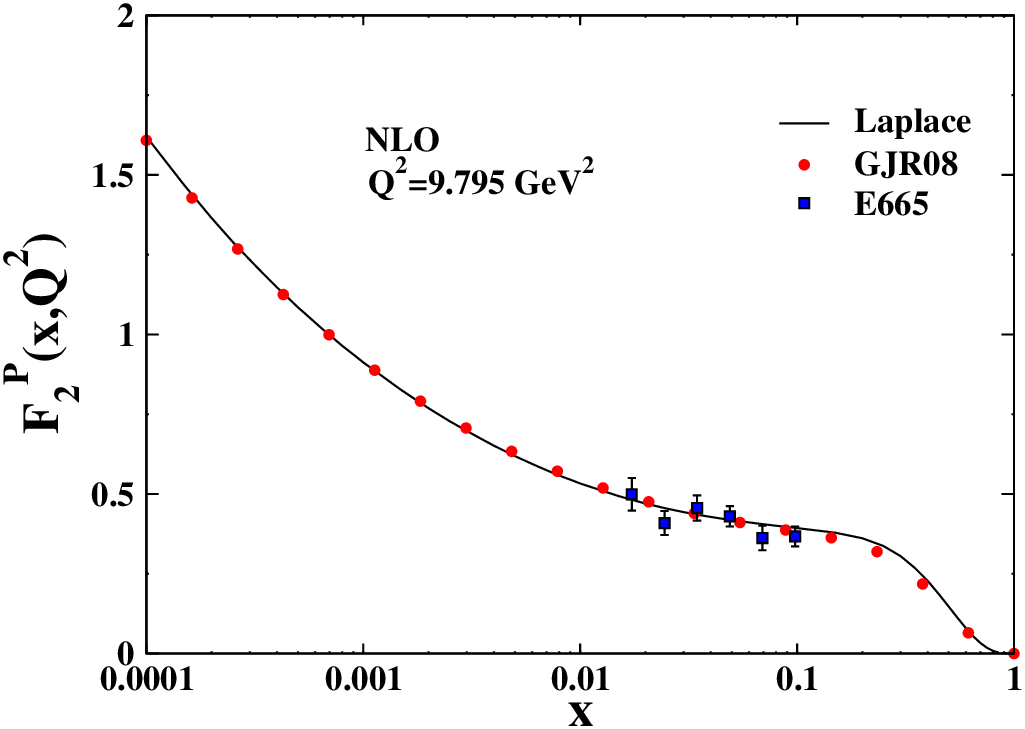}
		%\vspace{-1cm}
		\caption{(Color online) The next-to-leading order approximation of the total proton structure function, $F_2^{\rm p, total} (x, Q^2)$, as a function of $x$ at Q$^2$ = 9.795 GeV$^2$. The input distributions are obtained from the {\tt GJR08} model~\cite{Gluck:2007ck}. Here the straight line represents our result, using the Laplace transform technique, and the red circles represent the proton structure function arising from the {\tt GJR08} global QCD analysis. A comparison with the E665 experimental data~\cite{Adams:1996gu} is also shown.}\label{fig:F2p-laplace-compare-with-the-data-Q9.795}
	\end{center}
\end{figure}
%--------------------------------
%--------------------------------
\begin{figure}[htb]
	%	\vspace*{0.5cm}
	\begin{center}
		\includegraphics[clip,width=0.45\textwidth]{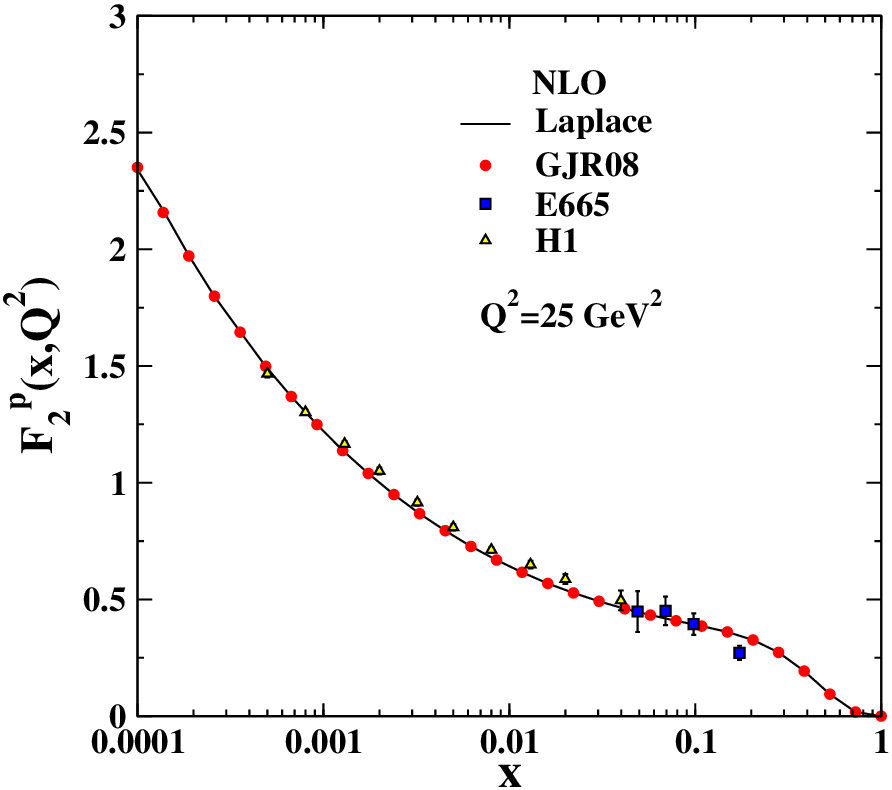}
		%\vspace{-1cm}
		\caption{(Color online) The next-to-leading order approximation of the total proton structure function, $F_2^{\rm p, total} (x, Q^2)$, as a function of $x$ at Q$^2$ = 25 GeV$^2$. The input distributions are obtained from the {\tt GJR08} model~\cite{Gluck:2007ck}. Here the straight line represents our result, using the Laplace transform technique, and the red circles represent the proton structure function arising from the {\tt GJR08} global QCD analysis. The square and up-triangle signs represent the the E665 experimental data~\cite{Adams:1996gu} and H1 inclusive deep inelastic neutral current data~\cite{Aaron:2009kv}, respectively.}\label{fig:F2p-laplace-compare-with-the-data-Q25}
	\end{center}
\end{figure}
%--------------------------------
%
		
		%
		%%%%%%%%%%%%%%%%%%%%%%%%%%%%%%%%%%%%%%%%%%    The results of Laplace transformation technique    %%%%%%%%%%%%%%%%%%%%%%%%%%%%%%%%%%%%%%%%%%%%%%%%%
		%
		\section{The results of Laplace transformation technique}\label{Sec:Results}

		In this section, we present our results that have been obtained for the parton distribution functions and proton structure function $F_2^p(x, Q^2)$ using the Laplace transformation technique to find an analytical solution for the DGLAP evolution equations.
		We obtain the valence quark distributions, $x u_v (x,Q^2)$ and $x d_v (x,Q^2)$, using  Eq.\eqref{eq:convolation-nonsinglet} and compare them with the next-to-leading order {\tt GJR08} results. Since the {\tt GJR08} Collaboration started their evolution at Q$_0^2$ = 2 GeV$^2$, we used $F_{\rm NS}^0$, $F_{\rm S}^0$ and $G^0$ constructed from their values at Q$_0^2$ in Eq.(\ref{eq:prton}). The results for the evolved non-singlet distributions are depicted in Fig.~\ref{fig:nonsingletQ9p795}. To double check and indicate the sufficient precision of our analysis, we have also used the  QCD evolution package, {\tt QCDnum}~\cite{Botje:2010ay} and linked it to the {\tt LHAPDF}~\cite{Buckley:2014ana} package for the {\tt GJR08} PDFs, which directly render the parton densities in $x$ space. As can be seen from the related figures, a good agreement between our results and the other ones exist. It indicates the evolution works well beyond the charm quark mass threshold, Q$^2$ > Q$_0^2$ ($\approx m_c^2$ = 2 GeV$^2$). In this figure the straight line represents the solution resulting from the Laplace transform technique and the red circles represent the valance quark distributions from {\tt GJR08} global QCD analysis. The dashed line indicates the results, arising out from {\tt QCDnum} evolution package. One can conclude that the agreement, over the large span of $0 < x < 1$, is quite striking. The accuracy of the present analysis has been investigated and is typically better than about 1 part in 10$^5$ at small and large values of Bjorken $x$ for the up-valence quark distribution $x u_v$. For the down-valence quark distribution $x d_v$, disagreements between our calculation and the {\tt GJR08} results are less than 1--2\% for $0 < x < 0.2$.
		
		In Fig.~\ref{fig:singletQ9p795}, the results for sea quark and singlet distributions are shown and compared with the next-to-leading order analysis of the {\tt GJR08} model as well as {\tt QCDnum} evolution package. However some researchers are reporting the singlet solution rather than the individual distribution for sea quarks, but following the technique which was
		introduced in~\cite{Furmanski:1980cm,Furmanski:1981ja}, it is possible to present separately the see quark distributions.
		The analytical solution for the gluon distribution, $G(x, Q^2) = x g(x, Q^2)$, is also shown. All distributions are obtained from Eq.\eqref{eq:NLODGLAPSspace} in $(s, \tau)$ space and then converted to the ($x$, Q$^2$) space, using the convolution integrals  in Eq.\eqref{eq:convolutions-final}. The results indicated by the solid line correspond to Q$^2$ = 10 GeV$^2$ and the ones indicated by the dashed line correspond to Q$^2$ = 20 GeV$^2$. The strange sea distribution $x s = x \overline{s}$ and its comparison with the next-to-leading order results of the {\tt GJR08} model are also shown in Fig.~\ref{fig:strangelaplace} at Q$^2$ = 10 GeV$^2$ and Q$^2$ = 20 GeV$^2$. This figure indicates that the obtained results from the present analysis based on the Laplace transform technique are in good agreement with the ones obtained by global QCD analysis of {\tt GJR08} for the parton distribution functions and also the obtained results from the QCD evolution package, {\tt QCDnum}. One can conclude from Figs.~\ref{fig:singletQ9p795} and ~\ref{fig:strangelaplace} that the agreements between our results and {\tt GJR08} global analysis are excellent over the entire range of momentum fraction-$x$ and the virtuality Q$^2$.
		We found slightly disagreements between $x$-space results calculated from the {\tt QCDnum} package and the {\tt GJR08} analysis which are 1.5--2\% for all parton species except for the gluon distribution. It is clear from the mentioned plots that, over the enormous Q$^2$ and $x$ spans, our analytic solutions are in satisfactory agreement with the {\tt GJR08} analysis.
		
		A detailed comparison has also been shown with the next-to-leading order results from the {\tt KKT12} global QCD analysis and depicted in Fig.~\ref{fig:kktQ100}. In this figure our analytical solution based on the Laplace transform technique is presented for sea and singlet distributions as well as for the gluon distribution $G(x, Q^2) = x g(x, Q^2)$ at Q$^2$ = 100 GeV$^2$. The analytical solution arises from Eq.\eqref{eq:NLODGLAPSspace} which is related to the {\tt KKT12} initial distributions at Q$_0^2$ = 2 GeV$^2$.
		
		The results of analytical solutions for all parton distribution functions clearly show significant agreement over a wide range of $x$ and Q$^2$ variables. The only serious disagreements which we found between our calculations and the {\tt KKT12} results are for $x u + x \bar{u}$ and $x d + x \bar{d}$ distributions, which are smaller than 2--2.5\% at $0.01 < x < 0.1$.
		
		As a numerical illustration of our analytical approaches at the next-to-leading order approximation of the total proton structure function, $F_2^p(x, Q^2)$, we compare our results with the {\tt GJR08} proton structure function and depict them in Figs.~\ref{fig:F2p-laplace-compare-with-the-data-Q9.795} and~\ref{fig:F2p-laplace-compare-with-the-data-Q25}. A comparison with E665 data at fixed-target experiments ~\cite{Adams:1996gu} and H1 inclusive deep inelastic neutral current data~\cite{Aaron:2009kv} are also shown there. The results for the total proton structure function $F_2^{\rm p, total} (x, Q^2)$ have been presented as a function of $x$ (both for large and small $x$) at Q$^2$ = 9.795 and 25 GeV$^2$. It is seen that our analytical solutions based on the inverse Laplace transform technique at the NLO approximation for the proton structure function over a wide range of $x$ and Q$^2$ values correspond well with the experimental data and the QCD analysis performed by {\tt GJR08} analysis. One can conclude that, in spite of small disagreement for the parton densities, we found a satisfactory agreement for the proton structure function over a wide range of $x$ and Q$^2$. The overall agreement is found to be 1 part in 10$^5$.
		
		Based on our obtained results for the next-to-leading order proton structure function, $F_2^p(x, Q^2)$, and its good agreement with other theoretical models as well as experimental data, one can evaluate the parton distributions functions at the input scale Q$_0^2$ by performing a global QCD fit to the all available and up-to-date DIS and hadron collision data, using the Jacobi polynomials approach. We plan to present our detailed QCD analysis based on the analytical calculation in the next section.

%
%%%%%%%%%%%%%%%%%%%%%%%%%%%%%%%%%%%%%%%%%%    DIS analysis    %%%%%%%%%%%%%%%%%%%%%%%%%%%%%%%%%%%%%%%%%%%%%%%%%
%
\section{ Jacobi polynomials technique for the DIS analysis }\label{Sec:Jacobi-polynomials}

Global analysis of deep-inelastic scattering (DIS) data in the framework of QCD provides us with new knowledge of hadron physics and serves as a test of reliability of our theoretical understanding of the hard scattering of leptons and hadrons. Various QCD analyses, both for polarized and un-polarized case, can be constructed using all available data from fixed-target experiments, DIS data and the precise data from hadron colliders. For further literature on various PDFs models, we refer the reader to review articles~\cite{Carrazza:2016htc,Harland-Lang:2014zoa,Mangano:2016jyj,Pennington:2016dpj,Dulat:2015mca,Jimenez-Delgado:2014twa,Carrazza:2015hva,Rojo:2015acz,DeRoeck:2011na,Accardi:2016qay,McNulty:2016xtv,Accardi:2016ndt}. The kinematics spanned by each DIS data set used in our fit are described in Secs.~\ref{Sec:data}.

We shall focus here on the non-singlet (NS) structure functions, $F_2^{\rm NS} (x, Q^2)$, with their corresponding Laplace $s$-space moments ${\cal M}^{\rm NS}(s, Q^2)$ in order to perform a QCD analysis of deep inelastic scattering data up to the next-to-leading order (NLO).
Based on a popular parametrization for the parton distribution functions (PDFs), we apply the Jacobi polynomial formalism.
We consider a wide range of DIS data corresponding the momentum transfer from low Q$_0^2 \gtrsim 2 \, {\rm GeV}^2$ to high Q$^2 \sim 30000 \, {\rm GeV}^2$ where the approach still works reasonably still works. In this section, we first give an introductory description  of the Jacobi polynomial approach, as the method of our QCD analysis for the non singlet (NS) structure functions and the procedure of the QCD fit to the data.

In the common $\overline{MS}$ factorization scheme, one can obtained the relevant $F_2$ structure function up to NLO from the combination of non-singlet, flavour singlet and gluon contributions of Eqs.(\ref{eq:F2pLaplacespace-singlet}--\ref{eq:F2pLaplacespace-nonsinglet}).

In Laplace $s$ space, the combinations of parton densities at the valence region $x \geq 0.3$ for the proton structure function ${\cal M}_2^{p}$ in NLO can be written as:
%--------------------------------
\begin{eqnarray} \label{Fp}
{\cal M}_2^p(s, \tau) & = & \left(\frac{4}{9} u_v(s)  + \frac{1}{9} d_v(s) \right)   \times   \nonumber \\
&& \left( 1 + \frac{\tau}{4 \pi} C_q^{(1)}(s)\right) e^{\tau \Phi_{\rm NS}(s)}
\end{eqnarray}
%--------------------------------

In the above region, the combinations of parton densities for the deuteron structure function ${\cal M}_2^{d}$ are also given by,

%--------------------------------
\begin{eqnarray} \label{Fd}
{\cal M}_2^d(s, \tau) & = & \frac{5}{18} \, (u_v(s) + d_v(s)) \times \,, \nonumber \\
&& \left( 1 + \frac{\tau}{4 \pi} C_q^{(1)}(s) \right) e^{\tau \Phi_{\rm NS}(s)}\,,
\end{eqnarray}
where $d = \frac{p + n}{2}$. In the region of $x \leq 0.3$ for the difference of proton ${\cal M}_2^p$ and deuteron ${\cal M}_2^d$ data, we use:

%--------------------------------
\begin{eqnarray} \label{FNS}
{\cal M}_2^{\rm NS}(s, \tau) &\equiv& 2 ( {\cal M}_2^p - {\cal M}_2^d)(s, \tau) \nonumber \\  & = & \left( \frac{1}{3} \,
(u_v - d_v)(s) + \frac{2} {3} \, (\bar u - \bar d)(s) \right) \times ~ \nonumber \\
&& \left( 1 + \frac{\tau}{4 \pi} C_q^{(1)}(s) \right)e^{\tau \Phi_{\rm NS} (s)}
\end{eqnarray}
%--------------------------------

Since sea quarks can not be neglected for $x$ smaller than about 0.3, in our calculation we suppose the $\bar d  - \bar u$ distribution from {\tt JR14}~\cite{Jimenez-Delgado:2014twa} at Q$_0^2$ = 2 GeV$^2$ to be
%--------------------------------
\begin{equation}
x(\bar{d} - \bar{u})(x, Q_0^2) = 37.0 x^{2.2}(1 - x)^{19.2} (1 + 2.1 \sqrt{x})\,,
\end{equation}
%--------------------------------

As we mentioned at the beginning of this section, the method we have employed is using the Jacobi polynomials expansion of the structure functions.
The details of the Jacobi polynomial approach are presented in our previous work~\cite{Khorramian:2009xz}. Here we outline a brief review of this method.
According to this approach, using the Jacobi polynomial moments $a_n (Q^2)$, one can reconstruct the structure function as,

%--------------------------------
\begin{equation} \label{xf}
xf(x, Q^2) = x^{\beta}(1 - x)^{\alpha} \sum_{n = 0}^{\rm N_{max}} a_n (Q^2) \Theta_n^{\alpha, \beta}(x) \,,
\end{equation}
where ${\rm N_{max}}$ is the number of polynomials and $\Theta_n^{\alpha, \beta}(x)$ are the Jacobi polynomials of order $n$,

%--------------------------------
\begin{equation} \label{Theta}
\Theta_n^{\alpha, \beta}(x) = \sum_{j = 0}^n \, c_j^{(n)}{(\alpha, \beta )}  \, x^j \,,
\end{equation}
in which $c_j^{(n)}{(\alpha, \beta )}$ are the coefficients that are expressed through $\Gamma$ functions and satisfy the orthogonality relation with the weight
$w^{\alpha, \beta} = x^{\beta}(1-x)^{\alpha}$ as follows

%--------------------------------
\begin{equation}\label{int}
\int_0^1 dx \, x^{\beta}(1 - x)^{\alpha} \Theta_m^{\alpha,	\beta}(x) \Theta_n ^{\alpha, \beta}(x) = \delta_{m n} \,.
\end{equation}
%--------------------------------

Using the above equations, we can relate the proton, neutron and non singlet structure functions with their Laplace $s$-space moments,

%--------------------------------
\begin{eqnarray}\label{Jacob}
F_2^{\rm p, d, NS}(x, Q^2) & = & x^{\beta}(1 - x)^{\alpha}
\sum_{n = 0}^{\rm N_{max}}\Theta_n ^{\alpha, \beta}(x) \nonumber \\ & \times & \sum_{j = 0}^{n} \, c_j^{(n)}{(\alpha, \beta)}
{\cal M}_2^{\rm p, d, NS}(s=j + 1, Q^2)\,,  \nonumber  \\
\end{eqnarray}
where ${\cal M}_2^{\rm p, d, NS}(s, Q^2)$ are the moments in Laplace $s$ space presented in Eqs.(\ref{Fp}--\ref{FNS}) for the proton, neutron and non-singlet structure functions.
Here the Q$^2$ dependence of the structure functions will be provided by the Q$^2$ dependence of their moments in the Laplace-$s$ space. We consider ${\rm N_{max}}$ to be 9, $\alpha$ to be 3.0 and $\beta$ to be 0.7 to achieve the fastest convergence of the above series~\cite{Khorramian:2009xz,Krivokhizhin:1987rz,Krivokhizhin:1990ct}.

%------------------------------------------------------------------
\subsection{ Method of the QCD analysis }\label{Sec:Method}
%------------------------------------------------------------------

In this section, we present the details of the analysis which our analysis is based.
We begin with a short discussion of the parametrization chosen for the various
flavour PDFs. We then present a detailed discussion on the data set used and kinematic cuts applied. The method of the minimizations also will be discussed.
Then we present the results of the analysis and describe the approach taken in this analysis.

%------------------------------------------------------------------
\subsection*{ PDF parametrizations }\label{Sec:parametrizations}
%------------------------------------------------------------------

For the parametrization of the PDFs at the input scale Q$^2_0$, chosen here to be 2 GeV$^2$, standard five-parameter form is adopted for valence parton species $f$,
%--------------------------------
\begin{eqnarray}
x u_v (x, Q_0^2) &=& {\cal N}_u x^{\alpha_u}	(1 - x)^{\beta_u} ( 1 + \gamma_u \sqrt{x} + \eta_u x) \,, \nonumber \\
x d_v (x, Q_0^2) &=& {\cal N}_d x^{\alpha_d}	(1 - x)^{\beta_d} ( 1 + \gamma_d \sqrt{x} + \eta_d x) \,. \nonumber \\
\end{eqnarray}
%--------------------------------
This form applies to the up-valence $x u_v = x u - x \bar u$  and down-valence  $x d_v = x d - x \bar d$  distributions.
The normalization factors, ${\cal N}_u$ and ${\cal N}_d$, will be fixed by $\int_0^1 u_v dx = 2$ and  $\int_0^1 d_v dx = 1$, respectively.
In the Laplace $s$ space, the normalizations ${\cal N}_u$ and ${\cal N}_d$ are fixed by ${\cal L} [e^{-v} u_v(e^{-v}); s=0] = 2$ and ${\cal L} [e^{-v} d_v(e^{-v}); s=0] = 1$, respectively.

%------------------------------------------------------------------
\subsection*{ Data sets } \label{Sec:data}
%------------------------------------------------------------------

Our valence PDFs are obtained by fitting to a global database of over 572 data points from a variety of high energy
scattering processes. The data sets used in this analysis are listed in Table.~\ref{tab:data}.
These include deep-inelastic scattering data from {\tt BCDMS}~\cite{Benvenuti:1989fm,Benvenuti:1989rh,Benvenuti:1989gs}, {\tt SLAC}~\cite{Whitlow:1991uw} and {\tt NMC}~\cite{Arneodo:1996qe,Arneodo:1995cq} experiments. The {\tt BCDMS} data were collected at CERN and both proton and deuterium targets were used in the same experiment. These
data sets facilitate flavor separation of PDFs at large $x$. The {\tt NMC} experiment was also performed at CERN. The {\tt NMC} data span lower values of $x$ and Q$^2$ and, due to the better coverage of the small-$x$ region, those data are also sensitive to the isospin asymmetry in the sea distribution.

The DIS data from {\tt H1}~\cite{H1} and {\tt ZEUS}~\cite{ZEUS} Collaborations are also included. New data sets from combined measurement of H1 and ZEUS Collaborations at HERA for the inclusive $e^\pm p$ scattering cross sections are also added~\cite{Aaron:2009aa}. As one can see from Table.~\ref{tab:data}, we use three data samples: for $F_2^p (x, Q^2)$ and $F_2^d (x, Q^2)$ in the valence quarks regions $x \geq 0.3$ and for $F_2^{\rm NS}(x, Q^2) = 2 (F_2^p(x, Q^2) - F_2^d(x, Q^2))$ in the region of $x < 0.3$.
Before the fitting process, we apply different cuts on data samples in order to widely eliminate the higher twist (HT) effects.
Cuts on the kinematic coverage of the DIS data have been made for Q$^2 > 4$ GeV$^2$ and on the hadronic mass of $W^2 > 12.5$.
An additional cut on the {\tt BCDMS} data ($y>0.35$) and on the {\tt NMC} data ($Q^2>8$ GeV$^2$) was also applied.
The DIS data used in our fit and the number of data points for each experiment after the cuts are listed in Table ~\ref{tab:data}. The numbers of reduced data points by the additional cuts are given in the fifth column of the table. This reduces the number of data points from 467 to 248 for $F_2^p (x, Q^2)$, from 232 to 159 for $F_2^d (x, Q^2)$ and from 208 to 165 for $F_2^{\rm NS} (x, Q^2)$.

%--------------------------------
\renewcommand{\arraystretch}{0.8}
\begin{table*}
	 \caption{  \label{tab:data}{ Data sets used in our analysis, with the corresponding number of data points (a) $F_2^p (x, Q^2)$, (b) $F_2^d (x, Q^2)$, and (c) $F_2^{\rm NS} (x, Q^2)$ for the non-singlet QCD analysis with their $x$ and Q$^2$ ranges. The name of different data sets, and range of $x$ and Q$^2$ are given in the three first columns. The normalization shifts are also listed in the last column. The details of corrections to data and the kinematic cuts applied on data are contained in the text. }}
	 	
	\begin{tabular}{l|c|c|c|c||c}
		\hline
		{\tt Experiment}  & $x$ & Q$^2, {\rm GeV}^2$ & $F_2^p$ & $F_2^p~{\rm cuts}$ &  ${\cal N}$  \\  \hline
		\hline {\tt BCDMS (100)}   & 0.35--0.75 &  11.75--75.00  & 51 & 29 	& 0.9984 \\
		{\tt BCDMS (120)}   & 0.35--0.75 &  13.25--75.00  &  59 &  32 &  0.9968      \\
		{\tt BCDMS (200)}   & 0.35--0.75 &  32.50--137.50  &  50 &  28 & 	0.9986   \\
		{\tt BCDMS (280)}   & 0.35--0.75 &  43.00--230.00  &  49 &  26 & 	1.005    \\
		{\tt NMC (comb)}    & 0.35--0.50 &  7.00--65.00  &  15 &  14 & 	0.9996       \\
		{\tt SLAC (comb)}   & 0.30--0.62 &  7.30--21.39  &  57 &  57 &  1.0000       \\
		{\tt H1 (hQ2)}      & 0.40--0.65 &  200--30000  &  26 &  26 & 	1.0015       \\
		{\tt ZEUS (hQ2)}    & 0.40--0.65 &  650--30000  &  15 &  15 &   1.0000       \\
		{\tt H1 (comb)}    &  0.40--0.65 &  90--30000  &  145 &  21 & 	1.0000        \\
		\hline
		{\bf Proton}      &              &                    & 467 & 248 &          \\
		\hline
	\end{tabular}
	\\
	\vspace{0.1cm} \text{(a)  $F_2^p (x, Q^2)$ data points.}   \\ \vspace{0.2cm}
	\begin{tabular}{l|c | c |c |c ||c}
		\hline
		{\tt Experiment}  & $x$ & Q$^2, {\rm GeV}^2$ & $F_2^d$ & $F_2^d~{\rm cuts}$ & ${\cal N}$  \\   \hline
		\hline
		{\tt BCDMS (120)}   & 0.35--0.75 & 13.25--99.00   &  59 &  32 &    1.0069 \\
		{\tt BCDMS (200)}   & 0.35--0.75 & 32.50--137.50  &  50 &  28 &    1.0048 \\
		{\tt BCDMS (280)}   & 0.35--0.75 & 43.00--230.00  &  49 &  26 &    1.0038 \\
		{\tt NMC (comb)}    & 0.35--0.50 & 7.00--65.00    &  15 &  14 &    0.9987 \\
		{\tt SLAC (comb)}   & 0.30--0.62 & 10.00--21.40   &  59 &  59 &    0.9961 \\
		\hline
		{\bf Deuteron}    &              &                  & 232 & 159 &  \\
		\hline
	\end{tabular}
	\\
	\vspace{0.1cm} \text{(b)  $F_2^d (x, Q^2)$ data points.} \\  \vspace{0.2cm}
	\begin{tabular}{l|c| c |c |c ||c}
		\hline  {\tt Experiment}  & $x$ & Q$^2, {\rm GeV}^2$ & $F_2^{\rm NS}$ & $F_2^{\rm NS}~{\rm cuts}$
		& ${\cal N}$  \\
		\hline    \hline
		{\tt BCDMS (120)}  & 0.070--0.275 &  8.75--43.00    &  36 &  30 &     0.9987  \\
		{\tt BCDMS (200)}  & 0.070--0.275 &  17.00--75.00   &  29 &  28 &     0.9929  \\
		{\tt BCDMS (280)}  & 0.100--0.275 &  32.50--115.50  &  27 &  26 &     0.9997  \\
		{\tt NMC (comb)}   & 0.013--0.275 &  4.50--65.00    &  88 &  53 &     1.0002  \\
		{\tt SLAC (comb)}  & 0.153--0.293 &  4.18--5.50     &  28 &  28 &     1.0010  \\
		\hline
		{\bf Non-singlet} &               &                 & 208 & 165 &    \\
		\hline
	\end{tabular}
	\\
	\vspace{0.1cm} \text{(c) $F_2^{\rm NS} (x, Q^2)$ data points. \vspace{0.5cm}}
\end{table*}
%--------------------------------

%------------------------------------------------------------------
\subsection*{ Statistical procedures } \label{Sec:minimization}
%------------------------------------------------------------------

Agreement between the data sets and our theory predictions is quantified by the following $\chi^2_{\rm global}$ functional:

%--------------------------------
\begin{eqnarray}
\chi^2_{\rm global} = \sum_{n=1}^{N_{\rm exp}} w_n \, \chi^2_n\,, \label{eq:Chi1}
\end{eqnarray}
in which
%--------------------------------
\begin{eqnarray}
\chi_n^2 & = & \left( \frac{1 - {\cal N}_n} {\Delta{\cal N}_n}\right)^2 + \sum_{i=1}^{N^{\rm data}} \left( \frac{{\cal N}_n \, F_{2,i}^{\rm Data} - F_{2,i}^{\rm Theory})^2}{{\cal N}_n \, \Delta	F_{2,i}^{\rm Data}} \right)^2 \label{eq:Chi2}
\end{eqnarray}
%--------------------------------
where $F_{2}^{\rm Data}$ and $F_{2}^{\rm Theory}$ stand for the measurements and theory predictions, respectively. $\Delta	F_{2}^{\rm Data}$ is the measurements uncertainty (statistical and systematic combined in quadrature) and $i$ stands for $i$th data point in the fit.
$\Delta{\cal N}_n$ is the experimental normalization uncertainty and ${\cal N}_n$ is the overall normalization
factor which should be obtained from the fit to the data and then kept fixed.
The minimization of the above $\chi^2_{\rm global}$ value to determine the best fit parameters of the valence parton distributions is done using the CERN program {\tt MINUIT} \cite{James:1975dr}.
The value of $\chi^2 / {\rm n.d.f}$ computed according to Eq.\eqref{eq:Chi1} for the used data sets is given in Table~\ref{tab:fitresults}. The description quality is good enough for all data. This value is comparable to 1, therefore, the data can be easily accommodated in our fit.
The uncertainties on the observables and on the PDFs throughout this paper, are computed using well-known Hessian error propagation, as outlined in
Refs.~\cite{Pumplin:2001ct,Martin:2002aw,Martin:2009iq,Accardi:2016qay,Owens:2012bv,Accardi:2016ndt,Khanpour:2016pph,Shahri:2016uzl}, with $\Delta \chi^2 = 5.86$, which corresponds to a 68\% confidence level (C.L.) in the ideal Gaussian statistics.

%------------------------------------------------------------------
\subsection{ Target mass corrections (TMCs) } \label{Sec:TMC}
%------------------------------------------------------------------

It is important to consider all sources of corrections in a QCD analysis which may contribute to a comparable magnitude, such as target mass corrections (TMCs)~\cite{Georgi:1976ve,Mirjalili:2012zz}.
In this section, we will focus on the target mass corrections, which formally are subleading $1/Q^2$ corrections to leading twist structure functions.
Their effects are important at large value of $x$ and moderate $Q^2$, which coincides with the region where parton distribution
functions (PDFs) are not very well determined. Consequently, a reliable perturbative QCD based analysis which includes data in the low-$Q^2$
region, demands an accurate description of the TMCs. To study the effect of TMCs, we follow the method presented in Refs.~\cite{Georgi:1976ve,Gluck:2006yz,Steffens:2012jx,Khorramian:2009xz} to
determine the analytical form in Laplace $s$-space. The moments of flavor non singlet structure functions in the presence of TMCs
and in the Laplace $s$ space have the following form
%--------------------------------
\begin{eqnarray}\label{equ:TMC}
&& {\cal M}_{2,{\rm TMC}}^{k}(s, Q^2)     \equiv
{\cal L} [{\cal M}_{2, {\rm TMC}}^k(e^{-v}, Q^2; s)]\,   \nonumber  \\
&& = {\cal M}_2^k(s, Q^2)+\frac{s (s-1)}{s+2}\left(\frac{m_N^2}{Q^2}\right) \, {\cal M}_2^{
	k}(s+2, Q^2)   \nonumber \\
&&+\frac{\Gamma(s+3)^2}{2\Gamma(s-1)\Gamma(s+5)}\left(\frac{m_N^2}{Q^2}\right)^2 \, {\cal M}_2^{
	k}(s+4, Q^2)   \nonumber  \\
&&+\frac{\Gamma(s+4)\Gamma(s+5)}{6\Gamma(s-1)\Gamma(s+7)}\left(\frac{m_N^2}{Q^2}\right)^3 \, {\cal M}_2^{
	k}(s+6, Q^2)   \nonumber  \\
&&+\frac{\Gamma(s+5)\Gamma(s+7)}{24\Gamma(s-1)\Gamma(s+9)}\left(\frac{m_N^2}{Q^2}\right)^4 \, {\cal M}_2^k(s+8, Q^2)  \nonumber  \\
&&+ {\cal{O}}\left( \frac{m_N^2}{Q^2}\right)^5 \,,
\end{eqnarray}
%--------------------------------
where higher powers $(m_{\rm N}^2/Q^2)^n$ ($n \geqslant 2$) are negligible for the relevant $x < 0.8$ region. Consequently we can neglect these higher order parts. By inserting Eq.(\ref{equ:TMC}) into Eq.\eqref{xf}, one can obtain
%--------------------------------
\begin{eqnarray} \label{eg1JacobTMC}
F_2^{k, {\rm TMC}}(x,Q^2) & = & x^{\beta}(1 - x)^{\alpha} \sum_{n = 0}^{\rm N_{max}} \Theta_n ^{\alpha, \beta}(x)    \nonumber   \\
& \times & \sum_{j = 0}^n \, c_{j}^{(n)}{(\alpha, \beta)} \, {\cal M}_{2, {\rm TMC}}^k (j+1, Q^2)\,. \nonumber \\
 \end{eqnarray}
%--------------------------------
In this equation ${\cal M}_{2, {\rm TMC}}^k(j+1, Q^2)$ are the moments determined by Eq.\eqref{equ:TMC}. The effects of TMCs on the PDFs and the corresponding observables will be illustrated in Sec.~\ref{Sec:fitresults}.

%------------------------------------------------------------------
\subsection{  Higher twist (HT) corrections } \label{Sec:HT}
%------------------------------------------------------------------

In addition to the important role played by TMCs at large values of $x$ and moderate $Q^2$, the effects of higher twist (HT) corrections are also significant~\cite{Abt:2016vjh,Jimenez-Delgado:2013boa,Leader:2006xc,Nath:2016phi,Wei:2016far}.
Consequently, in the context of parton distribution analyses, the study of higher twists is also important in its own right.
In addition to the kinematic cuts ($Q^2 \geq 4$ GeV$^2$, $W^2 \geq 12.5$ GeV$^2$) we apply in our analysis, we also take into account higher twist corrections to the proton $F_2^p(x, Q^2)$ and deuteron structure functions $F_2^d(x, Q^2)$ for the kinematic region $Q^2 \geq 4 {\rm GeV}^2, 4 < W^2 < 12.5 {\rm GeV}^2$. For this purpose, we extrapolate our QCD fit results to this region.

In practice, higher twist contributions are usually parameterized independently from the leading twist one
with some function of $x$, which is typically polynomial in $x$. In the region where the power corrections are non-negligible for the case of the DIS data, they are defined within an entirely phenomenologically motivated ansatz, as follows:

%--------------------------------
\begin{eqnarray}\label{equ:HTC}
F_2^{\rm HT}(x, Q^2) = {\cal O}_{\rm TMC}[F_2^{\rm TMC}(x, Q^2)] \cdot \left( 1 + \frac{h(x, Q^2)}{Q^2[{\rm GeV}^2]} \right)\,, \nonumber \\
\end{eqnarray}
%--------------------------------

where $F_2^{\rm TMC}(x,Q^2)$ are given by Eq.\eqref{eg1JacobTMC}. In the above equation, the operation ${\cal O}_{\rm TMC} [...]$ denotes taking the target mass corrections of
the twist--2 contributions to the respective structure function.
As we mentioned, the coefficients $h(x, Q^2)$ are determined in bins of $x$ and Q$^2$ and are then averaged over Q$^2$. The $x$-shape of the higher twist contributions is defined by
the following expression,

%--------------------------------
\begin{equation}
h(x) = \alpha \left (\frac{x^\beta}{1 - x} - \gamma \right) \,.
\end{equation}
%--------------------------------

This choice of $h(x, Q^2)$ provides sufficient flexibility of the higher twist terms with respect to the data analyzed. To perform higher twist QCD analysis of the non-singlet world data, we consider the $Q^2 \geq 4 \, {\rm GeV}^2, 4 < W^2 < 12.5 \, {\rm GeV}^2$ cuts.
The parameter values of the $h(x)$ function were fitted to the data simultaneously with the valence PDFs parameters and the value of $\Lambda_{\rm QCD}$. The corresponding parameter values are presented in Table.~\ref{tab:ht}.
One can see the sensitivity of the fit to the higher twist terms in Sec.~\ref{Sec:fitresults}.

%------------------------------------------------------------------
\subsection{ Results of QCD fit }\label{Sec:fitresults}
%------------------------------------------------------------------

In this section, we present the results of our global QCD analysis which is based on the analytical solution based on Laplace transform technique and Jacobi polynomials approach.
The parameter values of the next-to-leading order non-singlet QCD fit at the input scale of Q$_0^2 = 2 \, {\rm GeV}^2$ are presented in Table.~\ref{tab:fitresults}.
The parameters values without error have been fixed after the first minimization since the present data do not constrain these parameters well enough. From the table, one can finds rather stable PDF central values.
The value for $\alpha_{\rm s}^{\rm N_f=4}(Q_0^2)$  has also been obtained from the fit. This result can be expressed in terms of $\alpha_s(M_Z^2)$, which is correspond to $\alpha_s(M_Z^2) = 0.1173 \pm 0.0011$.
%--------------------------------
\begin{table}
	\begin{tabular}{c | c | c }
		\hline
		\multicolumn{3}{c}{ Next-to-leading order (NLO) fit }   \\ \hline
		\hline $u_v$      & $\alpha_u$         &  0.7108 $\pm$ 0.1295  \\
		& $\beta_u$                    &  3.3595  $\pm$ 0.027   \\
		& $\gamma_u  $                  &  0.2979               \\
		& $\eta_u$                    &  1.3440                 \\
		\hline $d_v$     & $\alpha_d$          &  0.9467 $\pm$ 0.0261  \\
		& $\beta_d$                      &  2.8468  $\pm$ 0.3130    \\
		& $\gamma_d  $                    &  1.1004                 \\
		& $\eta_d$                      &  -1.1330                \\
		\hline \multicolumn{2}{c|}{ $\alpha_{\rm s}^{\rm N_f=4}(Q_0^2)$}
		& 0.3521 $\pm$ 0.0139 \\
		\hline \hline \multicolumn{2}{c|}{ $\chi^2 / {\rm n.d.f}$ } & 521.303/563 = 0.92  \\
		\hline \hline
	\end{tabular}
	\caption{\label{tab:fitresults}{\sf Parameter values of the NLO non-singlet QCD fit at Q$_0^2 = 2 \, {\rm GeV}^2$. The
			parameters values without error have been fixed after the first minimization. }}
\end{table}
%--------------------------------

\begin{figure}[htb]
	\begin{center}
		\includegraphics[clip,width=0.46\textwidth]{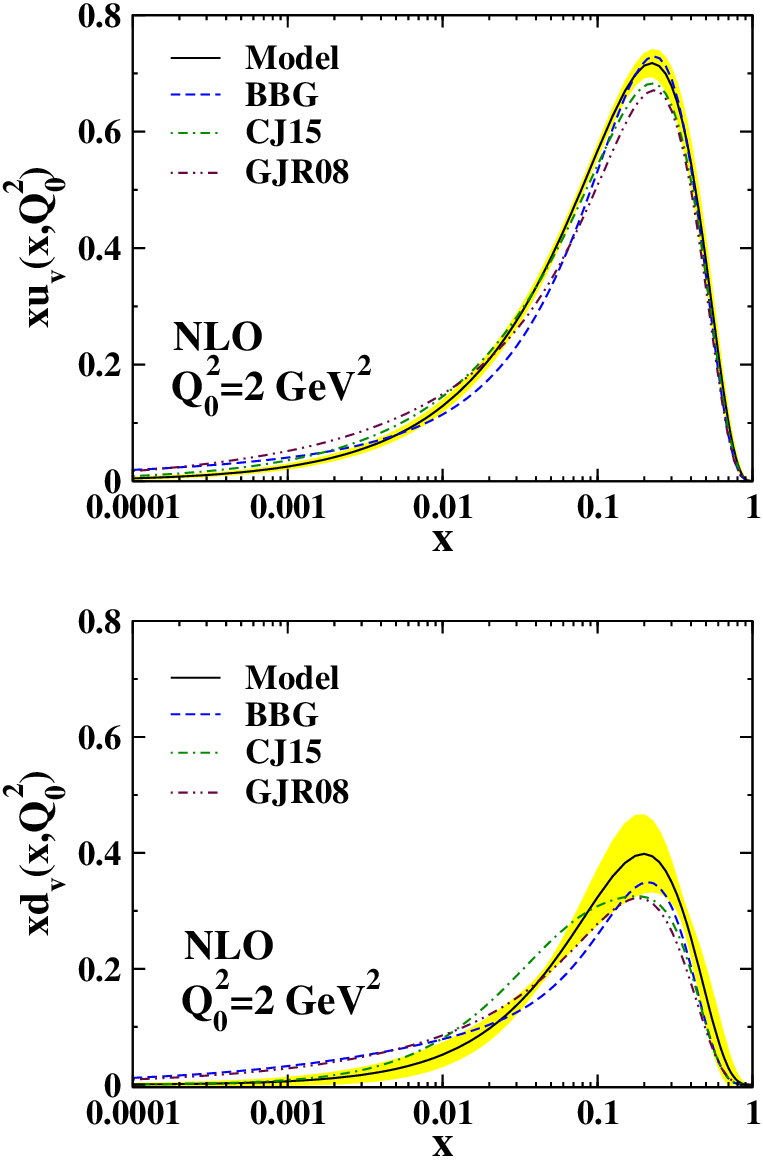}
		%\vspace{-1cm}
		\caption{(Color online) The parton densities $xu_v$ and $xd_v$ at the input scale Q$^2_0= 2 \, {\rm GeV}^2$. The uncertainties of our PDFs (yellow band)
		 correspond to a 68\% confidence level (C.L.) with $\Delta \chi^2_{\text{global}} =5.86$. The dashed
			line is the {\tt BBG} PDF~\cite{Blumlein:2006be},the dashed-dotted line is the {\tt CJ15} PDF~\cite{Accardi:2016qay} and dashed-double-dotted line
			is the result from {\tt GJR08}~\cite{Gluck:2008gs}}\label{fig:partonvalnceQ0}
	\end{center}
\end{figure}

The obtained valence-quarks PDFs themselves are displayed in Fig.~\ref{fig:partonvalnceQ0} at the input scale of Q$_0^2 = 2 \, {\rm GeV}^2$ along with their $\Delta \chi^2_{\text{global}} =5.86$ (68 \% C.L.) uncertainty bands computed with the Hessian approach, for $xu_v (x, Q_0^2)$ and $xd_v (x, Q_0^2)$. For comparison, we also show the results from {\tt BBG}~\cite{Blumlein:2006be}, {\tt GJR08}~\cite{Gluck:2008gs} and up-to-date results from {\tt CJ15}~\cite{Accardi:2016qay} PDFs.
%------------------------------------------------------------------
\begin{table}
	\begin{tabular}{|c|c|c|c|}
		\hline \multicolumn{4}{|c|}{\rm NLO}   \\ \hline \hline
     	$h(x)$	& $ \alpha = 1.089$ & $ \beta = 1.132$ & $\gamma = 0.960$  \\  \hline
\end{tabular}
\caption{\label{tab:ht} {\sf Parameter values of the NLO HT fit at Q$_0^2 = 2 \, {\rm GeV}^2$. } }
\end{table}
%------------------------------------------------------------------
%------------------------------------------------------------------
%------------------------------------------------------------------

\begin{figure}[htb]
	\begin{center}
		\includegraphics[clip,width=0.46\textwidth]{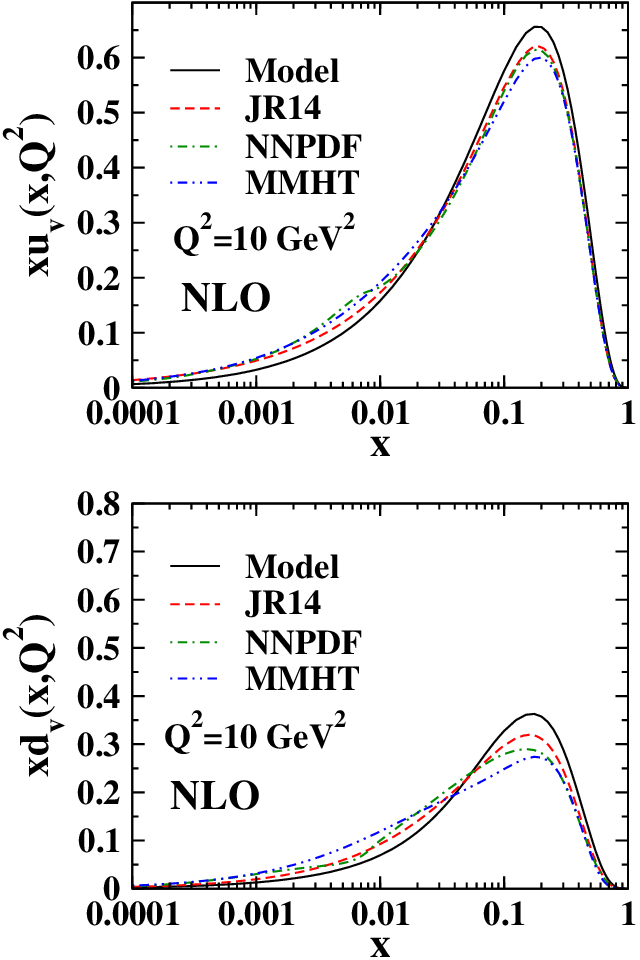}
		\caption{(Color online) The parton densities $xu_v (x, Q^2)$ and $xd_v (x, Q^2)$ at the
		    scale of Q$^2= 10 \, {\rm GeV}^2$. The dashed line is the {\tt JR14} PDFs~\cite{Jimenez-Delgado:2014twa},the dashed-dotted line is the
			{\tt NNPDF2.3} model~\cite{Ball:2012cx}, and the dashed-double-dotted line is the results from the {\tt MMHT14} group~\cite{Harland-Lang:2014zoa}. }\label{fig:partonvalnceQ10}
	\end{center}
\end{figure}

For the higher value of Q$^2$ (=10 GeV$^2$), we plot our $xu_v (x, Q^2)$ and $xd_v (x, Q^2)$ parton densities in Fig.~\ref{fig:partonvalnceQ10}.
The valence-quark densities from several recent representative NLO global parametrizations including {\tt JR14}~\cite{Jimenez-Delgado:2014twa}, {\tt NNPDF2.3}~\cite{Ball:2012cx} and {\tt MMHT14}~\cite{Harland-Lang:2014zoa} are also shown for comparison. As this plot shows, the results of our analysis and from different parametrizations are in good agreement.
%------------------------------------------------------------------
%--------------------------------

\begin{figure}[htb]
	\begin{center}
		\includegraphics[clip,width=0.47\textwidth]{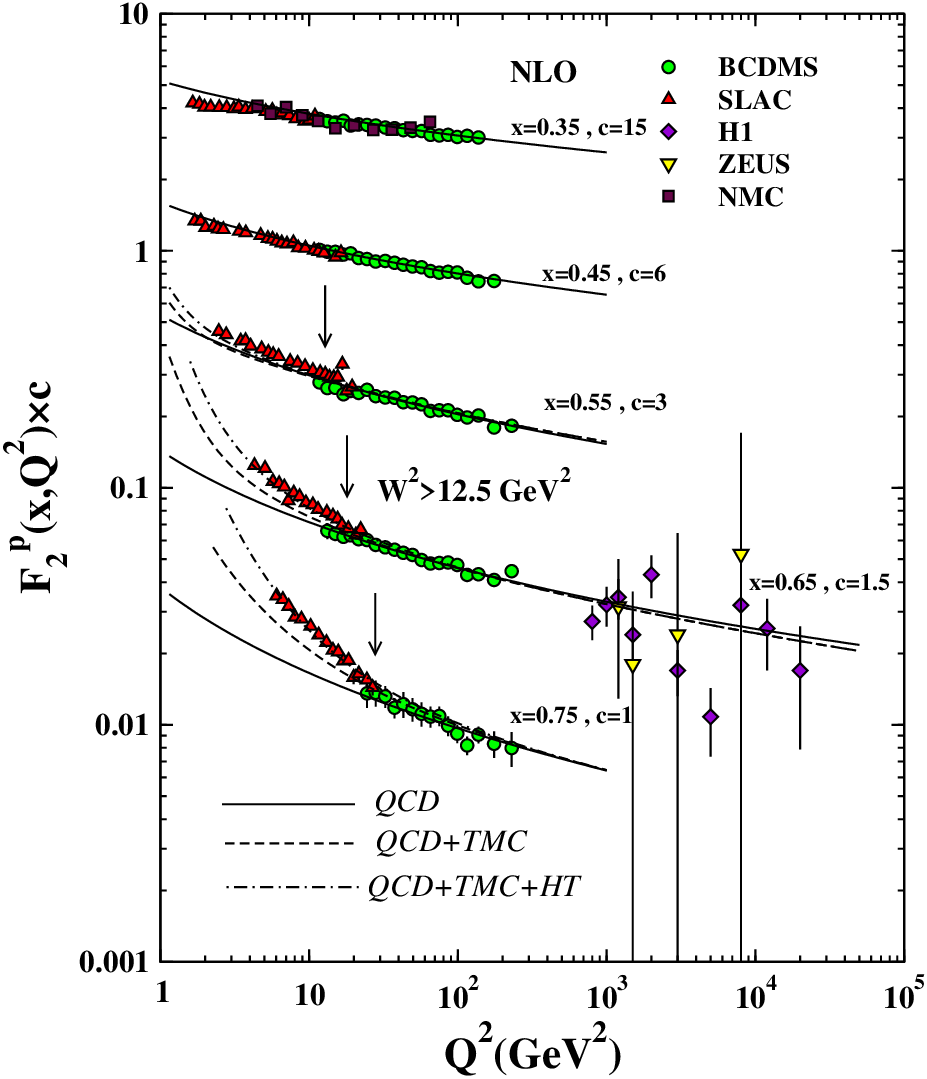}
		\caption{ (Color online) Comparison of proton structure function $F^p_2$ data from {\tt BCDMS}, {\tt SLAC}, {\tt NMC}, {\tt H1} and {\tt ZEUS} with our theory predictions, as a function of Q$^2$ for fixed values of $x$. The pure QCD fit in next-to-leading order is shown as a solid line, the contributions from target mass corrections (TMCs)
			are shown as dashed line, and the higher twist (HT) correction is shown as a dashed-dotted line. \label{fig:f2p}}
	\end{center}
\end{figure}

The quality of the fit to the data is illustrated in Fig.~\ref{fig:f2p}, where the inclusive proton $F_2^p$ structure functions
from {\tt BCDMS}, {\tt SLAC}, {\tt NMC}, {\tt H1} and {\tt ZEUS} are compared with our next-to-leading order fit as a function of Q$^2$ at approximately constant values of $x$.
The data have been scaled by a factor $c$, from $c=1$ for $x=0.75$ to $c=15$ for $x=0.35$. The vertical arrowed line in the plot indicates the regions with $W^2 > 12.5 \, {\rm GeV}^2$.

%--------------------------------
%--------------------------------

\begin{figure}[htb]
	\begin{center}
		\includegraphics[clip,width=0.47\textwidth]{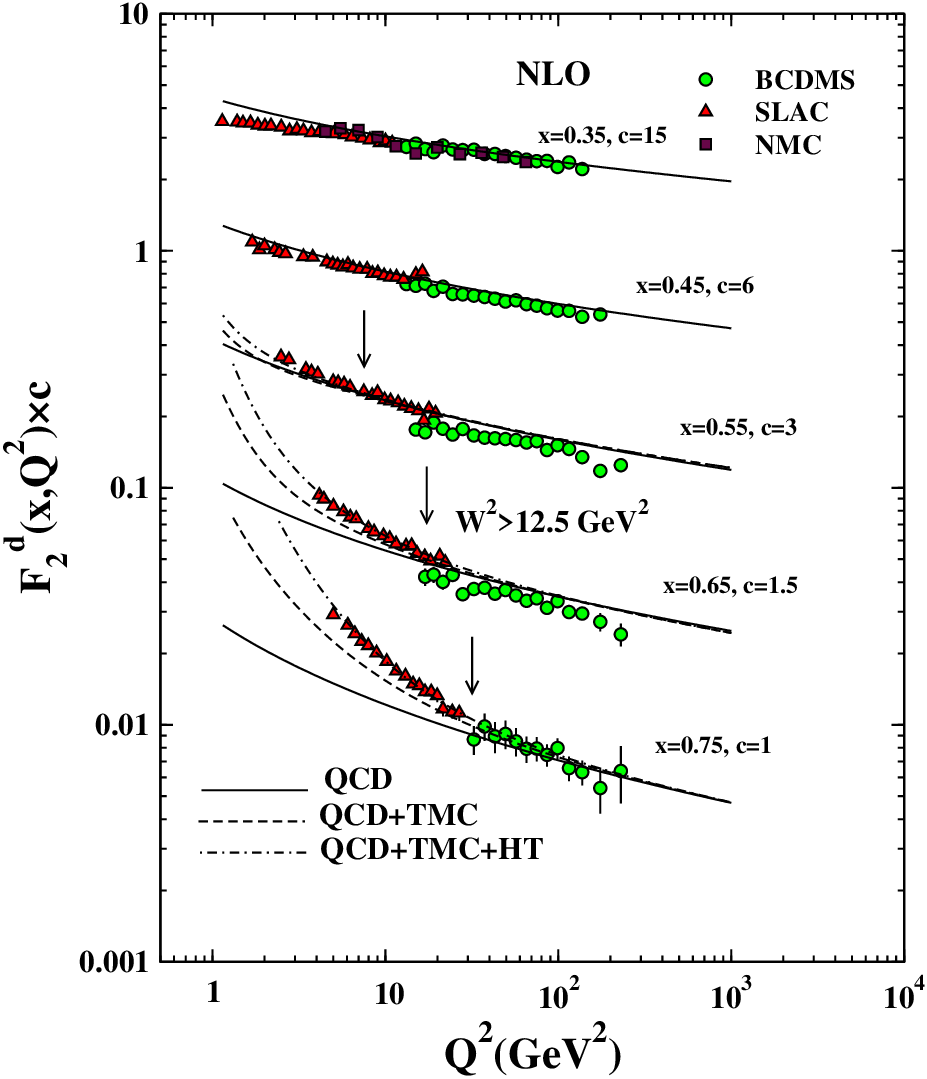}
		\caption{(Color online) Comparison of the structure function $F^d_2$ data from  {\tt BCDMS}, {\tt SLAC} and {\tt NMC} with our theory predictions,
as a function of Q$^2$ for the fixed values of $x$. The pure QCD fit in next-to-leading order is shown by a solid line, the contributions from target mass corrections (TMCs)
			shown as a dashed line, and the higher twist (HT)correction is shown as a dashed-dotted line.  \label{fig:f2d} }
	\end{center}
\end{figure}

In Fig.~\ref{fig:f2d}, detailed comparisons of the deuteron structure function $F^d_2$ data from the {\tt BCDMS}, {\tt SLAC} and {\tt NMC} experiments are shown with the theory predictions of our fit. The results have been plotted as a function of Q$^2$ with the corresponding $x$ ranges. The data have been
scaled by a factor $c$, from $c=1$ for $x=0.75$ to $c=15$ for $x=0.35$.

%--------------------------------
%--------------------------------

\begin{figure}[htb]
	\begin{center}
		\includegraphics[clip,width=0.47\textwidth]{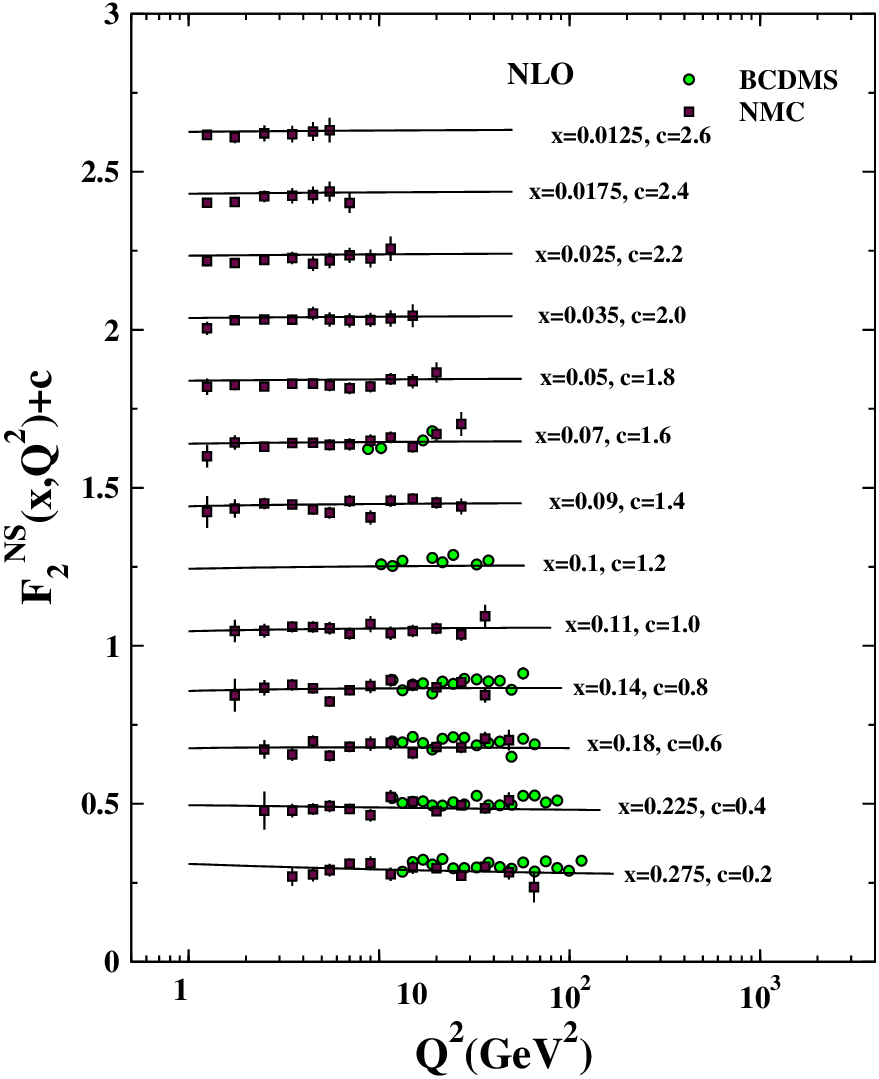}
		\caption{(Color online) Comparison of {\tt BCDMS} and {\tt NMC} data for the non-singlet structure function $F^{\rm NS}_2$ with our QCD predictions at next-to-leading order.  }\label{fig:f2ns}
	\end{center}
\end{figure}

Comparisons to data from {\tt BCDMS} and {\tt NMC} experiments for the non-singlet structure function $F^{\rm NS}_2$ are shown in Fig.~\ref{fig:f2ns}.
The data have been scaled by a factor $c$, from $c=0.2$ for $x=0.275$ to $c=2.6$ for $x=0.0125$.

%--------------------------------
%--------------------------------

One can see that our theory predictions based on analytical solutions using Laplace transform and Jacobi polynomials provide a very good description of the data.
When the effects of target mass corrections (TMCs) and higher-twist  (HT) corrections are included, the agreement between theory prediction and the data become strikingly better.
Figures.~\ref{fig:f2p} and \ref{fig:f2d} clearly present this result. The agreements between the theory prediction for $F^p_2$ and $F^d_2$ structure functions (including TMCs and HT) and data, over several decades of Q$^2$ and $x$, are also excellent.

%
%%%%%%%%%%%%%%%%%%%%%%%%%%%%%%%%%%%%%%%%%%    Summary and conclusion    %%%%%%%%%%%%%%%%%%%%%%%%%%%%%%%%%%%%%%%%%%%%%%%%%
%
\section{Summary and conclusion}\label{Sec:Summary}

We presented the next-to-leading order decoupled analytical evolution equations for singlet $F_{\rm S}(x, Q^2)$, gluon $G(x, Q^2)$ and non-singlet $F_{\rm NS}(x, Q^2)$ distributions, arising from the coupled DGLAP evolution equations in the Laplace s-space. We then rendered the results for valance quark distributions $x u_v$ and $x d_v$, the anti-quark distributions $x (\overline{d} + \overline{u})$ and $x\Delta=x(\overline{d}-\overline{u})$, the strange sea distribution $x s = x \overline{s}$ and the gluon distribution $x g$ initiated from {\tt KKT12} and {\tt GJR08} input parton distributions at Q$_0^2$ = 2 GeV$^2$. In this work, we also calculated the proton structure function $F_2^p(x, Q^2)$ using directly the Laplace transform technique, derived from corresponding analytical solutions for singlet $F_2^{\rm S}(x, Q^2)$, $F_2^{\rm G}(x, Q^2)$ and non-singlet $F_2^{\rm NS}(x, Q^2)$ structure functions. To determine the proton structure function at any arbitrary Q$^2$ scale, we only need to know the initial distributions for singlet, gluon and non-singlet distributions at the input scale Q$_0^2$.
The method presented in this analysis enable us to achieve strictly analytical solution for parton densities and structure function in terms of the $x$ variable.
We observed that the general solutions are in satisfactory agreements with the available experimental data and other parametrization models. In further research activities we  hope to report the results of the Laplace transform technique to get analytical solutions for heavy quark contributions of the proton structure function. Extension of the current result to the higher next-to-next-to-leading order (NNLO) approximation is also a valuable task to pursue in future.

We also applied our approach to extract the initial valence-quarks densities $xu_v$ and $xd_v$ from fit to DIS data for the non-singlet sector. The Laplace transform technique and Jacobi polynomial approach were used to performed the analysis. When using this approach, the target mass corrections (TMCs) and higher twist (HT) effects are taken into account in the analysis. The obtained results are in satisfactory agreements with the DIS data and other phenomenological models. We hope to apply these techniques to a global fit of the experimental neutrino-nucleon structure function $xF_3(x, Q^2)$ data in order to determine at the NLO approximation, the valence-quarks distributions $xu_v$ and $xd_v$, which can be used for the interpretation of results from future neutrino experiments.

In summary, there are various numerical methods to solve the DGLAP evolution equations to obtain the quarks and gluon
parton distribution functions. In this paper we have shown that the methods of the Laplace transforms technique are also the reliable and alternative schemes to obtain the analytical solution of these equations. The advantage of using such a technique is that it enables us to achieve strictly analytical solutions for the proton distribution functions $F_2^p (x, Q^2)$ in terms of the Bjorken-$x$ variable and virtuality Q$^2$.

%
%%%%%%%%%%%%%%%%%%%%%%%%%%%%%%%%%%%%%%%%%%    Acknowledgments    %%%%%%%%%%%%%%%%%%%%%%%%%%%%%%%%%%%%%%%%%%%%%%%%%
%
\section*{Acknowledgments}

The authors would like to thank Andrei Kataev, Loyal Durand and F. Taghavi-Shahri for reading the manuscript and for fruitful discussion and critical remarks.
A. M. acknowledges Yazd University for facilities provided to do this project. Hamzeh Khanpour is indebted the University of Science and Technology of Mazandaran and the School of Particles and Accelerators, Institute for Research in Fundamental Sciences (IPM) for financial supoprt of this research. Hamzeh Khanpour also is grateful for the hospitality of the Theory Division at CERN where this work was completed.

\clearpage

%
%%%%%%%%%%%%%%%%%%%%%%%%%%%%%%%%%%%%%%%%%%    Appendix A    %%%%%%%%%%%%%%%%%%%%%%%%%%%%%%%%%%%%%%%%%%%%%%%%%
%
\section*{Appendix A: The Laplace transforms of splitting functions at the NLO approximation}\label{Sec:AppendixA}

We present here  the Laplace transforms of the splitting functions  for quark and gluon sectors, denoted by $\Phi^{\rm  NLO}$ and $\Theta^{\rm  NLO}$  respectively at the
next-to-leading order approximation which we used in Eqs.(\ref{eq:spliLOandNLO}) and (\ref{eq:fi-nonsinglet}). We fixed the usual quadratic Casimir operators to their exact values, using $C_A = 3$, ${\rm T_F} = f$ and $C_F = 4/3$. The $\psi(s)$ is defined by $\psi(s) = \frac{d} {ds} {\rm ln} \Gamma(s)$ and $\gamma _E = 0.577216$ is the Euler-Lagrange constant.

\begin{widetext}
	
\begin{eqnarray*}
&&\Phi_{\rm{NS},qq}^{\rm{NLO}} = \hspace{14 cm}\text{(A.1)}\\
&&C_F T_F\left(-\frac{2}{3 (1+s)^2}-\frac{2}{9 (1+s)}-\frac{2}{3 (2+s)^2}+\frac{22}{9 (2+s)}+\frac{20 \left(\gamma _E+\psi (s+1)\right)}{9}+\frac{4}{3} \psi '(s+1)\right)+\\
&&C_F{}^2 \left(-\frac{1}{(1+s)^3}-\frac{5}{1+s}-\frac{1}{(2+s)^3}+\frac{2}{(2+s)^2}+\frac{5}{2+s}+\right.\\
&&\frac{2 \left(\gamma _E+\frac{1}{1+s}+\psi (s+1)-(1+s) \psi '(s+2)\right)}{(1+s)^2}+
\frac{2 \left(\gamma _E+\frac{1}{2+s}+\psi (s+2)-(2+s) \psi '(s+3)\right)}{(2+s)^2}-\\
&&\left.4 \left(\left(\gamma _E+\psi (s+1)\right) \psi '(s+1)-\frac{1}{2} \psi ''(s+1)\right)+3 \psi '(s+1)\right)+\\
&&C_A C_F \left(-\frac{1}{(1+s)^3}+\frac{5}{6 (1+s)^2}+\frac{53}{18 (1+s)}+\frac{\pi ^2}{6 (1+s)}-\frac{1}{(2+s)^3}+\right.\\
&&\frac{5}{6 (2+s)^2}-\frac{187}{18 (2+s)}+\frac{\pi ^2}{6 (2+s)}-\frac{67 \left(\gamma _E+\psi (s+1)\right)}{9}+\\
&&\left.\frac{1}{3} \pi ^2 \left(\gamma _E+\psi (s+1)\right)-\frac{11}{3} \psi '(s+1)-\psi ''(s+1)\right)\,,
\end{eqnarray*}

\begin{eqnarray*}
&&\Phi_{\rm{NS},q\bar{q}}^{\rm{NLO}}=\hspace{14 cm}\text{(A.2)}\\
&&C_F \left(-\frac{C_A }{2}+C_F \right) \\
&&\left(-\frac{2}{(1+s)^3}-\frac{2}{(1+s)^2}+\frac{4}{1+s}+\frac{\pi ^2}{3 (1+s)}+\frac{2}{(2+s)^3}-\frac{2}{(2+s)^2}-\frac{8}{2+s}-\frac{\pi
^2}{3 (2+s)}+\frac{5}{3+s}-\right.\\
&&\frac{13}{9 (4+s)}+\frac{25}{36 (5+s)}-\frac{41}{100 (6+s)}+\frac{4}{25 (7+s)}-\left(\gamma _E+\psi \left(2+\frac{s}{2}\right)\right)-\\
&&\frac{1}{3} \pi ^2 \left(-\left(\gamma _E+\psi \left(\frac{1+s}{2}\right)\right)+\left(\gamma _E+\psi \left(1+\frac{s}{2}\right)\right)\right)+\left(\gamma
_E+\psi \left(\frac{3+s}{2}\right)\right)+\\
&&4 \left(-\left(\gamma _E+\psi \left(1+\frac{s}{2}\right)\right)+\left(\gamma _E+\psi \left(\frac{3+s}{2}\right)\right)\right)+\frac{4}{9} \left(-\left(\gamma
_E+\psi \left(2+\frac{s}{2}\right)\right)+\left(\gamma _E+\psi \left(\frac{5+s}{2}\right)\right)\right)+\\
&&\frac{1}{4} \left(-\left(\gamma _E+\psi \left(3+\frac{s}{2}\right)\right)+\left(\gamma _E+\psi \left(\frac{5+s}{2}\right)\right)\right)+\frac{4}{25}
\left(-\left(\gamma _E+\psi \left(3+\frac{s}{2}\right)\right)+\left(\gamma _E+\psi \left(\frac{7+s}{2}\right)\right)\right)+\\
&&\frac{1}{(1+s)^3}\left(-8+(1+s) \text{Ln}(16)+2 (1+s) \psi \left(1+\frac{s}{2}\right)-2 (1+s) \psi \left(\frac{1+s}{2}\right)-(1+s)^2 \psi '\left(1+\frac{s}{2}\right)+\right.\\
&&\left.(1+s)^2 \psi '\left(\frac{1+s}{2}\right)\right)-
\frac{0.99984}{(2+s)^3}\left(\frac{16}{(1+s)^2}+\frac{12 s}{(1+s)^2}+(2+s) \text{Ln}(16)-2 (2+s) \psi \left(1+\frac{s}{2}\right)+2 (2+s) \psi
\left(\frac{1+s}{2}\right)+\right.\allowdisplaybreaks[1]\\
&&\left.(2+s)^2 \psi '\left(1+\frac{s}{2}\right)-(2+s)^2 \psi '\left(\frac{1+s}{2}\right)\right)-\\
&&\frac{1.9702}{(3+s)^3}\left(\frac{164+284 s+188 s^2+60 s^3+8 s^4}{(1+s)^2 (2+s)^2}-4 (3+s) \text{Ln}(2)-2 (3+s) \psi \left(1+\frac{s}{2}\right)+\right.\\
&&\left.2 (3+s) \psi \left(\frac{1+s}{2}\right)+(3+s)^2 \psi '\left(1+\frac{s}{2}\right)-(3+s)^2 \psi '\left(\frac{1+s}{2}\right)\right)-\\
&&\frac{1.801}{(4+s)^3} \left(\frac{2176+4392 s+3504 s^2+1408 s^3+288 s^4+24 s^5}{(1+s)^2 (2+s)^2 (3+s)^2}+4 (4+s) \text{Ln}(2)-\right.\\
&&\left.2 (4+s) \psi \left(1+\frac{s}{2}\right)+2 (4+s) \psi \left(\frac{1+s}{2}\right)+(4+s)^2 \psi '\left(1+\frac{s}{2}\right)-(4+s)^2 \psi
'\left(\frac{1+s}{2}\right)\right)-\\
&&\frac{1.3242}{(5+s)^3}\left(\left.\left(57328+146144 s+162160 s^2+103728 s^3+42144 s^4+11160 s^5+1880 s^6+184 s^7+8 s^8\right)\right/\right.\\
&&\left((1+s)^2 (2+s)^2 (3+s)^2 (4+s)^2\right)-4 (5+s) \text{Ln}(2)-2 (5+s) \psi \left(1+\frac{s}{2}\right)+\\
&&\left.2 (5+s) \psi \left(\frac{1+s}{2}\right)+(5+s)^2 \psi '\left(1+\frac{s}{2}\right)-(5+s)^2 \psi '\left(\frac{1+s}{2}\right)\right)-\\
&&\frac{0.6348}{(6+s)^2} \left(\text{Ln}(16)-2\psi \left(4+\frac{s}{2}\right)+2 \psi \left(\frac{7+s}{2}\right)+(6+s) \psi '\left(4+\frac{s}{2}\right)-(6+s)
\psi '\left(\frac{7+s}{2}\right)\right)+\\
&&\frac{0.1398}{(7+s)^2}\left(\text{Ln}(16)+2\psi \left(4+\frac{s}{2}\right)-2 \psi \left(\frac{9+s}{2}\right)-(7+s) \psi '\left(4+\frac{s}{2}\right)+(7+s)\psi
'\left(\frac{9+s}{2}\right)\right)+\\
&&\left.\frac{1}{4} \left(\psi''\left(\frac{2+s}{2}\right)- \psi''\left(\frac{1+s}{2}\right)\right)\right)\,,
\end{eqnarray*}

\begin{eqnarray*}
&&\Phi_{qq}^{\rm{NLO}}=\hspace{14 cm}\text{(A.3)}\\
&&C_F T_F \left(\frac{40}{9 s}-\frac{4}{(1+s)^3}-\frac{8}{3 (1+s)^2}-\frac{38}{9 (1+s)}-\frac{4}{(2+s)^3}-\frac{32}{3 (2+s)^2}+\frac{130}{9 (2+s)}-\frac{16}{3
(3+s)^2}-\right.\\
&&\left.\frac{112}{9 (3+s)}+\frac{20 \left(\gamma _E+\psi (s+1)\right)}{9}+\frac{4}{3} \psi '(s+1)\right)+\\
&&C_F^2 \left(\frac{1}{(1+s)^3}-\frac{2}{(1+s)^2}-\frac{1}{1+s}-\frac{\pi ^2}{3 (1+s)}-\frac{2.999}{(2+s)^3}+\frac{8.289}{2+s}+\frac{\text{{``}3.9404{''}}}{(3+s)^3}-\frac{11.481}{3+s}-\right.\\
&&\frac{3.602}{(4+s)^3}+\frac{15.25}{4+s}+\frac{2.648}{(5+s)^3}-\frac{14.225}{5+s}-\frac{1.269}{(6+s)^3}+\frac{10.472}{6+s}+\frac{0.279}{(7+s)^3}-\frac{5.776}{7+s}+\\
&&\frac{2.548}{8+s}-\frac{1.0411}{9+s}+\frac{0.4327}{10+s}-\frac{0.136}{11+s}+\frac{0.0224}{12+s}+\\
&&\frac{1}{(1+s)^3}\left(-8+(1+s)\text{Ln}(16)+2 (1+s) \psi \left(1+\frac{s}{2}\right)-2 (1+s) \psi \left(\frac{1+s}{2}\right)-\right.\\
&&\left.(1+s)^2 \psi '\left(1+\frac{s}{2}\right)+(1+s)^2 \psi '\left(\frac{1+s}{2}\right)\right)-\\
&&\frac{0.999}{(2+s)^3} \left(\frac{16}{(1+s)^2}+\frac{12 s}{(1+s)^2}+(2+s)\text{Ln}(16)-2 (2+s) \psi \left(1+\frac{s}{2}\right)+2 (2+s) \psi
\left(\frac{1+s}{2}\right)+\right.\\
&&\left.(2+s)^2 \psi '\left(1+\frac{s}{2}\right)-(2+s)^2 \psi '\left(\frac{1+s}{2}\right)\right)-\\
&&\frac{1.9702}{(3+s)^3}\left(\frac{164+284 s+188 s^2+60 s^3+8 s^4}{(1+s)^2 (2+s)^2}-4 (3+s) \text{Ln}[2]-2 (3+s) \psi \left(1+\frac{s}{2}\right)+\right.\\
&&\left.2 (3+s) \psi \left(\frac{1+s}{2}\right)+(3+s)^2 \psi '\left(1+\frac{s}{2}\right)-(3+s)^2 \psi '\left(\frac{1+s}{2}\right)\right)-\allowdisplaybreaks[1]\\
&&\frac{1.801}{(4+s)^3}\left(\frac{2176+4392 s+3504 s^2+1408 s^3+288 s^4+24 s^5}{(1+s)^2 (2+s)^2 (3+s)^2}+4 (4+s) \text{Ln}[2]-\right.\\
&&\left.2 (4+s) \psi \left(1+\frac{s}{2}\right)+2 (4+s) \psi \left(\frac{1+s}{2}\right)+(4+s)^2 \psi '\left(1+\frac{s}{2}\right)-(4+s)^2 \psi
'\left(\frac{1+s}{2}\right)\right)-\\
&&\frac{1.3242 }{(5+s)^3}\left(\frac{57328+146144 s+162160 s^2+103728 s^3+42144 s^4+11160 s^5+1880 s^6+184 s^7+8 s^8}{(1+s)^2 (2+s)^2 (3+s)^2 (4+s)^2}-\right.\\
&&4 (5+s) \text{Ln}[2]-2 (5+s) \psi \left(1+\frac{s}{2}\right)+2 (5+s) \psi \left(\frac{1+s}{2}\right)+(5+s)^2 \psi '\left(1+\frac{s}{2}\right)-\\
&&\left.(5+s)^2 \psi '\left(\frac{1+s}{2}\right)\right)+\frac{2 \left(\gamma _E+\frac{1}{1+s}+\psi (s+1)-(1+s) \psi '(s+2)\right)}{(1+s)^2}+\\
&&\frac{2 \left(\gamma _E+\frac{1}{2+s}+\psi (s+2)-(2+s) \psi '(s+3)\right)}{(2+s)^2}-\\
&&\frac{0.6348}{(6+s)^2} \left(\text{Ln}(16)-2 \psi \left(4+\frac{s}{2}\right)+2 \psi \left(\frac{7+s}{2}\right)+(6+s)\psi '\left(4+\frac{s}{2}\right)-(6+s)
\psi '\left(\frac{7+s}{2}\right)\right)+\\
&&\frac{0.1398}{(7+s)^2} \left(\text{Ln}(16)+2 \psi \left(4+\frac{s}{2}\right)-2 \psi \left(\frac{9+s}{2}\right)-(7+s)\psi '\left(4+\frac{s}{2}\right)+(7+s)
\psi '\left(\frac{9+s}{2}\right)\right)-\\
&&\left.4 \left(\left(\gamma _E+\psi (s+1)\right) \psi '(s+1)-\frac{1}{2} \psi ''(s+1)\right)+3 \psi '(s+1)\right)+\\
&&C_A C_F \left(-\frac{2}{(1+s)^3}+\frac{11}{6 (1+s)^2}+\frac{17}{18 (1+s)}+\frac{\pi ^2}{3 (1+s)}-\frac{0.00016}{(2+s)^3}+\frac{11}{6 (2+s)^2}-\frac{10.39}{2+s}-\right.\\
&&\frac{1.97}{(3+s)^3}+\frac{5.74}{3+s}+\frac{1.801}{(4+s)^3}-\frac{7.625}{4+s}-\frac{1.32}{(5+s)^3}+\frac{7.112}{5+s}+\frac{0.634}{(6+s)^3}-\frac{5.23}{6+s}-\frac{0.139}{(7+s)^3}+\\
&&\frac{2.88}{7+s}-\frac{1.274}{8+s}+\frac{0.5205}{9+s}-\frac{0.216}{10+s}+\frac{0.068}{11+s}-\frac{0.0112}{12+s}-\frac{67 \left(\gamma _E+\psi
(s+1)\right)}{9}+\\
&&\frac{1}{3} \pi ^2 \left(\gamma _E+\psi (s+1)\right)-\\
&&\frac{1}{2 (1+s)^3}\left(-8+(1+s)\text{Ln}(16)+2 (1+s) \psi \left(1+\frac{s}{2}\right)-2 (1+s) \psi \left(\frac{1+s}{2}\right)-\right.\\
&&\left.(1+s)^2 \psi '\left(1+\frac{s}{2}\right)+(1+s)^2 \psi '\left(\frac{1+s}{2}\right)\right)+\\
&&\frac{0.49992}{(2+s)^3}\left(\frac{16}{(1+s)^2}+\frac{12 s}{(1+s)^2}+(2+s)\text{Ln}(16)-2 (2+s) \psi \left(1+\frac{s}{2}\right)+2 (2+s) \psi
\left(\frac{1+s}{2}\right)+\right.\\
&&\left.(2+s)^2 \psi '\left(1+\frac{s}{2}\right)-(2+s)^2 \psi '\left(\frac{1+s}{2}\right)\right)+\\
&&\frac{0.9851}{(3+s)^3} \left(\frac{164+284 s+188 s^2+60 s^3+8 s^4}{(1+s)^2 (2+s)^2}-4 (3+s) \text{Ln}[2]-2 (3+s) \psi \left(1+\frac{s}{2}\right)+\right.\\
&&\left.2 (3+s) \psi \left(\frac{1+s}{2}\right)+(3+s)^2 \psi '\left(1+\frac{s}{2}\right)-(3+s)^2 \psi '\left(\frac{1+s}{2}\right)\right)+\\
&&\frac{0.9005}{(4+s)^3} \left(\frac{2176+4392 s+3504 s^2+1408 s^3+288 s^4+24 s^5}{(1+s)^2 (2+s)^2 (3+s)^2}+4 (4+s) \text{Ln}[2]-\right.\\
&&\left.2 (4+s) \psi \left(1+\frac{s}{2}\right)+2 (4+s) \psi \left(\frac{1+s}{2}\right)+(4+s)^2 \psi '\left(1+\frac{s}{2}\right)-(4+s)^2 \psi
'\left(\frac{1+s}{2}\right)\right)+\\
&&\frac{0.6621}{(5+s)^3}\left(\frac{57328+146144 s+162160 s^2+103728 s^3+42144 s^4+11160 s^5+1880 s^6+184 s^7+8 s^8}{(1+s)^2 (2+s)^2 (3+s)^2 (4+s)^2}-\right.\\
&&4 (5+s) \text{Ln}[2]-2 (5+s) \psi \left(1+\frac{s}{2}\right)+2 (5+s) \psi \left(\frac{1+s}{2}\right)+(5+s)^2 \psi '\left(1+\frac{s}{2}\right)-\\
&&\left.(5+s)^2 \psi '\left(\frac{1+s}{2}\right)\right)+\frac{0.3174}{(6+s)^2}\left(\text{Ln}(16)-2 \psi \left(4+\frac{s}{2}\right)+2 \psi \left(\frac{7+s}{2}\right)+(6+s)\psi '\left(4+\frac{s}{2}\right)-(6+s)
\psi '\left(\frac{7+s}{2}\right)\right)-\\
&&\frac{0.0699}{(7+s)^2}\left(\text{Ln}(16)+2 \psi \left(4+\frac{s}{2}\right)-2 \psi \left(\frac{9+s}{2}\right)-(7+s)\psi '\left(4+\frac{s}{2}\right)+(7+s)
\psi '\left(\frac{9+s}{2}\right)\right)-\left.\frac{11}{3} \psi '(s+1)-\psi ''(s+1)\right)\,,
\end{eqnarray*}

\begin{eqnarray*}
&&\Theta_f^{\rm{NLO}}=\hspace{14 cm}\text{(A.4)}\\
&&C_F T_F \left(-\frac{2}{(1+s)^3}+\frac{1}{(1+s)^2}+\frac{14}{1+s}-\frac{2 \pi ^2}{3 (1+s)}+\frac{4}{(2+s)^3}-\frac{4}{(2+s)^2}-\frac{29}{2+s}+\right.\\
&&\frac{4 \pi ^2}{3 (2+s)}+\frac{20}{3+s}-\frac{4 \pi ^2}{3 (3+s)}+\frac{4 \left(\gamma _E+\psi (s+1)\right)}{1+s}-\frac{4 \left(\gamma _E+\psi
(s+2)\right)}{1+s}-\\
&&\frac{8 \left(\gamma _E+\psi (s+2)\right)}{2+s}+\frac{8 \left(\gamma _E+\psi (s+3)\right)}{3+s}+\frac{2 \left(\pi ^2+6 \left(\gamma _E+\psi
(s+1)\right){}^2+6 \psi '(s+1)\right)}{6+6 s}-\\
&&\frac{4 \left(\pi ^2+6 \left(\gamma _E+\psi (s+2)\right){}^2+6 \psi '(s+2)\right)}{12+6 s}+\\
&&\left.\frac{4 \left(\pi ^2+6\left(\gamma _E+\psi (s+3)\right){}^2+6 \psi '(s+3)\right)}{18+6 s}\right)+\\
&&C_A T_F \left(\frac{40}{9 s}-\frac{4}{(1+s)^3}-\frac{2}{(1+s)^2}-\frac{4}{1+s}-\frac{8}{(2+s)^3}-\frac{16}{(2+s)^2}+\frac{54}{2+s}-\frac{4 \pi
^2}{3 (2+s)}-\right.\\
&&\frac{88}{3 (3+s)^2}-\frac{373}{9 (3+s)}+\frac{58}{9 (4+s)}-\frac{49}{36 (5+s)}+\frac{247}{450 (6+s)}-\frac{9}{50 (7+s)}+\\
&&\frac{8}{25 (8+s)}+\frac{8 \left(\gamma _E+\psi (s+3)\right)}{2+s}-\frac{8 \left(\gamma _E+\psi (s+4)\right)}{3+s}+\\
&&\frac{1}{(1+s)^3}\left(-8+(1+s) \text{Ln}(16)+2 (1+s)\psi \left(1+\frac{s}{2}\right)-2 (1+s)\psi \left(\frac{1+s}{2}\right)-\right.\\
&&\left.(1+s)^2 \psi '\left(1+\frac{s}{2}\right)+(1+s)^2\psi '\left(\frac{1+s}{2}\right)\right)+\\
&&\frac{1}{(2+s)^3}2 \left(\frac{16}{(1+s)^2}+\frac{12 s}{(1+s)^2}+(2+s) \text{Ln}(16)-2 (2+s)\psi \left(1+\frac{s}{2}\right)+\right.\\
&&\left.2 (2+s)\psi \left(\frac{1+s}{2}\right)+(2+s)^2 \psi '\left(1+\frac{s}{2}\right)-(2+s)^2\psi '\left(\frac{1+s}{2}\right)\right)-\\
&&\frac{1}{(3+s)^3}2 \left(\frac{164+284 s+188 s^2+60 s^3+8 s^4}{(1+s)^2 (2+s)^2}-4 (3+s) \text{Ln}(2)-\right.\\
&&2 (3+s)\psi \left(1+\frac{s}{2}\right)+2 (3+s)\psi \left(\frac{1+s}{2}\right)+(3+s)^2 \psi '\left(1+\frac{s}{2}\right)-\\
&&\left.(3+s)^2\psi '\left(\frac{1+s}{2}\right)\right)-\frac{2 \left(\pi ^2+6\left(\gamma _E+\psi (s+2)\right){}^2-6 \psi '(s+2)\right)}{6+6 s}+\\
&&\frac{4 \left(\pi ^2+6 \left(\gamma _E+\psi (s+3)\right){}^2-6 \psi '(s+3)\right)}{12+6 s}-\\
&&\left.\frac{4 \left(\pi ^2+6 \left(\gamma _E+\psi (s+4)\right){}^2-6\psi '(s+4)\right)}{18+6 s}\right)\,,
\end{eqnarray*}

\begin{eqnarray*}
&&\Theta_g^{\rm{NLO}}=\hspace{14 cm}\text{(A.5)}\\
&&C_F T_F \left(-\frac{40}{9 s}+\frac{40}{9 (1+s)}-\frac{32}{9 (2+s)}+\frac{8 \left(\gamma _E+\psi (s+1)\right)}{3 s}-\frac{8\left(\gamma _E+\psi
(s+2)\right)}{3 (1+s)}+\right.\\
&&\left.\frac{4 \left(\gamma _E+\psi (s+3)\right)}{3 (2+s)}\right)+\\
&&C_F^2 \left(-\frac{2}{(1+s)^3}-\frac{2}{(1+s)^2}-\frac{5}{2 (1+s)}+\frac{1}{(2+s)^3}-\frac{7}{2 (2+s)^2}-\frac{7}{2 (2+s)}+\right.\\
&&\frac{6 \left(\gamma _E+\psi (s+1)\right)}{s}-\frac{6 \left(\gamma _E+\psi (s+2)\right)}{1+s}+\frac{5 \left(\gamma _E+\psi (s+3)\right)}{2+s}-\\
&&\frac{\pi ^2+6 \left(\gamma _E+\psi (s+1)\right){}^2-6 \psi '(s+1)}{3 s}+\frac{2 \left(\pi ^2+6 \left(\gamma _E+\psi (s+2)\right){}^2-6 \psi '(s+2)\right)}{6+6 s}-\\
&&\left.\frac{\pi ^2+6\left(\gamma _E+\psi (s+3)\right){}^2-6 \psi '(s+3)}{12+6 s}\right)+\\
&&C_A C_F \left(\frac{1}{s}+\frac{4}{(1+s)^3}+\frac{12}{(1+s)^2}-\frac{17}{9 (1+s)}+\frac{2 \pi ^2}{3 (1+s)}+\frac{2}{(2+s)^3}+\frac{5}{(2+s)^2}+\right.\\
&&\frac{10}{9 (2+s)}+\frac{8}{3 (3+s)^2}+\frac{31}{9 (3+s)}+\frac{11}{36 (4+s)}-\frac{119}{900 (5+s)}-\frac{7}{200 (6+s)}-\\
&&\frac{2}{25 (7+s)}-\frac{22 \left(\gamma _E+\psi (s+1)\right)}{3 s}+\frac{22 \left(\gamma _E+\psi (s+2)\right)}{3 (1+s)}-\frac{17 \left(\gamma
_E+\psi (s+3)\right)}{3 (2+s)}-\\
&&\frac{1}{s^2}\left(\text{Ln}(16)-2 \psi \left(1+\frac{s}{2}\right)+2 \psi \left(\frac{1+s}{2}\right)+s \psi '\left(1+\frac{s}{2}\right)-s \psi
'\left(\frac{1+s}{2}\right)\right)-\\
&&\frac{1}{(1+s)^3}\left(-8+(1+s) \text{Ln}(16)+2 (1+s) \psi \left(1+\frac{s}{2}\right)-2 (1+s) \psi \left(\frac{1+s}{2}\right)-\right.\\
&&\left.(1+s)^2 \psi '\left(1+\frac{s}{2}\right)+(1+s)^2 \psi '\left(\frac{1+s}{2}\right)\right)-\\
&&\frac{1}{2 (2+s)^3}\left(\frac{16}{(1+s)^2}+\frac{12 s}{(1+s)^2}+(2+s) \text{Ln}(16)-2 (2+s) \psi \left(1+\frac{s}{2}\right)+\right.\\
&&\left.2 (2+s) \psi \left(\frac{1+s}{2}\right)+(2+s)^2 \psi '\left(1+\frac{s}{2}\right)-(2+s)^2 \psi '\left(\frac{1+s}{2}\right)\right)+\\
&&\frac{\pi ^2+6\left(\gamma _E+\psi (s+1)\right){}^2-6 \psi '(s+1)}{3 s}-\frac{4 \left(1+\gamma _E s+s (\psi (s)-s \psi '(s+1))\right)}{s^3}-\\
&&\frac{2 \left(\pi ^2+6 \left(\gamma _E+\psi (s+2)\right){}^2-6 \psi '(s+2)\right)}{6+6 s}+\frac{4 \left(\gamma _E+\frac{1}{1+s}+\psi (s+1)-(1+s) \psi '(s+2)\right)}{(1+s)^2}+\\
&&\frac{\pi ^2+6 \left(\gamma _E+\psi (s+3)\right){}^2-6 \psi '(s+3)}{12+6 s}-\left.\frac{2 \left(\gamma _E+\frac{1}{2+s}+\psi (s+2)-(2+s) \psi '(s+3)\right)}{(2+s)^2}\right)\,,
\end{eqnarray*}

\begin{eqnarray*}
&&\Phi_g^{\rm{NLO}}=\hspace{14 cm}\text{(A.6)}\\
&&C_FT_F\left(\frac{4}{3 s}-\frac{4}{(1+s)^3}+\frac{6}{(1+s)^2}-\frac{16}{1+s}-\frac{4}{(2+s)^3}+\frac{10}{(2+s)^2}+\frac{8}{2+s}+\frac{20}{3
(3+s)}\right) +\\
&&C_A T_F \left(-\frac{46}{9 s}+\frac{4}{3 (1+s)^2}+\frac{58}{9 (1+s)}+\frac{4}{3 (2+s)^2}-\frac{38}{9 (2+s)}+\frac{46}{9 (3+s)}+\right.\\
&&\left.\frac{20 \left(\gamma _E+\psi '(s+1)\right)}{9}\right)+\\
&&C_A{}^2 \left(\frac{2}{(1+s)^3}+\frac{25}{3 (1+s)^2}-\frac{97}{18 (1+s)}+\frac{\pi ^2}{1+s}+\frac{\text{{``}6.{''}}}{(2+s)^3}-\frac{11}{3 (2+s)^2}-\right.\\
&&\frac{2.476}{2+s}-\frac{\pi ^2}{3 (2+s)}-\frac{2.029}{(3+s)^3}+\frac{44}{3 (3+s)^2}-\frac{7.395}{3+s}+\frac{\pi ^2}{3 (3+s)}-\\
&&\frac{1.801}{(4+s)^3}+\frac{4.708}{4+s}+\frac{1.324}{(5+s)^3}-\frac{6.564}{5+s}-\frac{0.635}{(6+s)^3}+\frac{4.926}{6+s}+\frac{0.139}{(7+s)^3}-\\
&&\frac{2.878}{7+s}+\frac{1.114}{8+s}-\frac{0.521}{9+s}+\frac{0.217}{10+s}-\frac{0.0683}{11+s}+\frac{0.0112}{12+s}-\frac{67 \left(\gamma _E+\psi
'(s+1)\right)}{9}+\\
&&\frac{1}{3} \pi ^2 \left(\gamma _E+\psi '(s+1)\right)-\frac{1}{s^2}\left(\text{Ln}(16)-2 \psi \left(1+\frac{s}{2}\right)+2 \psi \left(\frac{1+s}{2}\right)+s \psi '\left(1+\frac{s}{2}\right)-s \psi
'\left(\frac{1+s}{2}\right)\right)-\\
&&\frac{1}{(1+s)^3}\left(-8+(1+s) \text{Ln}(16)+2 (1+s) \psi \left(1+\frac{s}{2}\right)-2 (1+s) \psi \left(\frac{1+s}{2}\right)-\right.\\
&&\left.(1+s)^2 \psi '\left(1+\frac{s}{2}\right)+(1+s)^2 \psi '\left(\frac{1+s}{2}\right)\right)-\\
&&\frac{1.99992}{(2+s)^3} \left(\frac{16+12 s}{(1+s)^2}+(2+s) \text{Ln}(16)-2 (2+s) \psi \left(1+\frac{s}{2}\right)+\right.\\
&&\left.2 (2+s) \psi \left(\frac{1+s}{2}\right)+(2+s)^2 \psi '\left(1+\frac{s}{2}\right)-(2+s)^2 \psi '\left(\frac{1+s}{2}\right)\right)+\\
&&\frac{0.0149}{(3+s)^3}\left(\frac{164+284 s+188 s^2+60 s^3+8 s^4}{(1+s)^2 (2+s)^2}-4 (3+s) \text{Ln}(2)-\right.\\
&&2 (3+s) \psi \left(1+\frac{s}{2}\right)+2 (3+s) \psi \left(\frac{1+s}{2}\right)+(3+s)^2 \psi '\left(1+\frac{s}{2}\right)-\\
&&\left.(3+s)^2 \psi '\left(\frac{1+s}{2}\right)\right)-\frac{0.9005}{(4+s)^3}\left(\frac{2176+4392 s+3504 s^2+1408 s^3+288 s^4+24 s^5}{(1+s)^2 (2+s)^2 (3+s)^2}+\right.\\
&&4 (4+s) \text{Ln}(2)-2 (4+s) \psi \left(1+\frac{s}{2}\right)+2 (4+s) \psi \left(\frac{1+s}{2}\right)+\left.(4+s)^2 \psi '\left(1+\frac{s}{2}\right)-(4+s)^2 \psi '\left(\frac{1+s}{2}\right)\right)-\\
&&\frac{0.6621}{(5+s)^3} \left(\left(57328+146144 s+162160 s^2+103728 s^3+42144 s^4+11160 s^5+\right.\right.  \\
&&\left.1880 s^6+184 s^7+8 s^8\right)/\left((1+s)^2 (2+s)^2 (3+s)^2 (4+s)^2\right)-\\
&&4 (5+s) \text{Ln}(2)-2 (5+s) \psi \left(1+\frac{s}{2}\right)+2 (5+s) \psi \left(\frac{1+s}{2}\right)+\\
&&\left.(5+s)^2 \psi '\left(1+\frac{s}{2}\right)-(5+s)^2 \psi '\left(\frac{1+s}{2}\right)\right)-\frac{4 \left(1+\gamma _E s+s (\psi (s)-s \psi '(s+1))\right)}{s^3}+\\
&&\frac{8 \left(\gamma _E+\frac{1}{1+s}+\psi '(s+1)-(1+s) \psi '(s+2)\right)}{(1+s)^2}-\frac{4 \left(\gamma _E+\frac{1}{2+s}+\psi (s+2)-(2+s) \psi '(s+3)\right)}{(2+s)^2}+\\
&&\frac{4 \left(\gamma _E+\frac{1}{3+s}+\psi (s+3)-(3+s) \psi '(s+4)\right)}{(3+s)^2}-\\
&&\frac{0.3174}{(6+s)^2} \left(\text{Ln}(16)-2 \psi \left(4+\frac{s}{2}\right)+2 \psi \left(\frac{7+s}{2}\right)+(6+s) \psi '\left(4+\frac{s}{2}\right)-\right.\allowdisplaybreaks[1]\\
&&\left.(6+s) \psi '\left(\frac{7+s}{2}\right)\right)+\frac{0.0699}{(7+s)^2} \left(\text{Ln}(16)+2 \psi \left(4+\frac{s}{2}\right)-2 \psi \left(\frac{9+s}{2}\right)-(7+s) \psi '\left(4+\frac{s}{2}\right)+\right.\\
&&\left.(7+s) \psi '\left(\frac{9+s}{2}\right)\right)-4 \left(\left(\gamma _E+\psi '(s+1)\right) \psi '(s+1)-\frac{1}{2} \psi ''(s+1)\right)-\left.\psi ''(s+1)\right)\,.
\end{eqnarray*}

\end{widetext}

%
%%%%%%%%%%%%%%%%%%%%%%%%%%%%%%%%%%%%%%%%%%    Appendix B    %%%%%%%%%%%%%%%%%%%%%%%%%%%%%%%%%%%%%%%%%%%%%%%%%
%
\section*{Appendix B: The coefficient functions of singlet and non-singlet distributions in the Laplace s space at the NLO approximation}\label{Sec:AppendixB}
We present here the Laplace transformed for the coefficient functions of the singlet and gluon distributions which we used in Eq.\eqref{eq:NLODGLAPSspace}.

\begin{widetext}

\begin{eqnarray*}
&&k_{ff}=\hspace{14 cm}\text{(B.1)}\\
&&\left(e^{\frac{1}{2} \tau  \left(-2 b_1+\Phi _f+\Phi _g-R\right)} \right.\\
&&\left(b_1 \left(a_1 \Theta _f \Theta _g \left(-\Phi _f-\Phi _g+R\right)+a_1 e^{\tau  R} \Theta _f \Theta _g \left(\Phi _f+\Phi _g+R\right)+b_1{}^2
e^{b_1 \tau } \left(-\left(-1+e^{\tau  R}\right) \left(\Phi _f-\Phi _g\right)-R-e^{\tau  R} R\right)+\right.\right.\\
&&e^{\tau  \left(b_1+R\right)} \left(\Phi _f{}^3+\Phi _f{}^2 \left(-3 \Phi _g+R\right)+\left(\Phi _g{}^2-\left(-4+a_1\right) \Theta _f \Theta
_g\right) \left(-\Phi _g+R\right)+\Phi _f \left(3 \Phi _g{}^2+\left(4+a_1\right) \Theta _f \Theta _g-2 \Phi _g R\right)\right)+\\
&&\left.e^{b_1 \tau } \left(-\Phi _f{}^3+\left(\Phi _g{}^2-\left(-4+a_1\right) \Theta _f \Theta _g\right) \left(\Phi _g+R\right)+\Phi _f{}^2 \left(3
\Phi _g+R\right)-\Phi _f \left(3 \Phi _g{}^2+\left(4+a_1\right) \Theta _f \Theta _g+2 \Phi _g R\right)\right)\right)+\\
&&4 a_1 e^{\frac{1}{2} \tau  \left(b_1+R\right)} \left(R \left(-b_1{}^2 \Phi _f+\Phi _f \left(\Phi _f-\Phi _g\right){}^2+\left(3 \Phi _f-\Phi
_g\right) \Theta _f \Theta _g\right) \text{Cosh}\left[\frac{1}{2} \tau  R\right] \text{Sinh}\left[\frac{b_1 \tau }{2}\right]+\right.\\
&&\left.\left.\left.\left.\left(b_1{}^2 \Theta _f \Theta _g \text{Cosh}\left[\frac{b_1 \tau }{2}\right]+\left(\Phi _f{}^2-\Phi _f \Phi _g+\Theta
_f \Theta _g\right) \left(-b_1{}^2+R^2\right) \text{Sinh}\left[\frac{b_1 \tau }{2}\right]\right) \text{Sinh}\left[\frac{1}{2} \tau  R\right]\right)\right)\right)\right/\\
&&\left(2 b_1 R \left(-b_1{}^2+R^2\right)\right),
\end{eqnarray*}

\begin{eqnarray*}
&&k_{fg}=\hspace{14 cm}\text{(B.2)}\\
&&\left(e^{\frac{1}{2} \tau  \left(-2 b_1+\Phi _f+\Phi _g-R\right)} \right.\\
&&\left(b_1 \Theta _f\left(-2 b_1{}^2 e^{b_1 \tau } \left(-1+e^{\tau  R}\right)+e^{\tau  \left(b_1+R\right)} \left(2 \Phi _f{}^2-\left(4+a_1\right)
\Phi _f \Phi _g+\left(2+a_1\right) \Phi _g{}^2+2 \left(4+a_1\right) \Theta _f\Theta _g-a_1 \Phi _g R\right)-\right.\right.\\
&&e^{b_1 \tau } \left(2 \Phi _f{}^2-\left(4+a_1\right) \Phi _f \Phi _g+\left(2+a_1\right) \Phi _g{}^2+2 \left(4+a_1\right) \Theta _f\Theta _g+a_1
\Phi _g R\right)+a_1 \left(-2 \Theta _f\Theta _g+\Phi _g \left(\Phi _f-\Phi _g+R\right)\right)+\\
&&\left.a_1 e^{\tau  R} \left(2 \Theta _f\Theta _g+\Phi _g \left(-\Phi _f+\Phi _g+R\right)\right)\right)+\\
&&4 a_1 e^{\frac{1}{2} \tau  \left(b_1+R\right)} \Theta _f\left(\left(-b_1{}^2+\Phi _f{}^2-\Phi _f \Phi _g+2 \Theta _f\Theta _g\right) R \text{Cosh}\left[\frac{1}{2}
\tau  R\right] \text{Sinh}\left[\frac{b_1 \tau }{2}\right]+\right.\\
&&\left.\left.\left.\left(b_1{}^2 \Phi _g \text{Cosh}\left[\frac{b_1 \tau }{2}\right]+\Phi _f \left(-b_1{}^2+R^2\right) \text{Sinh}\left[\frac{b_1
\tau }{2}\right]\right) \text{Sinh}\left[\frac{1}{2} \tau  R\right]\right)\right)\right)/\left(2 b_1 R \left(-b_1{}^2+R^2\right)\right),
\end{eqnarray*}

\begin{eqnarray*}
&&k_{gf}=\hspace{14 cm}\text{(B.3)}\\
&&\left(2 e^{\frac{1}{2} \left(-b_1+\Phi _f+\Phi _g\right) \tau } \Theta _g \left(-a_1 \left(\left(b_1+\Phi _f-\Phi _g\right) \left(b_1+\Phi _g\right)-2
\Theta _f \Theta _g\right) R \text{Cosh}\left[\frac{1}{2} \tau  R\right] \text{Sinh}\left[\frac{b_1 \tau }{2}\right]+\right.\right.\\
&&\left(b_1 \left(-\left(b_1+\Phi _f-\Phi _g\right) \left(b_1-\left(1+a_1\right) \Phi _f+\Phi _g\right)+2 \left(2+a_1\right) \Theta _f \Theta
_g\right) \text{Cosh}\left[\frac{b_1 \tau }{2}\right]-\right.\\
&&\left.\left.\left.\left(b_1+a_1 \Phi _g\right) \left(b_1{}^2-\left(\Phi _f-\Phi _g\right){}^2-4 \Theta _f \Theta _g\right) \text{Sinh}\left[\frac{b_1
\tau }{2}\right]\right) \text{Sinh}\left[\frac{1}{2} \tau  R\right]\right)\right)/\left(b_1 R \left(-b_1{}^2+R^2\right)\right),
\end{eqnarray*}

\begin{eqnarray*}
&&k_{gg}=\hspace{14 cm}\text{(B.4)}\\
&&\left(e^{\frac{1}{2} \tau  \left(-2 b_1+\Phi _f+\Phi _g-R\right)} \right.\\
&&\left(b_1 \left(a_1 \Theta _f \Theta _g \left(-\Phi _f-\Phi _g+R\right)+a_1 e^{\tau  R} \Theta _f \Theta _g \left(\Phi _f+\Phi _g+R\right)+b_1{}^2
e^{b_1 \tau } \left(\left(-1+e^{\tau  R}\right) \Phi _f+\Phi _g-R-e^{\tau  R} \left(\Phi _g+R\right)\right)+\right.\right.\\
&&e^{b_1 \tau } \left(\Phi _f{}^3-\Phi _g{}^3-\left(4+a_1\right) \Phi _g \Theta _f \Theta _g+\Phi _g{}^2 R-\left(-4+a_1\right) \Theta _f \Theta
_g R+\Phi _f{}^2 \left(-3 \Phi _g+R\right)+\right.\\
&&\left.\Phi _f \left(3 \Phi _g{}^2-\left(-4+a_1\right) \Theta _f \Theta _g-2 \Phi _g R\right)\right)+\\
&&e^{\tau  \left(b_1+R\right)} \left(-\Phi _f{}^3+\Phi _g{}^3+\left(4+a_1\right) \Phi _g \Theta _f \Theta _g+\Phi _g{}^2 R-\left(-4+a_1\right)
\Theta _f \Theta _g R+\Phi _f{}^2 \left(3 \Phi _g+R\right)+\right.\\
&&\left.\left.\Phi _f \left(-3 \Phi _g{}^2+\left(-4+a_1\right) \Theta _f \Theta _g-2 \Phi _g R\right)\right)\right)+\\
&&4 a_1 e^{\frac{1}{2} \tau  \left(b_1+R\right)} \left(R \left(-b_1{}^2 \Phi _g+\left(\Phi _f-\Phi _g\right){}^2 \Phi _g-\left(\Phi _f-3 \Phi
_g\right) \Theta _f \Theta _g\right) \text{Cosh}\left[\frac{1}{2} \tau  R\right] \text{Sinh}\left[\frac{b_1 \tau }{2}\right]+\right.\\
&&\left.\left.\left.\left.\left(b_1{}^2 \Theta _f \Theta _g \text{Cosh}\left[\frac{b_1 \tau }{2}\right]+\left(b_1{}^2-\left(\Phi _f-\Phi _g\right){}^2-4
\Theta _f \Theta _g\right) \left(\left(\Phi _f-\Phi _g\right) \Phi _g-\Theta _f \Theta _g\right) \text{Sinh}\left[\frac{b_1 \tau }{2}\right]\right)
\text{Sinh}\left[\frac{1}{2} \tau  R\right]\right)\right)\right)\right/\\
&&\left(2 b_1 R \left(-b_1{}^2+R^2\right)\right) \,,
\end{eqnarray*}

where $R$ is defined as,
\begin{eqnarray*}
R = \sqrt{\left(\Phi _f-\Phi _g\right){}^2+4 \Theta _f \Theta _g}\,. \hspace{8 cm}\text{(B.5)}
\end{eqnarray*}
\end{widetext}

%%%%%%%%%%%%%%%%%%%%%%%%%%%%%%%%%%%%%%%%%%%%%%%%%%%%%%%%%%%%%%%%%%%%%%%%%%%%%%%%%%%%%%%%%%%%%%%%%%%%%%%%%%%%%%%%%%
%%%%%%%%%%%%%%%%%%%%%%%%%%%%%%%%%%%%%%%%%%%%%%%%%%%%%%%%%%%%%%%%%%%%%%%%%%%%%%%%%%%%%%%%%%%%%%%%%%%%%%%%%%%%%%%%%%

%
% BibTeX users please use
% \bibliographystyle{}
% \bibliography{}

\begin{thebibliography}{}


%\cite{Dokshitzer:1977sg}
\bibitem{Dokshitzer:1977sg}
Y.~L.~Dokshitzer,
%``Calculation of the Structure Functions for Deep Inelastic Scattering and e+ e- Annihilation by Perturbation Theory in Quantum Chromodynamics.,''
Sov.\ Phys.\ JETP {\bf 46}, 641 (1977)
[Zh.\ Eksp.\ Teor.\ Fiz.\  {\bf 73}, 1216 (1977)].
%%CITATION = SPHJA,46,641;%%
%2803 citations counted in INSPIRE as of 08 Dec 2015



%\cite{Gribov:1972ri}
\bibitem{Gribov:1972ri}
V.~N.~Gribov and L.~N.~Lipatov,
%``Deep inelastic e p scattering in perturbation theory,''
Sov.\ J.\ Nucl.\ Phys.\  {\bf 15}, 438 (1972)
[Yad.\ Fiz.\  {\bf 15}, 781 (1972)].
%%CITATION = SJNCA,15,438;%%
%3162 citations counted in INSPIRE as of 08 Dec 2015



%\cite{Lipatov:1974qm}
\bibitem{Lipatov:1974qm}
L.~N.~Lipatov,
%``The parton model and perturbation theory,''
Sov.\ J.\ Nucl.\ Phys.\  {\bf 20}, 94 (1975)
[Yad.\ Fiz.\  {\bf 20}, 181 (1974)].
%%CITATION = SJNCA,20,94;%%
%1171 citations counted in INSPIRE as of 08 Dec 2015




%\cite{Altarelli:1977zs}
\bibitem{Altarelli:1977zs}
G.~Altarelli and G.~Parisi,
%``Asymptotic Freedom in Parton Language,''
\href{http://dx.doi.org/10.1016/0550-3213(77)90384-4}{{\rm Nucl.\ Phys.\ B} {\bfseries 126}, 298 (1977)}.
%Nucl.\ Phys.\ B {\bf 126}, 298 (1977).
%doi:10.1016/0550-3213(77)90384-4
%%CITATION = doi:10.1016/0550-3213(77)90384-4;%%
%5469 citations counted in INSPIRE as of 08 Dec 2015



%\cite{Abramowicz:2016ztw}
\bibitem{Abramowicz:2016ztw}
H.~Abramowicz {\it et al.} [ZEUS Collaboration],
%``Combined QCD and electroweak analysis of HERA data,''
\href{http://dx.doi.org/10.1103/PhysRevD.93.092002}{{\rm Phys.\ Rev.\ D} {\bfseries 93}, no. 9, 092002 (2016)}.
%Phys.\ Rev.\ D {\bf 93}, no. 9, 092002 (2016)
%doi:10.1103/PhysRevD.93.092002
%[arXiv:1603.09628 [hep-ex]].
%%CITATION = doi:10.1103/PhysRevD.93.092002;%%
%5 citations counted in INSPIRE as of 23 Sep 2016



%\cite{Abt:2016zth}
\bibitem{Abt:2016zth}
I.~Abt, A.~M.~Cooper-Sarkar, B.~Foster, C.~Gwenlan, V.~Myronenko, O.~Turkot and K.~Wichmann,
%``Combined Electroweak and QCD Fit to HERA Data,''
\href{http://link.aps.org/doi/10.1103/PhysRevD.94.052007}{{\rm Phys.\ Rev.\ D} {\bfseries 94}, no. 5, 052007 (2016)}.
%Phys.\ Rev.\ D {\bf 94}, no. 5, 052007 (2016)
%doi:10.1103/PhysRevD.94.052007
%[arXiv:1604.05083 [hep-ex]].
%%CITATION = doi:10.1103/PhysRevD.94.052007;%%


%\cite{Abramowicz:2015mha}
\bibitem{Abramowicz:2015mha}
H.~Abramowicz {\it et al.} [H1 and ZEUS Collaborations],
%``Combination of measurements of inclusive deep inelastic ${e^{\pm }p}$ scattering cross sections and QCD analysis of HERA data,''
\href{http://dx.doi.org/10.1140/epjc/s10052-015-3710-4}{{\rm Eur.\ Phys.\ J. C} {\bfseries 75}, no. 12, 580 (2015)}.
%Eur.\ Phys.\ J.\ C {\bf 75}, no. 12, 580 (2015).
%doi:10.1140/epjc/s10052-015-3710-4
%[arXiv:1506.06042 [hep-ex]].
%%CITATION = doi:10.1140/epjc/s10052-015-3710-4;%%
%26 citations counted in INSPIRE as of 25 Dec 2015



%\cite{Aaron:2009kv}
\bibitem{Aaron:2009kv}
F.~D.~Aaron {\it et al.}  [H1 Collaboration],
%``A Precision Measurement of the Inclusive ep Scattering Cross Section at
%HERA,''
\href{http://dx.doi.org/10.1140/epjc/s10052-009-1169-x}{{\rm Eur.\ Phys.\ J.\ C} {\bfseries 64}, no. 12, 561 (2009)}.
%Eur.\ Phys.\ J.\  C {\bf 64}, 561 (2009).
%[arXiv:0904.3513 [hep-ex]].
%%CITATION = EPHJA,C64,561;%%


%\cite{Aaron:2009bp}
\bibitem{Aaron:2009bp}
F.~D.~Aaron {\it et al.} [H1 Collaboration],
%``Measurement of the Inclusive ep Scattering Cross Section at Low Q^2 and x at HERA,''
\href{http://dx.doi.org/10.1140/epjc/s10052-009-1128-6}{{\rm Eur.\ Phys.\ J.\ C} {\bfseries 63}, no. 12, 625 (2009)}.
%Eur.\ Phys.\ J.\ C {\bf 63}, 625 (2009).
%doi:10.1140/epjc/s10052-009-1128-6
%[arXiv:0904.0929 [hep-ex]].
%%CITATION = doi:10.1140/epjc/s10052-009-1128-6;%%
%103 citations counted in INSPIRE as of 25 Dec 2015



%\cite{Aaron:2009aa}
\bibitem{Aaron:2009aa}
F.~D.~Aaron {\it et al.} [H1 and ZEUS Collaborations],
%``Combined Measurement and QCD Analysis of the Inclusive e+- p Scattering Cross Sections at HERA,''
\href{http://dx.doi.org/10.1007/JHEP01(2010)109}{{\rm JHEP} {\bfseries 1001}, 109 (2010)}.
%JHEP {\bf 1001}, 109 (2010)
%doi:10.1007/JHEP01(2010)109
%[arXiv:0911.0884 [hep-ex]].
%%CITATION = doi:10.1007/JHEP01(2010)109;%%
%757 citations counted in INSPIRE as of 24 Aug 2016


%\cite{Adolph:2015saz}
\bibitem{Adolph:2015saz}
C.~Adolph {\it et al.} [COMPASS Collaboration],
%``The spin structure function $g_1^{\rm p}$ of the proton and a test of the Bjorken sum rule,''
\href{http://dx.doi.org/10.1016/j.physletb.2015.11.064}{{\rm Phys.\ Lett.\ B} {\bfseries 753}, 18 (2016)}.
%Phys.\ Lett.\ B {\bf 753}, 18 (2016).
%doi:10.1016/j.physletb.2015.11.064
%[arXiv:1503.08935 [hep-ex]].
%%CITATION = doi:10.1016/j.physletb.2015.11.064;%%
%6 citations counted in INSPIRE as of 25 Dec 2015




%\cite{Aaltonen:2008eq}
\bibitem{Aaltonen:2008eq}
T.~Aaltonen {\it et al.} [CDF Collaboration],
%``Measurement of the Inclusive Jet Cross Section at the Fermilab Tevatron p anti-p Collider Using a Cone-Based Jet Algorithm,''
\href{http://dx.doi.org/10.1103/PhysRevD.78.052006}{{\rm Phys.\ Rev.\ D} {\bfseries 78}, 052006 (2008)}.
[\href{http://dx.doi.org/10.1103/PhysRevD.79.119902}{{\rm Phys.\ Rev.\ D} {\bfseries 79}, 119902(E) (2009)}].
%Phys.\ Rev.\ D {\bf 78}, 052006 (2008)
%[Phys.\ Rev.\ D {\bf 79}, 119902 (2009)].
%doi:10.1103/PhysRevD.79.119902, 10.1103/PhysRevD.78.052006
%[arXiv:0807.2204 [hep-ex]].
%%CITATION = doi:10.1103/PhysRevD.79.119902, 10.1103/PhysRevD.78.052006;%%
%168 citations counted in INSPIRE as of 25 Dec 2015


%\cite{Abazov:2008ae}
\bibitem{Abazov:2008ae}
V.~M.~Abazov {\it et al.} [D0 Collaboration],
%``Measurement of the inclusive jet cross-section in $p \bar{p}$ collisions at $s^{(1/2)}$ =1.96-TeV,''
\href{http://dx.doi.org/10.1103/PhysRevLett.101.062001}{{\rm Phys.\ Rev.\ Lett.} {\bfseries 101}, 062001 (2008)}.
%Phys.\ Rev.\ Lett.\  {\bf 101}, 062001 (2008).
%doi:10.1103/PhysRevLett.101.062001
%[arXiv:0802.2400 [hep-ex]].
%%CITATION = doi:10.1103/PhysRevLett.101.062001;%%
%230 citations counted in INSPIRE as of 25 Dec 2015


%\cite{Abulencia:2007ez}
\bibitem{Abulencia:2007ez}
A.~Abulencia {\it et al.} [CDF Collaboration],
%``Measurement of the Inclusive Jet Cross Section using the {\boldmath $k_{\rm T}$} algorithmin{\boldmath $p\overline{p}$} Collisions at{\boldmath $\sqrt{s}$} = 1.96 TeV with the CDF II Detector,''
\href{http://dx.doi.org/10.1103/PhysRevD.75.092006}{{\rm Phys.\ Rev.\ D} {\bfseries 75}, 092006 (2007)}.\\
\href{http://dx.doi.org/10.1103/PhysRevD.75.119901}{{\rm Phys.\ Rev.\ D} {\bfseries 75}, 119901(E) (2007)}.
%Phys.\ Rev.\ D {\bf 75}, 092006 (2007)
%[Phys.\ Rev.\ D {\bf 75}, 119901 (2007)].
%doi:10.1103/PhysRevD.75.119901, 10.1103/PhysRevD.75.092006
%[hep-ex/0701051].
%%CITATION = doi:10.1103/PhysRevD.75.119901, 10.1103/PhysRevD.75.092006;%%
%164 citations counted in INSPIRE as of 25 Dec 2015


%\cite{Abbott:2000ew}
\bibitem{Abbott:2000ew}
B.~Abbott {\it et al.} [D0 Collaboration],
%``Inclusive jet production in $p\bar{p}$ collisions,''
\href{http://dx.doi.org/10.1103/PhysRevLett.86.1707}{{\rm Phys.\ Rev.\ Lett.} {\bfseries 86}, 1707 (2001)}.
%Phys.\ Rev.\ Lett.\  {\bf 86}, 1707 (2001).
%doi:10.1103/PhysRevLett.86.1707
%[hep-ex/0011036].
%%CITATION = doi:10.1103/PhysRevLett.86.1707;%%
%98 citations counted in INSPIRE as of 25 Dec 2015


%\cite{Onengut:2005kv}
\bibitem{Onengut:2005kv}
G.~Onengut {\it et al.} [CHORUS Collaboration],
%``Measurement of nucleon structure functions in neutrino scattering,''
\href{http://dx.doi.org/10.1016/j.physletb.2005.10.062}{{\rm Phys.\ Lett.\ B} {\bfseries 632}, 65 (2006)}.
%Phys.\ Lett.\ B {\bf 632}, 65 (2006).
%doi:10.1016/j.physletb.2005.10.062
%%CITATION = doi:10.1016/j.physletb.2005.10.062;%%
%84 citations counted in INSPIRE as of 25 Dec 2015

%\cite{Tzanov:2005kr}
\bibitem{Tzanov:2005kr}
M.~Tzanov {\it et al.} [NuTeV Collaboration],
%``Precise measurement of neutrino and anti-neutrino differential cross sections,''
\href{http://dx.doi.org/10.1103/PhysRevD.74.012008}{{\rm Phys.\ Rev.\ D} {\bfseries 74}, 012008 (2006)}.
%Phys.\ Rev.\ D {\bf 74}, 012008 (2006).
%doi:10.1103/PhysRevD.74.012008
%[hep-ex/0509010].
%%CITATION = doi:10.1103/PhysRevD.74.012008;%%
%127 citations counted in INSPIRE as of 25 Dec 2015


%\cite{Collaboration:2010ry}
\bibitem{Collaboration:2010ry}
F.~D.~Aaron {\it et al.} [H1 Collaboration],
%``Measurement of the Inclusive e{\pm}p Scattering Cross Section at High Inelasticity y and of the Structure Function $F_L$,''
\href{http://dx.doi.org/10.1140/epjc/s10052-011-1579-4}{{\rm Eur.\ Phys.\ J.\ C} {\bfseries 71}, 1579 (2011)}.
%Eur.\ Phys.\ J.\ C {\bf 71}, 1579 (2011).
%doi:10.1140/epjc/s10052-011-1579-4
%[arXiv:1012.4355 [hep-ex]].
%%CITATION = doi:10.1140/epjc/s10052-011-1579-4;%%
%82 citations counted in INSPIRE as of 25 Dec 2015


%\cite{Harland-Lang:2014zoa}
\bibitem{Harland-Lang:2014zoa}
L.~A.~Harland-Lang, A.~D.~Martin, P.~Motylinski and R.~S.~Thorne,
%``Parton distributions in the LHC era: MMHT 2014 PDFs,''
\href{http://dx.doi.org/10.1140/epjc/s10052-015-3397-6}{{\rm Eur.\ Phys.\ J.\ C} {\bfseries 75}, no. 5, 204 (2015)}.
%Eur.\ Phys.\ J.\ C {\bf 75}, no. 5, 204 (2015).
%[arXiv:1412.3989 [hep-ph]].
%%CITATION = ARXIV:1412.3989;%%
%14 citations counted in INSPIRE as of 24 Apr 2015


%\cite{Khanpour:2012tk}
\bibitem{Khanpour:2012tk}
H.~Khanpour, A.~N.~Khorramian and S.~A.~Tehrani,
%``New parton distributions in fixed flavour factorization scheme from recent deep-inelastic-scattering data,''
\href{http://dx.doi.org/10.1088/0954-3899/40/4/045002}{{\rm J.\ Phys.\ G} {\bfseries 40}, 045002 (2013)}.
%J.\ Phys.\ G {\bf 40} , 045002 (2013) .
%[arXiv:1205.5194 [hep-ph]].
%%CITATION = ARXIV:1205.5194;%%
%4 citations counted in INSPIRE as of 31 Mar 2015


%\cite{Alekhin:2012ig}
\bibitem{Alekhin:2012ig}
S.~Alekhin, J.~Blumlein and S.~Moch,
%``Parton Distribution Functions and Benchmark Cross Sections at NNLO,''
\href{http://dx.doi.org/10.1103/PhysRevD.86.054009}{{\rm Phys.\ Rev.\ D} {\bfseries 86},  054009 (2012)}.
%Phys.\ Rev.\ D {\bf 86}, 054009 (2012).
%%  [arXiv:1202.2281 [hep-ph]].
%%CITATION = ARXIV:1202.2281;%%
%201 citations counted in INSPIRE as of 24 Apr 2015


%\cite{::2014uva}
\bibitem{::2014uva}
P.~Belov {\it et al.}  [HERAFitter developers' Team Collaboration],
%``Parton distribution functions at LO, NLO and NNLO with correlated uncertainties between orders,''
\href{http://dx.doi.org/10.1140/epjc/s10052-014-3039-4}{{\rm Eur.\ Phys.\ J.\ C} {\bfseries 74}, no. 10, 3039 (2014)}.
%Eur.\ Phys.\ J.\ C {\bf 74}, no. 10, 3039 (2014).
%%  [arXiv:1404.4234 [hep-ph]].
%%CITATION = ARXIV:1404.4234;%%
%7 citations counted in INSPIRE as of 24 Apr 2015


%\cite{Buckley:2014ana}
\bibitem{Buckley:2014ana}
  A.~Buckley, J.~Ferrando, S.~Lloyd, K.~Nordstr\"{o}m, B.~Page, M.~R\"{u}fenacht, M.~Sch\"{o}nherr and G.~Watt,
  %``LHAPDF6: parton density access in the LHC precision era,''
  \href{http://dx.doi.org/10.1140/epjc/s10052-015-3318-8}{{\rm Eur.\ Phys.\ J.\ C} {\bfseries 75}, 132 (2015)}.
 % Eur.\ Phys.\ J.\ C {\bf 75}, 132 (2015).
 % doi:10.1140/epjc/s10052-015-3318-8
 % [arXiv:1412.7420 [hep-ph]].
  %%CITATION = doi:10.1140/epjc/s10052-015-3318-8;%%
  %46 citations counted in INSPIRE as of 10 Jan 2016

%\cite{Ball:2014uwa}
\bibitem{Ball:2014uwa}
R.~D.~Ball {\it et al.} [NNPDF Collaboration],
%``Parton distributions for the LHC Run II,''
 \href{http://dx.doi.org/10.1007/JHEP04(2015)040}{{\rm JHEP} {\bfseries 1504}, 040 (2015)}.
%JHEP {\bf 1504}, 040 (2015).
%doi:10.1007/JHEP04(2015)040
%[arXiv:1410.8849 [hep-ph]].
%%CITATION = doi:10.1007/JHEP04(2015)040;%%
%117 citations counted in INSPIRE as of 09 Dec 2015


%\cite{Martin:2009iq}
\bibitem{Martin:2009iq}
A.~D.~Martin, W.~J.~Stirling, R.~S.~Thorne and G.~Watt,
%``Parton distributions for the LHC,''
 \href{http://dx.doi.org/10.1140/epjc/s10052-009-1072-5}{{\rm Eur.\ Phys.\ J.\ C} {\bfseries 63}, 189 (2009)}.
%Eur.\ Phys.\ J.\ C {\bf 63}, 189 (2009).
%doi:10.1140/epjc/s10052-009-1072-5
%[arXiv:0901.0002 [hep-ph]].
%%CITATION = doi:10.1140/epjc/s10052-009-1072-5;%%
%2976 citations counted in INSPIRE as of 09 Dec 2015


%\cite{JimenezDelgado:2008hf}
\bibitem{JimenezDelgado:2008hf}
P.~Jimenez-Delgado and E.~Reya,
%``Dynamical NNLO parton distributions,''
\href{http://dx.doi.org/10.1103/PhysRevD.79.074023}{{\rm Phys.\ Rev.\ D} {\bfseries 79},  074023 (2009)}.
%Phys.\ Rev.\ D {\bf 79}, 074023 (2009).
%doi:10.1103/PhysRevD.79.074023
%[arXiv:0810.4274 [hep-ph]].
%%CITATION = doi:10.1103/PhysRevD.79.074023;%%
%192 citations counted in INSPIRE as of 09 Dec 2015


%\cite{Block:2010du}
\bibitem{Block:2010du}
  M.~M.~Block, L.~Durand, P.~Ha and D.~W.~McKay,
  %``Decoupling the NLO coupled DGLAP evolution equations: an analytic solution to pQCD,''
   \href{http://dx.doi.org/10.1140/epjc/s10052-010-1413-4}{{\rm Eur.\ Phys.\ J.\ C} {\bfseries 69}, 425 (2010)}.
 % Eur.\ Phys.\ J.\ C {\bf 69}, 425 (2010)
 % doi:10.1140/epjc/s10052-010-1413-4
 % [arXiv:1005.2556 [hep-ph]].
  %%CITATION = doi:10.1140/epjc/s10052-010-1413-4;%%
  %9 citations counted in INSPIRE as of 18 Nov 2015


%\cite{Zarei:2015jvh}
\bibitem{Zarei:2015jvh}
M.~Zarei, F.~Taghavi-Shahri, S.~Atashbar Tehrani and M.~Sarbishei,
%``Fragmentation functions of the pion, kaon, and proton in the NLO approximation: Laplace transform approach,''
\href{http://dx.doi.org/10.1103/PhysRevD.92.074046}{{\rm Phys.\ Rev.\ D} {\bfseries 92},  074046 (2015)}.
%Phys.\ Rev.\ D {\bf 92}, 074046 (2015).
%doi:10.1103/PhysRevD.92.074046
%%CITATION = doi:10.1103/PhysRevD.92.074046;%%



%\cite{Block:2011xb}
\bibitem{Block:2011xb}
  M.~M.~Block, L.~Durand, P.~Ha and D.~W.~McKay,
  %``Applications of the leading-order Dokshitzer-Gribov-Lipatov-Altarelli-Parisi evolution equations to the combined HERA data on deep inelastic scattering,''
  \href{http://dx.doi.org/10.1103/PhysRevD.84.094010}{{\rm Phys.\ Rev.\ D} {\bfseries 84}, 094010 (2011)}.
 % Phys.\ Rev.\ D {\bf 84}, 094010 (2011)
%  doi:10.1103/PhysRevD.84.094010
%  [arXiv:1108.1232 [hep-ph]].
  %%CITATION = doi:10.1103/PhysRevD.84.094010;%%
  %9 citations counted in INSPIRE as of 19 Nov 2015





  %\cite{Block:2010fk}
\bibitem{Block:2010fk}
  M.~M.~Block, L.~Durand, P.~Ha and D.~W.~McKay,
  %``An Analytic solution to LO coupled DGLAP evolution equations: a new pQCD tool,''
   \href{http://dx.doi.org/10.1103/PhysRevD.83.054009}{{\rm Phys.\ Rev.\ D} {\bfseries 83}, 054009 (2011)}.
 % Phys.\ Rev.\ D {\bf 83}, 054009 (2011)
%  doi:10.1103/PhysRevD.83.054009
%  [arXiv:1010.2486 [hep-ph]].
  %%CITATION = doi:10.1103/PhysRevD.83.054009;%%
  %11 citations counted in INSPIRE as of 19 Nov 2015




%\cite{Block:2009en}
\bibitem{Block:2009en}
  M.~M.~Block,
  %``A New numerical method for obtaining gluon distribution functions G(x,Q**2) = xg(x,Q**2), from the proton structure function F(2)**gamma p(x,Q**2),''
  \href{http://dx.doi.org/10.1140/epjc/s10052-009-1195-8}{{\rm Eur.\ Phys.\ J.\ C} {\bfseries 65}, 1 (2010)}.
 % Eur.\ Phys.\ J.\ C {\bf 65}, 1 (2010)
%  doi:10.1140/epjc/s10052-009-1195-8
%  [arXiv:0907.4790 [hep-ph]].
  %%CITATION = doi:10.1140/epjc/s10052-009-1195-8;%%
  %18 citations counted in INSPIRE as of 19 Nov 2015




%\cite{Block:2010ti}
\bibitem{Block:2010ti}
  M.~M.~Block,
  %``Addendum to: `A new numerical method for obtaining gluon distribution functions $G(x,Q^2)=xg(x,Q^2)$, from the proton structure function $F_2^{\gamma p}(x,Q^2)$.',''
   \href{http://dx.doi.org/10.1140/epjc/s10052-010-1374-7}{{\rm Eur.\ Phys.\ J.\ C} {\bfseries 68}, 683 (2010)}.
  %Eur.\ Phys.\ J.\ C {\bf 68}, 683 (2010)
 % doi:10.1140/epjc/s10052-010-1374-7
 % [arXiv:1004.3585 [hep-ph]].
  %%CITATION = doi:10.1140/epjc/s10052-010-1374-7;%%
  %6 citations counted in INSPIRE as of 19 Nov 2015




%\cite{Block:2007pg}
\bibitem{Block:2007pg}
  M.~M.~Block, L.~Durand and D.~W.~McKay,
  %``Analytic derivation of the leading-order gluon distribution function G(x,Q**2) = xg(x,Q**2) from the proton structure function F(2)**p(x,Q**2),''
  \href{http://dx.doi.org/10.1103/PhysRevD.77.094003}{{\rm Phys.\ Rev.\ D} {\bfseries 77}, 094003 (2008)}.
%  Phys.\ Rev.\ D {\bf 77}, 094003 (2008)
%  doi:10.1103/PhysRevD.77.094003
%  [arXiv:0710.3212 [hep-ph]].
  %%CITATION = doi:10.1103/PhysRevD.77.094003;%%
  %26 citations counted in INSPIRE as of 18 Nov 2015




  %\cite{Block:2008xc}
\bibitem{Block:2008xc}
  M.~M.~Block, L.~Durand and D.~W.~McKay,
  %``Analytic treatment of leading-order parton evolution equations: Theory and tests,''
  \href{http://dx.doi.org/10.1103/PhysRevD.79.014031}{{\rm Phys.\ Rev.\ D} {\bfseries 79}, 014031 (2009)}.
 % Phys.\ Rev.\ D {\bf 79}, 014031 (2009)
%  doi:10.1103/PhysRevD.79.014031
 % [arXiv:0808.0201 [hep-ph]].
  %%CITATION = doi:10.1103/PhysRevD.79.014031;%%
  %14 citations counted in INSPIRE as of 18 Nov 2015


  %\cite{AtashbarTehrani:2013qea}
\bibitem{AtashbarTehrani:2013qea}
  S.~Atashbar Tehrani, F.~Taghavi-Shahri, A.~Mirjalili and M.~M.~Yazdanpanah,
  %``NLO analytical solutions to the polarized parton distributions, based on the Laplace transformation,''
  \href{http://dx.doi.org/10.1103/PhysRevD.87.114012}{{\rm Phys.\ Rev.\ D} {\bfseries 87}, no. 11, 114012 (2013)}.
  \href{http://dx.doi.org/10.1103/PhysRevD.88.039902}{{\rm Phys.\ Rev.\ D} {\bfseries 88},  no. 3, 039902(E) (2013)}.
 % Phys.\ Rev.\ D {\bf 87}, no. 11, 114012 (2013)
 % [Phys.\ Rev.\ D {\bf 88}, no. 3, 039902 (2013)].
%  doi:10.1103/PhysRevD.87.114012, 10.1103/PhysRevD.88.039902
  %%CITATION = doi:10.1103/PhysRevD.87.114012, 10.1103/PhysRevD.88.039902;%%
  %1 citations counted in INSPIRE as of 19 Nov 2015


%\cite{Boroun:2015cta}
\bibitem{Boroun:2015cta}
G.~R.~Boroun, S.~Zarrin and F.~Teimoury,
%``Decoupling of the DGLAP evolution equations by Laplace method,''
\href{http://dx.doi.org/10.1140/epjp/i2015-15214-2}{{\rm Eur.\ Phys.\ J.\ Plus} {\bfseries 130}, no. 10, 214 (2015)}.
%Eur.\ Phys.\ J.\ Plus {\bf 130}, no. 10, 214 (2015).
%doi:10.1140/epjp/i2015-15214-2
%[arXiv:1509.06520 [hep-ph]].
%%CITATION = doi:10.1140/epjp/i2015-15214-2;%%


%\cite{Boroun:2014dka}
\bibitem{Boroun:2014dka}
G.~R.~Boroun and B.~Rezaei,
%``Decoupling of the DGLAP evolution equations at next-to-next-to-leading order (NNLO) at low-x,''
\href{http://dx.doi.org/10.1140/epjc/s10052-013-2412-z}{{\rm Eur.\ Phys.\ J.\ C} {\bfseries 73}, 2412 (2013)}.
%Eur.\ Phys.\ J.\ C {\bf 73}, 2412 (2013).
%doi:10.1140/epjc/s10052-013-2412-z
%[arXiv:1402.0164 [hep-ph]].
%%CITATION = doi:10.1140/epjc/s10052-013-2412-z;%%
%5 citations counted in INSPIRE as of 09 Dec 2015




%\cite{Gluck:2007ck}
\bibitem{Gluck:2007ck}
M.~Gluck, P.~Jimenez-Delgado and E.~Reya,
%``Dynamical parton distributions of the nucleon and very small-x physics,''
\href{http://dx.doi.org/10.1140/epjc/s10052-007-0462-9}{{\rm Eur.\ Phys.\ J.\ C} {\bfseries 53}, 355 (2008)}.
%Eur.\ Phys.\ J.\ C {\bf 53}, 355 (2008)
%  [arXiv:0709.0614 [hep-ph]].
%%CITATION = ARXIV:0709.0614;%%
%186 citations counted in INSPIRE as of 29 Oct 2015


%\cite{Vermaseren:2005qc}
\bibitem{Vermaseren:2005qc}
J.~A.~M.~Vermaseren, A.~Vogt and S.~Moch,
%``The Third-order QCD corrections to deep-inelastic scattering by photon exchange,''
\href{http://dx.doi.org/10.1016/j.nuclphysb.2005.06.020}{{\rm Nucl.\ Phys.\ B} {\bfseries 724}, 3 (2005)}.
%Nucl.\ Phys.\ B {\bf 724}, 3 (2005)
% doi:10.1016/j.nuclphysb.2005.06.020
% [hep-ph/0504242].
%%CITATION = doi:10.1016/j.nuclphysb.2005.06.020;%%
%263 citations counted in INSPIRE as of 19 Nov 2015





%\cite{Curci:1980uw}
\bibitem{Curci:1980uw}
G.~Curci, W.~Furmanski and R.~Petronzio,
%``Evolution of Parton Densities Beyond Leading Order: The Nonsinglet Case,''
\href{http://dx.doi.org/10.1016/0550-3213(80)90003-6}{{\rm Nucl.\ Phys.\ B} {\bfseries 175}, 27 (1980)}.
%Nucl.\ Phys.\ B {\bf 175}, 27 (1980).
%  doi:10.1016/0550-3213(80)90003-6
%%CITATION = doi:10.1016/0550-3213(80)90003-6;%%
%918 citations counted in INSPIRE as of 19 Nov 2015



 %\cite{Furmanski:1980cm}
\bibitem{Furmanski:1980cm}
W.~Furmanski and R.~Petronzio,
%``Singlet Parton Densities Beyond Leading Order,''
\href{http://dx.doi.org/10.1016/0370-2693(80)90636-X}{{\rm Phys.\ Lett.\ B} {\bfseries 97}, 437 (1980)}.
%Phys.\ Lett.\ B {\bf 97}, 437 (1980).
% doi:10.1016/0370-2693(80)90636-X
%%CITATION = doi:10.1016/0370-2693(80)90636-X;%%
%517 citations counted in INSPIRE as of 19 Nov 2015



%\cite{Botje:2010ay}
\bibitem{Botje:2010ay}
M.~Botje,
%``QCDNUM: Fast QCD Evolution and Convolution,''
\href{http://dx.doi.org/10.1016/j.cpc.2010.10.020}{{\rm Comput.\ Phys.\ Commun.} {\bfseries 182}, 490 (2011)}.
%Comput.\ Phys.\ Commun.\  {\bf 182}, 490 (2011)
%doi:10.1016/j.cpc.2010.10.020
%[arXiv:1005.1481 [hep-ph]].
%%CITATION = doi:10.1016/j.cpc.2010.10.020;%%
%97 citations counted in INSPIRE as of 10 Jan 2016



%\cite{Floratos:1981hs}
\bibitem{Floratos:1981hs}
E.~G.~Floratos, C.~Kounnas and R.~Lacaze,
%``Higher Order QCD Effects in Inclusive Annihilation and Deep Inelastic Scattering,''
\href{http://dx.doi.org/10.1016/0550-3213(81)90434-X}{{\rm Nucl.\ Phys.\ B} {\bfseries 192}, 417 (1981)}.
%Nucl.\ Phys.\ B {\bf 192}, 417 (1981).
%doi:10.1016/0550-3213(81)90434-X
%%CITATION = doi:10.1016/0550-3213(81)90434-X;%%
%283 citations counted in INSPIRE as of 31 Dec 2015



%\cite{Gluck:2006pm}
\bibitem{Gluck:2006pm}
M.~Gluck, C.~Pisano and E.~Reya,
%``The curvature of F2(p)(x,Q**2) as a probe of the range of validity of
%perturbative QCD evolutions in the small-x region,''
\href{http://dx.doi.org/10.1140/epjc/s10052-006-0189-z}{{\rm Eur.\ Phys.\ J.\ C} {\bfseries 50}, 29 (2007)}.
%Eur.\ Phys.\ J.\  C {\bf 50}, 29 (2007).
%[arXiv:hep-ph/0610060].
%%CITATION = EPHJA,C50,29;%%



%\cite{Gluck:2004fi}
\bibitem{Gluck:2004fi}
M.~Gluck, C.~Pisano and E.~Reya,
%``Probing the perturbative NLO parton evolution in the small-x region,''
\href{http://dx.doi.org/10.1140/epjc/s2005-02167-3}{{\rm Eur.\ Phys.\ J.\ C} {\bfseries 40}, 515 (2005)}.
%Eur.\ Phys.\ J.\  C {\bf 40}, 515 (2005).
%[arXiv:hep-ph/0412049].
%%CITATION = EPHJA,C40,515;%%


%\cite{Gluck:2008gs}
\bibitem{Gluck:2008gs}
M.~Gluck, P.~Jimenez-Delgado, E.~Reya and C.~Schuck,
%``On the role of heavy flavour parton distributions at high energy
%colliders,''
\href{http://dx.doi.org/10.1016/j.physletb.2008.04.063}{{\rm Phys.\ Lett.\ B} {\bfseries 664}, 133 (2008)}.
%Phys.\ Lett.\  B {\bf 664}, 133 (2008).
%[arXiv:0801.3618 [hep-ph]].
%%CITATION = PHLTA,B664,133;%%


%\cite{Gluck:1993dpa}
\bibitem{Gluck:1993dpa}
M.~Gluck, E.~Reya and M.~Stratmann,
%``Heavy quarks at high-energy colliders,''
\href{http://dx.doi.org/10.1016/0550-3213(94)00131-6}{{\rm Nucl.\ Phys.\ B} {\bfseries 422}, 37 (1994)}.
%Nucl.\ Phys.\  B {\bf 422}, 37 (1994).
%%CITATION = NUPHA,B422,37;%%



%\cite{Laenen:1992cc}
\bibitem{Laenen:1992cc}
E.~Laenen, S.~Riemersma, J.~Smith and W.~L.~van Neerven,
%``On The Heavy Quark Content Of The Nucleon,''
\href{http://dx.doi.org/10.1016/0370-2693(92)91053-C}{{\rm Phys.\ Lett.\ B} {\bfseries 291}, 325 (1992)}.
%Phys.\ Lett.\  B {\bf 291}, 325 (1992).
%%CITATION = PHLTA,B291,325;%%


%\cite{Riemersma:1994hv}
\bibitem{Riemersma:1994hv}
S.~Riemersma, J.~Smith and W.~L.~van Neerven,
%``Rates for inclusive deep inelastic electroproduction of charm quarks at
%HERA,''
\href{http://dx.doi.org/10.1016/0370-2693(95)00036-K}{{\rm Phys.\ Lett.\ B} {\bfseries 347}, 143 (1995)}.
%Phys.\ Lett.\  B {\bf 347}, 143 (1995).
%[arXiv:hep-ph/9411431].
%%CITATION = PHLTA,B347,143;%%


%\cite{Laenen:1992zk}
\bibitem{Laenen:1992zk}
E.~Laenen, S.~Riemersma, J.~Smith and W.~L.~van Neerven,
%``Complete O (alpha-s) corrections to heavy flavour structure functions in
%electroproduction,''
\href{http://dx.doi.org/10.1016/0550-3213(93)90201-Y}{{\rm Nucl.\ Phys.\ B} {\bfseries 392}, 162 (1993)}.
%Nucl.\ Phys.\  B {\bf 392}, 162 (1993).
%%CITATION = NUPHA,B392,162;%%




%\cite{Furmanski:1981ja}
\bibitem{Furmanski:1981ja}
W.~Furmanski and R.~Petronzio,
%``A Method of Analyzing the Scaling Violation of Inclusive Spectra in Hard Processes,''
\href{http://dx.doi.org/10.1016/0550-3213(82)90398-4}{{\rm Nucl.\ Phys.\ B} {\bfseries 195}, 237 (1982)}.
%Nucl.\ Phys.\ B {\bf 195}, 237 (1982).
%doi:10.1016/0550-3213(82)90398-4
%%CITATION = doi:10.1016/0550-3213(82)90398-4;%%
%96 citations counted in INSPIRE as of 10 Jan 2016



%\cite{Adams:1996gu}
\bibitem{Adams:1996gu}
M.~R.~Adams {\it et al.} [E665 Collaboration],
%``Proton and deuteron structure functions in muon scattering at 470-GeV,''
 \href{http://dx.doi.org/10.1103/PhysRevD.54.3006}{{\rm Phys.\ Rev.\ D} {\bfseries 54}, 3006 (1996)}.
%Phys.\ Rev.\ D {\bf 54}, 3006 (1996).
%doi:10.1103/PhysRevD.54.3006
%%CITATION = doi:10.1103/PhysRevD.54.3006;%%
%319 citations counted in INSPIRE as of 12 Dec 2015




%\cite{Carrazza:2016htc}
\bibitem{Carrazza:2016htc}
S.~Carrazza, S.~Forte, Z.~Kassabov and J.~Rojo,
%``Specialized minimal PDFs for optimized LHC calculations,''
\href{http://dx.doi.org/10.1140/epjc/s10052-016-4042-8}{{\rm Eur.\ Phys.\ J.\ C} {\bfseries 76},  no. 4, 205 (2016)}.
%Eur.\ Phys.\ J.\ C {\bf 76}, no. 4, 205 (2016)
%doi:10.1140/epjc/s10052-016-4042-8
%[arXiv:1602.00005 [hep-ph]].
%%CITATION = doi:10.1140/epjc/s10052-016-4042-8;%%
%6 citations counted in INSPIRE as of 28 Aug 2016



%\cite{Mangano:2016jyj}
\bibitem{Mangano:2016jyj}
M.~L.~Mangano {\it et al.},
%``Physics at a 100 TeV pp collider: Standard Model processes,''
\href{https://arxiv.org/abs/1607.01831}{arXiv:1607.01831 [hep-ph]}.
%arXiv:1607.01831 [hep-ph].
%%CITATION = ARXIV:1607.01831;%%
%5 citations counted in INSPIRE as of 28 Aug 2016


%\cite{Pennington:2016dpj}
\bibitem{Pennington:2016dpj}
M.~R.~Pennington,
%``Evolving images of the proton: Hadron physics over the past 40 years,''
\href{http://dx.doi.org/10.1088/0954-3899/43/5/054001}{{\rm J.\ Phys.\ G} {\bfseries 43},  054001 (2016)}.
%J.\ Phys.\ G {\bf 43}, 054001 (2016)
%doi:10.1088/0954-3899/43/5/054001
%[arXiv:1604.01441 [hep-ph]].
%%CITATION = doi:10.1088/0954-3899/43/5/054001;%%
%2 citations counted in INSPIRE as of 28 Aug 2016



%\cite{Dulat:2015mca}
\bibitem{Dulat:2015mca}
S.~Dulat {\it et al.},
%``New parton distribution functions from a global analysis of quantum chromodynamics,''
\href{http://dx.doi.org/10.1103/PhysRevD.93.033006}{{\rm Phys.\ Rev.\ D} {\bfseries 93}, no. 3, 033006 (2016)}.
%Phys.\ Rev.\ D {\bf 93}, no. 3, 033006 (2016)
%doi:10.1103/PhysRevD.93.033006
%[arXiv:1506.07443 [hep-ph]].
%%CITATION = doi:10.1103/PhysRevD.93.033006;%%
%194 citations counted in INSPIRE as of 28 Aug 2016


%\cite{Jimenez-Delgado:2014twa}
\bibitem{Jimenez-Delgado:2014twa}
P.~Jimenez-Delgado and E.~Reya,
%``Delineating parton distributions and the strong coupling,''
\href{http://dx.doi.org/10.1103/PhysRevD.89.074049}{{\rm Phys.\ Rev.\ D} {\bfseries 89}, no. 7, 074049 (2014)}.
%Phys.\ Rev.\ D {\bf 89}, no. 7, 074049 (2014)
%doi:10.1103/PhysRevD.89.074049
%[arXiv:1403.1852 [hep-ph]].
%%CITATION = doi:10.1103/PhysRevD.89.074049;%%
%55 citations counted in INSPIRE as of 28 Aug 2016






%\cite{Carrazza:2015hva}
\bibitem{Carrazza:2015hva}
S.~Carrazza, J.~I.~Latorre, J.~Rojo and G.~Watt,
%``A compression algorithm for the combination of PDF sets,''
\href{http://dx.doi.org/10.1140/epjc/s10052-015-3703-3}{{\rm Eur.\ Phys.\ J.\ C} {\bfseries 75}, 474 (2015)}.
%Eur.\ Phys.\ J.\ C {\bf 75}, 474 (2015)
%doi:10.1140/epjc/s10052-015-3703-3
%[arXiv:1504.06469 [hep-ph]].
%%CITATION = doi:10.1140/epjc/s10052-015-3703-3;%%
%22 citations counted in INSPIRE as of 29 Aug 2016


%\cite{Rojo:2015acz}
\bibitem{Rojo:2015acz}
J.~Rojo {\it et al.},
%``The PDF4LHC report on PDFs and LHC data: Results from Run I and preparation for Run II,''
\href{http://dx.doi.org/10.1088/0954-3899/42/10/103103}{{\rm J.\ Phys.\ G} {\bfseries 42},  103103 (2015)}.
%J.\ Phys.\ G {\bf 42}, 103103 (2015)
%doi:10.1088/0954-3899/42/10/103103
%[arXiv:1507.00556 [hep-ph]].
%%CITATION = doi:10.1088/0954-3899/42/10/103103;%%
%35 citations counted in INSPIRE as of 29 Aug 2016


%\cite{DeRoeck:2011na}
\bibitem{DeRoeck:2011na}
A.~De Roeck and R.~S.~Thorne,
%``Structure Functions,''
\href{http://dx.doi.org/10.1016/j.ppnp.2011.06.001}{{\rm Prog.\ Part.\ Nucl.\ Phys.} {\bfseries 66},  727 (2011)}.
%Prog.\ Part.\ Nucl.\ Phys.\  {\bf 66}, 727 (2011)
%doi:10.1016/j.ppnp.2011.06.001
%[arXiv:1103.0555 [hep-ph]].
%%CITATION = doi:10.1016/j.ppnp.2011.06.001;%%
%46 citations counted in INSPIRE as of 29 Aug 2016



%\cite{McNulty:2016xtv}
\bibitem{McNulty:2016xtv}
R.~McNulty, R.~S.~Thorne and K.~Wichmann,
%``Summary of Structure Functions and PDFs Working Group,''
PoS DIS {\bf 2016}, 281 (2016),
\href{https://arxiv.org/abs/1608.08121}{arXiv:1608.08121 [hep-ph]}.
%arXiv:1608.08121 [hep-ph].
%%CITATION = ARXIV:1608.08121;%%




%\cite{Accardi:2016ndt}
\bibitem{Accardi:2016ndt}
A.~Accardi {\it et al.},
%``A Critical Appraisal and Evaluation of Modern PDFs,''
\href{http://dx.doi.org/10.1140/epjc/s10052-016-4285-4}{{\rm Eur.\ Phys.\ J.\ C} {\bfseries 76},  no. 8, 471 (2016)}.
%Eur.\ Phys.\ J.\ C {\bf 76}, no. 8, 471 (2016)
%doi:10.1140/epjc/s10052-016-4285-4
%[arXiv:1603.08906 [hep-ph]].
%%CITATION = doi:10.1140/epjc/s10052-016-4285-4;%%
%9 citations counted in INSPIRE as of 28 Aug 2016




%\cite{Accardi:2016qay}
\bibitem{Accardi:2016qay}
A.~Accardi, L.~T.~Brady, W.~Melnitchouk, J.~F.~Owens and N.~Sato,
%``Constraints on large-$x$ parton distributions from new weak boson production and deep-inelastic scattering data,''
\href{http://dx.doi.org/10.1103/PhysRevD.93.114017}{{\rm Phys.\ Rev.\ D} {\bfseries 93},  no. 11, 114017 (2016)}.
%Phys.\ Rev.\ D {\bf 93}, no. 11, 114017 (2016)
%doi:10.1103/PhysRevD.93.114017
%[arXiv:1602.03154 [hep-ph]].
%%CITATION = doi:10.1103/PhysRevD.93.114017;%%
%8 citations counted in INSPIRE as of 30 Aug 2016




%\cite{Khorramian:2009xz}
\bibitem{Khorramian:2009xz}
A.~N.~Khorramian, H.~Khanpour and S.~A.~Tehrani,
%``Nonsinglet parton distribution functions from the precise next-to-next-to-next-to leading order QCD fit,''
\href{http://dx.doi.org/10.1103/PhysRevD.81.014013}{{\rm Phys.\ Rev.\ D} {\bfseries 81}, 014013 (2010)}.
%Phys.\ Rev.\ D {\bf 81}, 014013 (2010)
%doi:10.1103/PhysRevD.81.014013
%[arXiv:0909.2665 [hep-ph]].
%%CITATION = doi:10.1103/PhysRevD.81.014013;%%
%19 citations counted in INSPIRE as of 05 Sep 2016






%\cite{Krivokhizhin:1987rz}
\bibitem{Krivokhizhin:1987rz}
V.~G.~Krivokhizhin, S.~P.~Kurlovich, V.~V.~Sanadze, I.~A.~Savin, A.~V.~Sidorov and N.~B.~Skachkov,
%``{QCD} Analysis of Singlet Structure Functions Using Jacobi Polynomials: The Description of the Method,''
\href{http://dx.doi.org/10.1007/BF01556164}{{\rm Z.\ Phys.\ C} {\bfseries 36}, 51 (1987)}.
%Z.\ Phys.\ C {\bf 36}, 51 (1987).
%doi:10.1007/BF01556164
%%CITATION = doi:10.1007/BF01556164;%%
%113 citations counted in INSPIRE as of 07 Sep 2016



%\cite{Krivokhizhin:1990ct}
\bibitem{Krivokhizhin:1990ct}
V.~G.~Krivokhizhin, S.~P.~Kurlovich, R.~Lednicky, S.~Nemecek, V.~V.~Sanadze, I.~A.~Savin, A.~V.~Sidorov and N.~B.~Skachkov,
\href{http://dx.doi.org/10.1007/BF01554485}{{\rm Z.\ Phys.\ C} {\bfseries 48}, 347 (1990)}.
%``Next-to-leading order QCD analysis of structure functions with the help of Jacobi polynomials,''
%Z.\ Phys.\ C {\bf 48}, 347 (1990).
%doi:10.1007/BF01554485
%%CITATION = doi:10.1007/BF01554485;%%
%88 citations counted in INSPIRE as of 07 Sep 2016



%%%%%%%%%%%%%%%%%%%%%%%%%%%%%%%%%%%%%%%%%   experimental  data  %%%%%%%%%%%%%%%%%%%%%%%%%


%\cite{Benvenuti:1989fm}
\bibitem{Benvenuti:1989fm}
A.~C.~Benvenuti {\it et al.} [BCDMS Collaboration],
%``A High Statistics Measurement of the Deuteron Structure Functions F2 (X, $Q^2$) and R From Deep Inelastic Muon Scattering at High $Q^2$,''
\href{http://dx.doi.org/10.1016/0370-2693(90)91231-Y}{{\rm Phys.\ Lett.\ B} {\bfseries 237}, 592 (1990)}.
%Phys.\ Lett.\ B {\bf 237}, 592 (1990).
%doi:10.1016/0370-2693(90)91231-Y
%%CITATION = doi:10.1016/0370-2693(90)91231-Y;%%
%414 citations counted in INSPIRE as of 04 Sep 2016




%\cite{Benvenuti:1989rh}
\bibitem{Benvenuti:1989rh}
A.~C.~Benvenuti {\it et al.} [BCDMS Collaboration],
%``A High Statistics Measurement of the Proton Structure Functions F(2) (x, Q**2) and R from Deep Inelastic Muon Scattering at High Q**2,''
\href{http://dx.doi.org/10.1016/0370-2693(89)91637-7}{{\rm Phys.\ Lett.\ B} {\bfseries 223}, 485 (1989)}.
%Phys.\ Lett.\ B {\bf 223}, 485 (1989).
%doi:10.1016/0370-2693(89)91637-7
%%CITATION = doi:10.1016/0370-2693(89)91637-7;%%
%694 citations counted in INSPIRE as of 04 Sep 2016




%\cite{Benvenuti:1989gs}
\bibitem{Benvenuti:1989gs}
A.~C.~Benvenuti {\it et al.} [BCDMS Collaboration],
%``A Comparison of the Structure Functions F2 of the Proton and the Neutron From Deep Inelastic Muon Scattering at High $Q^2$,''
\href{http://dx.doi.org/10.1016/0370-2693(90)91232-Z}{{\rm Phys.\ Lett.\ B} {\bfseries 237}, 599 (1990)}.
%Phys.\ Lett.\ B {\bf 237}, 599 (1990).
%doi:10.1016/0370-2693(90)91232-Z
%%CITATION = doi:10.1016/0370-2693(90)91232-Z;%%
%150 citations counted in INSPIRE as of 04 Sep 2016


%----------------------------------------------------------------------

%\cite{Whitlow:1991uw}
\bibitem{Whitlow:1991uw}
L.~W.~Whitlow, E.~M.~Riordan, S.~Dasu, S.~Rock and A.~Bodek,
%``Precise measurements of the proton and deuteron structure functions from a global analysis of the SLAC deep inelastic electron scattering cross-sections,''
\href{http://dx.doi.org/10.1016/0370-2693(92)90672-Q}{{\rm Phys.\ Lett.\ B} {\bfseries 282}, 475 (1992)}.
%Phys.\ Lett.\ B {\bf 282}, 475 (1992).
%doi:10.1016/0370-2693(92)90672-Q
%%CITATION = doi:10.1016/0370-2693(92)90672-Q;%%
%464 citations counted in INSPIRE as of 04 Sep 2016

%----------------------------------------------------------------------
%



%\cite{Arneodo:1996qe}
\bibitem{Arneodo:1996qe}
M.~Arneodo {\it et al.} [New Muon Collaboration],
%``Measurement of the proton and deuteron structure functions, F2(p) and F2(d), and of the ratio sigma-L / sigma-T,''
\href{http://dx.doi.org/10.1016/S0550-3213(96)00538-X}{{\rm Nucl.\ Phys.\ B} {\bfseries 483}, 3 (1997)}.
%Nucl.\ Phys.\ B {\bf 483}, 3 (1997)
%doi:10.1016/S0550-3213(96)00538-X
%[hep-ph/9610231].
%%CITATION = doi:10.1016/S0550-3213(96)00538-X;%%
%548 citations counted in INSPIRE as of 04 Sep 2016



%\cite{Arneodo:1995cq}
\bibitem{Arneodo:1995cq}
M.~Arneodo {\it et al.} [New Muon Collaboration],
%``Measurement of the proton and the deuteron structure functions, F2(p) and F2(d),''
\href{http://dx.doi.org/10.1016/0370-2693(95)01318-9}{{\rm Phys.\ Lett.\ B} {\bfseries 364}, 107 (1995)}.
%Phys.\ Lett.\ B {\bf 364}, 107 (1995)
%doi:10.1016/0370-2693(95)01318-9
%[hep-ph/9509406].
%%CITATION = doi:10.1016/0370-2693(95)01318-9;%%
%352 citations counted in INSPIRE as of 04 Sep 2016



\bibitem{H1}
%\cite{Adloff:2000qk}
%\bibitem{Adloff:2000qk}
C.~Adloff {\it et al.}  [H1 Collaboration],
%``Deep-inelastic inclusive e p scattering at low x and a determination of
%alpha(s),''
\href{http://dx.doi.org/10.1007/s100520100720}{{\rm Eur.\ Phys.\ J.\ C} {\bfseries 21}, 33 (2001)}.\\
%Eur.\ Phys.\ J.\ C {\bf 21} (2001) 33
%[arXiv:hep-ex/0012053];\\
%%CITATION = HEP-EX 0012053;%%
%\cite{Adloff:2003uh}
%\bibitem{Adloff:2003uh}
C.~Adloff {\it et al.}  [H1 Collaboration],
%``Measurement and QCD analysis of neutral and charged current cross  sections
%at HERA,''
\href{http://dx.doi.org/10.1140/epjc/s2003-01257-6}{{\rm Eur.\ Phys.\ J.\ C} {\bfseries 30}, 1 (2003)}.
%Eur.\ Phys.\ J.\ C {\bf 30} (2003) 1 [arXiv:hep-ex/0304003].
%%CITATION = HEP-EX 0304003;%%
%----------------------------------------------------------------------

\bibitem{ZEUS}
%\cite{Breitweg:1998dz}
%\bibitem{Breitweg:1998dz}
J.~Breitweg {\it et al.}  [ZEUS Collaboration],
%``ZEUS results on the measurement and phenomenology of F2 at low x and  low
%Q**2,''
\href{http://dx.doi.org/10.1007/s100529901084}{{\rm Eur.\ Phys.\ J.\ C} {\bfseries 7}, 609 (1999)}.\\
%Eur.\ Phys.\ J.\ C {\bf 7} (1999) 609
%[arXiv:hep-ex/9809005];\\
%%CITATION = HEP-EX 9809005;%%
%\cite{Chekanov:2001qu}
%\bibitem{Chekanov:2001qu}
S.~Chekanov {\it et al.}  [ZEUS Collaboration],
%``Measurement of the neutral current cross section and F2 structure  function
%for deep inelastic e+ p scattering at HERA,''
\href{http://dx.doi.org/10.1007/s100520100749}{{\rm Eur.\ Phys.\ J.\ C} {\bfseries 21}, 443 (2001)}.
%Eur.\ Phys.\ J.\ C {\bf 21} (2001) 443 [arXiv:hep-ex/0105090].




\bibitem{James:1975dr}
%\cite{James:1975dr}
%\bibitem{James:1975dr}
F.~James and M.~Roos,
%``Minuit: A System For Function Minimization And Analysis Of The Parameter
%Errors And Correlations,''
\href{http://dx.doi.org/10.1016/0010-4655(75)90039-9}{{\rm Comput.\ Phys.\ Commun.} {\bfseries 10}, 343 (1975)}.
%Comput.\ Phys.\ Commun.\  {\bf 10} (1975) 343;
%%CITATION = CPHCB,10,343;%%
%F.~James and M.~Roos, \textsc{Minuit}: Function Minimization and Error Analysis, CERN Program Library Long Writeup D506 (1994);
%F.~James and M.~Winkler, \textsc{Minuit} User's Guide: C++ Version (2004).
%%CITATION = doi:10.1016/0010-4655(75)90039-9;%%




%\cite{Pumplin:2001ct}
\bibitem{Pumplin:2001ct}
J.~Pumplin, D.~Stump, R.~Brock, D.~Casey, J.~Huston, J.~Kalk, H.~L.~Lai and W.~K.~Tung,
%``Uncertainties of predictions from parton distribution functions. 2. The Hessian method,''
\href{http://dx.doi.org/10.1103/PhysRevD.65.014013}{{\rm Phys.\ Rev.\ D} {\bfseries 65}, 014013 (2001)}.
%Phys.\ Rev.\ D {\bf 65}, 014013 (2001)
%doi:10.1103/PhysRevD.65.014013
%[hep-ph/0101032].
%%CITATION = doi:10.1103/PhysRevD.65.014013;%%
%282 citations counted in INSPIRE as of 04 Sep 2016




%\cite{Martin:2002aw}
\bibitem{Martin:2002aw}
A.~D.~Martin, R.~G.~Roberts, W.~J.~Stirling and R.~S.~Thorne,
%``Uncertainties of predictions from parton distributions. 1: Experimental errors,''
\href{http://dx.doi.org/10.1140/epjc/s2003-01196-2}{{\rm Eur.\ Phys.\ J.\ C} {\bfseries 28}, 455 (2003)}.
%Eur.\ Phys.\ J.\ C {\bf 28}, 455 (2003)
%doi:10.1140/epjc/s2003-01196-2
%[hep-ph/0211080].
%%CITATION = doi:10.1140/epjc/s2003-01196-2;%%
%483 citations counted in INSPIRE as of 04 Sep 2016





%\cite{Owens:2012bv}
\bibitem{Owens:2012bv}
J.~F.~Owens, A.~Accardi and W.~Melnitchouk,
%``Global parton distributions with nuclear and finite-$Q^2$ corrections,''
\href{http://dx.doi.org/10.1103/PhysRevD.87.094012}{{\rm Phys.\ Rev.\ D} {\bfseries 87},no. 9, 094012 (2013)}.
%Phys.\ Rev.\ D {\bf 87}, no. 9, 094012 (2013)
%doi:10.1103/PhysRevD.87.094012
%[arXiv:1212.1702 [hep-ph]].
%%CITATION = doi:10.1103/PhysRevD.87.094012;%%
%113 citations counted in INSPIRE as of 04 Sep 2016





%\cite{Khanpour:2016pph}
\bibitem{Khanpour:2016pph}
H.~Khanpour and S.~Atashbar Tehrani,
%``Global Analysis of Nuclear Parton Distribution Functions and Their Uncertainties at Next-to-Next-to-Leading Order,''
\href{http://dx.doi.org/10.1103/PhysRevD.93.014026}{{\rm Phys.\ Rev.\ D} {\bfseries 93},  no. 1, 014026 (2016)}.
% Phys.\ Rev.\ D {\bf 93}, no. 1, 014026 (2016)
% doi:10.1103/PhysRevD.93.014026
% [arXiv:1601.00939 [hep-ph]].
%%CITATION = doi:10.1103/PhysRevD.93.014026;%%
%3 citations counted in INSPIRE as of 09 Aug 2016




%\cite{Shahri:2016uzl}
\bibitem{Shahri:2016uzl}
F.~Taghavi-Shahri, H.~Khanpour, S.~Atashbar Tehrani and Z.~Alizadeh Yazdi,
%``Next-to-next-to-leading order QCD analysis of spin-dependent parton distribution functions and their uncertainties: Jacobi polynomials approach,''
\href{http://dx.doi.org/10.1103/PhysRevD.93.114024}{{\rm Phys.\ Rev.\ D} {\bfseries 93}, 114024 (2016)}.
%Phys.\ Rev.\ D {\bf 93}, no. 11, 114024 (2016)
%doi:10.1103/PhysRevD.93.114024
%[arXiv:1603.03157 [hep-ph]].
%%CITATION = doi:10.1103/PhysRevD.93.114024;%%
%2 citations counted in INSPIRE as of 14 Dec 2016



%\cite{Georgi:1976ve}
\bibitem{Georgi:1976ve}
H.~Georgi and H.~D.~Politzer,
%``Freedom at Moderate Energies: Masses in Color Dynamics,''
\href{http://dx.doi.org/10.1103/PhysRevD.14.1829}{{\rm Phys.\ Rev.\ D} {\bfseries 14}, 1829 (1976)}.
%Phys.\ Rev.\ D {\bf 14}, 1829 (1976).
%doi:10.1103/PhysRevD.14.1829
%%CITATION = doi:10.1103/PhysRevD.14.1829;%%
%974 citations counted in INSPIRE as of 05 Sep 2016





%\cite{Mirjalili:2012zz}
\bibitem{Mirjalili:2012zz}
A.~Mirjalili and M.~M.~Yazdanpanah,
%``Target mass correction on the $He^3$ polarized structure function,''
\href{http://dx.doi.org/10.1140/epja/i2012-12071-0}{{\rm Eur.\ Phys.\ J.\ A} {\bfseries 48}, 71 (2012)}.
%Eur.\ Phys.\ J.\ A {\bf 47}, 71 (2012).
%doi:10.1140/epja/i2012-12071-0
%%CITATION = doi:10.1140/epja/i2012-12071-0;%%
%2 citations counted in INSPIRE as of 22 Sep 2016



%\cite{Gluck:2006yz}
\bibitem{Gluck:2006yz}
M.~Gluck, E.~Reya and C.~Schuck,
%``Non-singlet QCD analysis of F(2)(x,Q**2) up to NNLO,''
\href{http://dx.doi.org/10.1016/j.nuclphysb.2006.07.015}{{\rm Nucl.\ Phys.\ B} {\bfseries 754}, 178 (2006)}.
%Nucl.\ Phys.\ B {\bf 754}, 178 (2006)
%doi:10.1016/j.nuclphysb.2006.07.015
%[hep-ph/0604116].
%%CITATION = doi:10.1016/j.nuclphysb.2006.07.015;%%
%46 citations counted in INSPIRE as of 05 Sep 2016




%\cite{Steffens:2012jx}
\bibitem{Steffens:2012jx}
F.~M.~Steffens, M.~D.~Brown, W.~Melnitchouk and S.~Sanches,
%``Parton distributions in the presence of target mass corrections,''
\href{http://dx.doi.org/10.1103/PhysRevC.86.065208}{{\rm Phys.\ Rev.\ C} {\bfseries 86}, 065208 (2012)}.
%Phys.\ Rev.\ C {\bf 86}, 065208 (2012)
%doi:10.1103/PhysRevC.86.065208
%[arXiv:1210.4398 [hep-ph]].
%%CITATION = doi:10.1103/PhysRevC.86.065208;%%
%14 citations counted in INSPIRE as of 05 Sep 2016



%\cite{Abt:2016vjh}
\bibitem{Abt:2016vjh}
I.~Abt, A.~M.~Cooper-Sarkar, B.~Foster, V.~Myronenko, K.~Wichmann and M.~Wing,
%``Study of HERA ep data at low Q$^2$ and low $x_{Bj}$ and the need for higher-twist corrections to standard perturbative QCD fits,''
\href{http://dx.doi.org/10.1103/PhysRevD.94.034032}{{\rm Phys.\ Rev.\ D} {\bfseries 94}, 034032 (2016)}.
%Phys.\ Rev.\ D {\bf 94}, no. 3, 034032 (2016)
%doi:10.1103/PhysRevD.94.034032
%[arXiv:1604.02299 [hep-ph]].
%%CITATION = doi:10.1103/PhysRevD.94.034032;%%
%11 citations counted in INSPIRE as of 12 Jan 2017



%\cite{Jimenez-Delgado:2013boa}
\bibitem{Jimenez-Delgado:2013boa}
P.~Jimenez-Delgado, A.~Accardi and W.~Melnitchouk,
%``Impact of hadronic and nuclear corrections on global analysis of spin-dependent parton distributions,''
\href{http://dx.doi.org/10.1103/PhysRevD.89.034025}{{\rm Phys.\ Rev.\ D} {\bfseries 89}, 034025 (2014)}.
%Phys.\ Rev.\ D {\bf 89}, no. 3, 034025 (2014)
%doi:10.1103/PhysRevD.89.034025
%[arXiv:1310.3734 [hep-ph]].
%%CITATION = doi:10.1103/PhysRevD.89.034025;%%
%33 citations counted in INSPIRE as of 12 Jan 2017



%\cite{Leader:2006xc}
\bibitem{Leader:2006xc}
E.~Leader, A.~V.~Sidorov and D.~B.~Stamenov,
%``Impact of CLAS and COMPASS data on Polarized Parton Densities and Higher Twist,''
\href{http://dx.doi.org/10.1103/PhysRevD.75.074027}{{\rm Phys.\ Rev.\ D} {\bfseries 75}, 074027 (2007)}.
%Phys.\ Rev.\ D {\bf 75}, 074027 (2007)
%doi:10.1103/PhysRevD.75.074027
%[hep-ph/0612360].
%%CITATION = doi:10.1103/PhysRevD.75.074027;%%
%93 citations counted in INSPIRE as of 12 Jan 2017



%\cite{Nath:2016phi}
\bibitem{Nath:2016phi}
N.~M.~Nath, A.~Mukharjee, M.~K.~Das and J.~K.~Sarma,
%``xF (3)(x,Q (2)) Structure Function and Gross-Llewellyn Smith Sum Rule with Nuclear Effect and Higher Twist Correction,''
\href{http://dx.doi.org/10.1088/0253-6102/66/6/663}{{\rm Commun.\ Theor.\ Phys.} {\bfseries 66}, 663 (2016)}.
%Commun.\ Theor.\ Phys.\  {\bf 66}, no. 6, 663 (2016).
%doi:10.1088/0253-6102/66/6/663
%%CITATION = doi:10.1088/0253-6102/66/6/663;%%



%\cite{Wei:2016far}
\bibitem{Wei:2016far}
S.~y.~Wei, Y.~k.~Song, K.~b.~Chen and Z.~t.~Liang,
%``Twist-4 contributions to semi-inclusive deeply inelastic scatterings with polarized beam and target,''
\href{https://arxiv.org/abs/1611.08688}{arXiv:1611.08688 [hep-ph]}.
%arXiv:1611.08688 [hep-ph].
%%CITATION = ARXIV:1611.08688;%%






%\cite{Blumlein:2006be}
\bibitem{Blumlein:2006be}
J.~Blumlein, H.~Bottcher and A.~Guffanti,
%``Non-singlet QCD analysis of deep inelastic world data at O(alpha(s)**3),''
\href{http://dx.doi.org/10.1016/j.nuclphysb.2007.03.035}{{\rm Nucl.\ Phys.\ B} {\bfseries 774}, 182 (2007)}.
%  Nucl.\ Phys.\ B {\bf 774}, 182 (2007)
%  doi:10.1016/j.nuclphysb.2007.03.035
%  [hep-ph/0607200].
%%CITATION = doi:10.1016/j.nuclphysb.2007.03.035;%%
%130 citations counted in INSPIRE as of 05 Sep 2016



%nnpdf
%\cite{Ball:2012cx}
\bibitem{Ball:2012cx}
R.~D.~Ball {\it et al.},
%``Parton distributions with LHC data,''
\href{http://dx.doi.org/10.1016/j.nuclphysb.2012.10.003}{{\rm Nucl.\ Phys.\ B} {\bfseries 867}, 244 (2013)}.
%Nucl.\ Phys.\ B {\bf 867}, 244 (2013).
%doi:10.1016/j.nuclphysb.2012.10.003
%[arXiv:1207.1303 [hep-ph]].
%%CITATION = doi:10.1016/j.nuclphysb.2012.10.003;%%
%602 citations counted in INSPIRE as of 15 Jul 2016





%----------------------------------------------------------------------

%%%%%%%%%%%%%%%%%%%%%%%%%%%%%%%%%%%%%%%%%%%%%%%%%%%%%%%%%%%%%%%%%%%%%%%%%


\end{thebibliography}
%
% Non-BibTeX users please use

\end{document}